%% file: Article_arXiv.tex
\documentclass[12pt,letterpaper]{article}
\pdfoutput=1 

\input{Structure.tex} 

\makeatletter
\normalsize
\makeatother

\DeclareFixedFont\trfont{OT1}{phv}{b}{sc}{11}

\title{ \centering\boldmath\LARGE\bfseries%
       Introductory Lectures on Extended Supergravities and Gaugings
       \bigskip
       }

\author[a,b]{Antonio Gallerati\,}
\author[a,c]{Mario Trigiante\,}
\bigskip
\affil[a]{\,Politecnico di Torino, Dipartimento DISAT -
       corso Duca degli Abruzzi 24, 10129 Torino.}
\affil[a]{\,Istituto Nazionale di Fisica Nucleare (INFN) - Sezione di Torino, Italy.
\\
\medskip}
\affil[b]{%
    \,\href{mailto:antonio.gallerati@polito.it}{\texttt{antonio.gallerati@polito.it}}
      }
\affil[c]{%
    \,\href{mailto:mario.trigiante@polito.it}{\texttt{mario.trigiante@polito.it}}
      }
\date{}

\begin{document}

\maketitle

\abstract{%
\noindent
In an ungauged supergravity theory, the presence of a \emph{scalar potential} is allowed only for the minimal $\N=1$ case. In extended supergravities, a non-trivial scalar potential can be introduced without explicitly breaking supersymmetry only through the so-called \emph{gauging procedure}. The latter consists in promoting a suitable global symmetry group to local symmetry to be gauged by the vector fields of the theory. Gauged supergravities provide a valuable approach to the study of superstring flux-compactifications and the construction of phenomenologically viable, string-inspired models. The aim of these lectures is to give a pedagogical introduction to the subject of gauged supergravities, covering just selected issues and discussing some of their applications.
}

\section{Introduction} \label{sec:1}
A long-standing problem of high energy theoretical physics is the formulation of a fundamental theory unifying the four interactions. Superstring theory in ten dimensions and M-theory in eleven seem to provide a promising theoretical framework where this unification could be achieved. However, there are many shortcomings originating from this theoretical formulation.\par
First of all, these kinds of theories are defined in dimensions $D>4$, and, since we live in a four-dimensional universe, a fundamental requirement for any predictable model is the presence of a mechanism of dimensional reduction from ten or eleven dimensions to four. Moreover, the non-perturbative dynamics of the theory is far from being understood, and there is no mechanism to select a vacuum state for our universe (i.e.\ it is not clear how to formulate a phenomenological viable description for the model). Finally, there are more symmetries than those observed experimentally. These models, in fact, encode Supersymmetry (SUSY), but our universe is not supersymmetric and its gauge interactions are well described, at our energy scales, by the Standard Model (SM). Therefore deriving a phenomenologically viable model from string/M-theory also requires the definition of suitable mechanisms of supersymmetry breaking.\par
\bigskip

\paragraph{Spontaneous compactification.}
The simplest way for deriving a four-dimensional theory from a higher dimensional one is through \emph{spontaneous compactification} which generalizes the original Kaluza-Klein (KK) compactification of five-dimensional general relativity on a circle. We consider the low-energy dynamics of superstring/M-theory on space-time solutions with geometry of the form
\begin{equation}
\Mink\times~\Msi\;,
\end{equation}
where $\Mink$ is the maximally symmetric four dimensional space-time with Lorentzian signature and $\Msi$ is a compact internal manifold.
The $D=10$ or $D=11$ fields, excitations of the microscopic fundamental theory, are expanded in normal modes ($Y_{(n)}$) on the internal manifold
\begin{equation}
\Phi(x^\mu,\,y^\alpha)\=\sum_{(n)} \Phi_{(n)}(x^\mu)\;Y_{(n)}(y^\alpha)\;,
\end{equation}
the coefficients $\Phi_{(n)}$ of this expansion describing massive fields in $\Mink$ with mass of the order of $\frac1R$, where $R$ is the ``size'' of the internal manifold $\Msi$. These are the Kaluza-Klein states, forming an infinite tower.\par
In many cases, a consistent truncation of the massless modes $\Phi_{(0)}$ is well described by a $D=4$ Supergravity theory (SUGRA), an effective field theory consistently describing superstring dynamics on the chosen background at energies $\Lambda$, where
\begin{equation}
\Lambda~\ll~\frac1R~\ll~\text{string scale}\;.
\end{equation}
The effective supergravity has $\Mink$ as vacuum solution, and its general features depend on the original microscopic theory and on the chosen compactification. In fact, the geometry of $\Msi$ affects the amount of supersymmetry of the low-energy SUGRA, as well as its internal symmetries.\par\medskip

\subparagraph{Internal manifold, compactification and dualities.}
According to the Kaluza-Klein procedure, the isometries of $\Msi$ induce gauge symmetries in the lower-dimensional theory gauged by the vectors originating from the metric in the reduction mechanism (KK vectors). The internal manifold $\Msi$ also affects the field content of the $D=4$ theory, which arrange in supermultiplets according to the residual (super)symmetry of the vacuum solution $\Mink$.\par
The compactification of superstring/M-theory on a \emph{Ricci-flat} internal manifold (like a torus or a Calabi Yau space) in the absence of fluxes of higher-order form field-strengths, yields, in the low-energy limit, an effective four-dimensional SUGRA, which involves the massless modes on $\Mink$. The latter is an ungauged theory, namely the vector fields are not minimally coupled to any other field of the theory. At the classical level, ungauged supergravity models feature an on-shell global symmetry group, which was conjectured to encode the known superstring/M-theory dualities \cite{Hull:1994ys}. The idea behind these dualities is that superstring/M-theory provide a redundant description for the same microscopic degrees of freedom: different compactifications of the theory turns out to define distinct descriptions of the same quantum physics. These descriptions are connected by dualities, which also map the correspondent low-energy description into one another. The global symmetry group $G$ of the classical $D=4$ supergravity is in part remnant of the symmetry of the original higher dimensional theory, i.e.\ invariance under reparametrizations in $\Msi$
\footnote{
in part they originate from gauge symmetries associated with the higher dimensional antisymmetric tensor fields
}.\par\medskip

\paragraph{Ungauged vs Gauged models.}
From a phenomenological point of view, extended supergravity models on four dimensional Minkowski vacua, obtained through ordinary Kaluza-Klein reduction on a Ricci-flat manifold, are not consistent with experimental observations. These models typically contain a certain number of massless scalar fields -- which are associated with the geometry of the internal manifold $\Msi$ -- whose vacuum expectation values (vevs) define a continuum of degenerate vacua. In fact, there is no scalar potential that encodes any scalar dynamics, so we cannot avoid the degeneracy. This turns into an intrinsic lack of predictiveness for the model, in addition to a field-content of the theory which comprises massless scalar fields coupled to gravity, whose large scale effects are not observed in our universe.\par
Another feature of these models, as we said above, is the absence of a internal local-symmetry gauged by the vector fields. This means that no matter field is charged under a gauge group, hence the name \emph{ungauged supergravity}.\par
\medskip
Realistic quantum field theory models in four dimensions, therefore, require the presence of a non-trivial scalar potential, which could solve (in part or completely) moduli-degeneracy problem and, on the other hand, select a vacuum state for our universe featuring desirable physical properties like, for instance
\begin{enumerate}[-,itemsep=1ex]
 \item introduce mass terms for the scalars;
 \item support the presence of some effective cosmological constant;
 \item etc.
\end{enumerate}
\smallskip
The phenomenologically uninteresting ungauged SUGRAs can provide a general framework for the construction of realistic model. In a $D=4$ extended supergravity model (i.e.\ having $\N>1$ susy), it is possible to introduce a scalar potential, without explicitly breaking supersymmetry, through the so-called \emph{gauging procedure} \cite{deWit:1982ig,deWit:2002vt,deWit:2003hq,deWit:2005hv,deWit:2005ub,Hull:1984vg,Hull:1984qz,Trigiante:2007ki,Samtleben:2008pe}. The latter can be seen as a \emph{deformation} of an ungauged theory and consists in promoting some suitable subgroup $\Gg$ of the global symmetry group of the Lagrangian to \emph{local} symmetry. This can be done by introducing minimal couplings for the vector fields, mass deformation terms and the scalar potential itself. The coupling of the (formerly abelian) vector fields to the new local gauge group gives us matter fields that are charged under this new local gauge symmetry.\par

In particular, in the presence of fluxes of higher-order form field-strengths across cycles of the internal manifold
\begin{equation}
\langle \, \int_{\Sigma_p} F_{(p)} \;\, \rangle ~\neq~ 0\;,
\end{equation}
the non-linear dynamics of the low lying modes (or of a consistent truncation thereof) is, in most cases, captured by a $D=4$ theory which is gauged.\par
The gauge group $\Gg$ of the lower dimensional SUGRA depends on the geometry of the internal manifold and on the possible internal fluxes
\begin{center}
\begin{tikzpicture}[>=latex]
\path (0,0) node [inner sep=3pt,outer sep=4pt] (G) {$\Gg$};
\path (G)--++(0-45:2.2cm) node [inner sep=3pt,outer sep=4pt] (M) {geom.\ of $\Msi$};
\path (G)--++(180+45:2.2cm) node [inner sep=3pt,outer sep=4pt] (F) {int. fluxes};
\draw[->] (G)--++(F);
\draw[->] (G)--++(M);
\draw[<->] (M)--++(F);
\end{tikzpicture}
\end{center}
The fluxes and the structure of the internal manifold, aside from the gauge symmetry, also induce masses and a scalar potential $V(\phi)$ (for reviews on flux-compactifications see \cite{Grana:2005jc,Blumenhagen:2006ci,Douglas:2006es}). These mass terms produce, in general, supersymmetry breaking already at the classical level (which is phenomenologically desirable) and the presence of a scalar potential lift the moduli degeneracy (already at the tree level) and may produce an effective cosmological constant term
\begin{center}
\begin{tikzpicture}[>=latex]
\path (0,0) node[draw,inner sep=8pt,outer sep=2.5pt,align=center] (MF)
            {geom.\ of $\Msi$\,, \\[1.2ex] int. fluxes};
\path (MF)--++(30:3cm) node [draw,ellipse,inner sep=3pt,outer sep=3pt] (m) {masses};
\path (MF)--++(-30:3cm) node [draw,ellipse,inner sep=3pt,outer sep=3pt] (V) {$V(\phi)$};
\path (m.east)--++(0:2cm) node [inner sep=3pt,outer sep=3pt] (sb) {SUSY breaking};
\path (V.east)--++(30:2cm) node [inner sep=3pt,outer sep=3pt] (sm) {scalar masses};
\path (V.east)--++(-30:2cm) node [inner sep=3pt,outer sep=3pt] (L) {cosm. constant};
\draw[->] (MF) to [out=80, in=180] (m.180);
\draw[->] (MF) to [out=-80, in=180] (V.180);
\draw[->] (m) to [out=0, in=180] (sb.180);
\draw[->] (V) to [out=60, in=180] (sm.180);
\draw[->] (V) to [out=-60, in=180] (L.180);
\end{tikzpicture}
\end{center}
Supergravity theories in $D$ dimensions are consistently defined independently of their higher-dimensional origin, and are totally defined by
\begin{enumerate}[$\circ$,itemsep=1ex]
  \item amount of supersymmetry;
  \item field content;
  \item local symmetry, gauged by the vector fields (feature of gauged SUGRAs).
\end{enumerate}
\medskip
When originating from superstring/M-theory compactifications, gauged SUGRAs offer a unique window on the perturbative low-energy dynamics of these theories, since they describe the full non-linear dynamics of the low lying modes. In general, there is a correspondence between vacua of the microscopic fundamental theory and vacua of the low-energy supergravity. However, there are several gauged SUGRAs whose superstring/M-theory origin is not known.\par
Gauged supergravities are obtained from ungauged ones, with the same field content and amount of SUSY, through the gauging previously mentioned procedure, which is well-defined and works provided the gauge group $\Gg$ satisfies some stringent conditions originating from the requirement of gauge invariance and supersymmetry.
\begin{center}
\begin{tikzpicture}[>=latex]
\path (0,0) node[draw,inner sep=4pt,outer sep=5pt,align=center] (SSM)
            {SS/M-theory \\[0.5ex] $D=10,\,11$};
\path (SSM.north east)--++(45:2.5cm) node [inner sep=2pt,outer sep=2pt,align=center] (f0)
            {\begin{minipage}{3cm}
              \begin{align*}
                \begin{sqcases}
                \;\;\Mink\times~\Msi\,,\\
                \;\;\mbox{Ricci flat}\,,\\
                \;\;{\rm flux}=0
                \end{sqcases}
             \end{align*}
            \end{minipage}};
\path (SSM.south east)--++(-45:2.5cm) node [inner sep=2pt,outer sep=2pt] (fne0)
            {\begin{minipage}{3cm}
              \begin{align*}
                \begin{sqcases}
                 \;\;\Mink\times~\Msi\,,\\[\jot]
                 \;\;{\rm flux}\neq0
                \end{sqcases}
             \end{align*}
            \end{minipage}};
\path (f0.east)--++(0:3.5cm) node [inner sep=2pt,outer sep=5pt,align=left] (US)
             {\begin{minipage}{3cm}
              \begin{align*}
                \begin{sqcases}
                 \;\;\text{\underline{Ungauged SUGRA}}\,,\\[\jot]
                 \quad\circ\;\;\parbox[t]{1.\textwidth}{\linespread{1.08}\selectfont{global symmetry group $G$ encoding SS/M-th. dualities}}
                \end{sqcases}
             \end{align*}
            \end{minipage}};
\path (fne0.east)--++(0:3.5cm) node [inner sep=2pt,outer sep=4pt] (GS)
             {\begin{minipage}{3cm}
              \begin{align*}
                \begin{sqcases}
                 \;\;\text{\underline{Gauged SUGRA}}\,,\\[\jot]
                 \quad\circ\quad\Gg\,,\\
                 \quad\circ\quad\text{masses}\,,\\
                 \quad\circ\quad V(\phi)\neq0
                 \end{sqcases}
             \end{align*}
            \end{minipage}};
\draw[->] (SSM.north) to [out=80, in=190] (f0.west);
\draw[->] (SSM.south) to [out=-80, in=170] (fne0.west);
\draw[->] (f0.east) to [out=0, in=180] node[pos=0.5,above]{\scriptsize $D$-dim} (US.west);
\draw[->] (fne0.east) to [out=0, in=180] node[pos=0.5,above]{\scriptsize $D$-dim} (GS.west);
\draw[->] (US.south) to node[pos=0.5,left,align=center]{{\scriptsize gauging} \\ {\scriptsize of $\Gg\in G$} } (GS.north);
\end{tikzpicture}
\end{center}
As mentioned above, gauging is the only known way to introduce a scalar potential in extended supergravities without an explicit breaking of the supersymmetry. However this procedure will in general break the global symmetry group of the ungauged theory. The latter indeed acts as a generalized electric-magnetic duality and is thus broken by the minimal couplings, which only involve the electric vector fields. As a consequence of this, in a gauged supergravity we loose track of the string/M-theory dualities, which were described by global symmetries of the original ungauged theories. \par
The drawback can be avoided using the {\em embedding tensor} formulation of the gauging procedure \cite{Cordaro:1998tx,Nicolai:2000sc,deWit:2002vt,deWit:2005ub,deWit:2007mt} in which all deformations involved by the gauging is encoded in a single object, the embedding tensor, which is itself covariant with respect to the global symmetries of the ungauged model. This allows to formally restore such symmetries at the level of the gauged field equations and Bianchi identities, provided the embedding tensor is transformed together with all the other fields. The global symmetries of the ungauged theory now act as equivalences between gauged supergravities. Since the embedding tensor encodes all background quantities in the compactification describing the fluxes and the structure of the internal manifold, the action of the global symmetry group on it allows to systematically study the effect of dualities on flux compactifications.\par\bigskip
\noindent
These lectures are organized as follows.\par
In Sect.\ \ref{sec:2} we briefly review the general structure of ungauged supergravity theories.\par
In Sect.\ \ref{sec:3} we discuss the gauging procedure in the electric symplectic frame and comment on the relation between the embedding tensor and the internal fluxes and the action on the latter of dualities. We end the section by discussing, as an example, the gauging of the maximal four dimensional theory.\par
In Sect.\ \ref{sec:4} we review a manifestly covariant formulation of the gauging procedure and introduce the notion of tensor hierarchy in higher dimensions.\par
A more complete and detailed recent review of gauged supergravities can be found in \cite{Trigiante:2016mnt}.

\section{Review of ungauged supergravities}\label{sec:2}
Let us recall some basic aspects of the extended ungauged $D=4$ supergravity.\par

\paragraph{Field content and bosonic action.}
The bosonic sector consists in the graviton $g_{\mu\nu}(x)$, $n_v$ vector fields $A^\Lambda_\mu(x)$, $n_s$ scalar fields $\phi^s(x)$ and is described by bosonic Lagrangian of the following general form
\footnote{
using the ``mostly minus'' convention and
\;$8\pi\GN=c=\hbar=1$
}
\begin{equation}
\frac{1}{e}\;\LB~=
\-\frac{R}{2}
\+\frac{1}{2}\,\Gm_{st}(\phi)\,\partial_\mu\phi^s\,\partial^\mu\phi^t
\+\frac{1}{4}\,\I_{\Lambda\Sigma}(\phi)\,F^\Lambda_{\mu\nu}\,F^{\Sigma\,\mu\nu}
\+\frac{1}{8\,e}\,\R_{\Lambda\Sigma}(\phi)\,\eps^{\mu\nu\rho\sigma}\,F^\Lambda_{\mu\nu} \,F^{\Sigma}_{\rho\sigma}\,,
\label{boslagr}
\end{equation}
where $e=\sqrt{|{\rm Det}(g_{\mu\nu})|}$ and the $n_v$ vector field strengths are defined as usual:
\begin{align}
F^\Lambda_{\mu\nu}=\partial_\mu A^\Lambda_\nu-\partial_\nu A^\Lambda_\mu\;.\\ \nn
\end{align}
\bigskip
Let us comment on the general characteristics of the above action.
\begin{enumerate}[$\circ$,itemsep=1.5ex]
\item{The scalar fields $\phi^s$ are described by a non-linear $\sigma$-model, that is they are coordinates of a non-compact, \emph{Riemannian} $n_s$-dimensional differentiable manifold (target space), named \emph{scalar manifold} and to be denoted by $\Mscal$. The positive definite metric on the manifold is $\Gm_{st}(\phi)$. The corresponding kinetic part of the Lagrangian density reads:
   \begin{equation}
     \Lscal=\frac{e}{2}\,\Gm_{st}(\phi)\,\partial_\mu
          \phi^s\partial^\mu \phi^t\,.
    \end{equation}
 The $\sigma$-model action is clearly invariant under the action of global (i.e.\ space-time independent) isometries of the scalar manifold.
 As we shall discuss below, the group $G$ can be promoted to a global symmetry group of the field equations and Bianchi identities (i.e.\ \emph{on-shell global symmetry group}) provided its (non-linear) action on the scalar fields is combined with an electric-magnetic duality transformation on the vector field strengths and their magnetic duals.
 }
 \item{The two terms containing the vector field strengths will be called vector kinetic terms. A general feature of supergravity theories is that the scalar fields are non-minimally coupled to the vector fields as they enter these terms through symmetric matrices $\I_{\Lambda\Sigma}(\phi),\,\R_{\Lambda\Sigma}(\phi)$ which contract the vector field strengths. The former $\I_{\Lambda\Sigma}(\phi)$ is negative definite and generalizes the $-1/g^2$ factor in the Yang-Mills kinetic term. The latter $\R_{\Lambda\Sigma}(\phi)$ generalizes the $\theta$-term.}
 \item{There is a ${\rm U}(1)^{n_v}$ gauge invariance associated with the vector fields:
     \begin{equation}
      A_\mu^\Lambda\rightarrow A_\mu^\Lambda+\partial_\mu\zeta^\Lambda\;;
     \end{equation}
     and all the fields are neutral with respect to this symmetry group.}
\item{There is no scalar potential. In an ungauged supergravity a scalar potential is allowed only for $\N=1$ (called the \emph{F-term potential}). In extended supergravities a non-trivial scalar potential can be introduced without explicitly breaking supersymmetry only through the \emph{gauging procedure}, which implies the introduction of a local symmetry group to be gauged by the vector fields of the theory and which will be extensively dealt with in the following.}
\end{enumerate}
The fermion part of the action is totally determined by supersymmetry once the bosonic one is given. Let us discuss in some detail the scalar sector and its mathematical description.

\subsection{Scalar sector and coset geometry}\label{ghsect}
As mentioned above the scalar fields $\phi^s$ are coordinates of a Riemannian scalar manifold $\Mscal$, with metric $\Gm_{st}(\phi)$. The isotropy group $H$ of $\Mscal$ has the general form
\begin{equation}
H\=H_{\rm R}\times H_{\rm matt}\,,\label{Hgroup}
\end{equation}
where $H_{\rm R}$ is the R--symmetry group and $H_{\rm matt}$ is a compact group acting on the matter fields. The gravitino and spin-$\frac12$ fields will transform in representations of the $H$ group. The maximal theory $\N=8$ describes the gravitational multiplet only and thus $H=H_{\rm R}=\SU(8)$. The isometry group $G$ of $\Mscal$ clearly defines the global symmetries of the scalar action.\par
In $\N>2$ theories the scalar manifold is constrained by supersymmetry to be homogeneous symmetric, namely to have the general form
\begin{equation}
\Mscal\=\frac{G}{H}\,,
\end{equation}
where $G$ is the semisimple non-compact Lie group of isometries and $H$ its maximal compact subgroup. Generic homogeneous spaces $\Mscal$ can always be written in the above form though $G$ need not be semisimple.
\begin{table}\label{tabl}
\begin{center}
{\scriptsize
  \renewcommand{\arraystretch}{1.7}
  \begin{tabular}{ | c|| c | c| c|c|} 
  \hline
  $\N$ & $\dfrac{G}{H}$ & $n_s$ &$n_v$& ${\Scr R}_v$\\[2ex]
  \hline\hline
  & & & & \\
  8 & $\frac{{\rm E}_{7(7)}}{{\rm SU}(8)}$ & 70 & 28 &{\bf 56} \\
  & & & & \\
  \hline
  & & & & \\
  6 & $\frac{\SO^*(12)}{{\rm U}(6)}$   & 30  &16& ${\bf 32}_c$\\
  & & & & \\
  \hline
  & & & & \\
  5 & $\frac{{\rm SU}(5,1)}{{\rm U}(5)}$  & 10 &10&{\bf 20}\\
  & & & & \\
  \hline
  & & & & \\
  4 & $\frac{{\rm SL}(2,\mathbb{R})}{{\rm SO}(2)}\times \frac{{\rm SO}(6,n)}{{\rm SO}(6)\times {\rm SO}(n)}$  & 6n+2 &n+6 & ${\bf (2,6+n)}$\\
  & & & & \\
  \hline
  & & & & \\
  3 & $\frac{{\rm SU}(3,n)}{{\rm S}[{\rm U}(3)\times{\rm U}(n)]}$  & 6n &3+n & $({\bf 3+n})+({\bf 3+n})'$\\
  & & & & \\
  \hline
\end{tabular}
\caption{Homogeneous symmetric scalar manifolds in $\mathcal{N}>2$ supergravities, their real dimensions $n_s$ and the number $n_v$ of vector fields.}}
\end{center}
\end{table}
\begin{table}
\begin{center}
{\scriptsize
  \renewcommand{\arraystretch}{1.7}
  \begin{tabular}{ | c|| c | c| c|c|} 
  \hline
  $\N$ & $\dfrac{G}{H}$ & $n_s$ &$n_v$&${\Scr R}_v$ \\[2ex]
  \hline\hline
  & & & & \\
  & $\frac{{\rm SU}(1,n)}{{\rm U}(n)}$  & 2n &n+1&$({\bf 1+n})+({\bf 1+n})'$\\
  & & & & \\
  & $\frac{{\rm SL}(2,\mathbb{R})}{{\rm SO}(2)}\times\frac{{\rm SO}(2,n-1)}{{\rm SO}(2)\times {\rm SO}(n-1)}$ & $2n$ &n+1 & ${\bf (2,n+1)}$\\
  & & & & \\
  & $\frac{{\rm SU}(1,1)}{{\rm U}(1)}$  & 2 &2&${\bf 4}$\\
  & & & & \\
   2, SK & $\frac{{\rm Sp}(6)}{{\rm U}(3)}$  & 12 & 7 &${\bf 14}'$\\
  & & & & \\
 & $\frac{{\rm SU}(3,3)}{{\rm S}[{\rm U}(3)\times{\rm U}(3)]}$ & 18  &10& ${\bf 20}$\\
  & & & & \\
  & $\frac{{\rm SO}^*(12)}{{\rm U}(6)}$ & 30 &16& ${\bf 32}_c$\\
  & & & & \\
  & $\frac{{\rm E}_{7(-25)}}{{\rm U}(1)\times {\rm E}_6}$  & 54 &28& ${\bf 56}$\\
  & & & & \\
  \hline
  & & & & \\
  & $\frac{{\rm SU}(2,n_H)}{{\rm S}[{\rm U}(2)\times {\rm U}(n_H)]}$  & $4n_H$ & &  \\
  & & & & \\
  & $\frac{{\rm SO}(4,n_H)}{{\rm SO }(4)\times {\rm SO}(n_H)}$  & $4n_H$ &  & \\
    & & & & \\
  & $\frac{{\rm G}_{2(2)}}{{\rm SU }(2)\times {\rm SU }(2)}$  & $8$ &  & \\
  & & & & \\
 & $\frac{{\rm F}_{4(+4)}}{{\rm SU}(2)\times {\rm USp}(6)}$  & $28$ &  & \\
    & & & & \\
   2, QK& $\frac{{\rm E}_{6(+2)}}{{\rm SU}(2)\times {\rm SU}(6)}$  & $40$ &  & \\
    & & & & \\
  & $\frac{{\rm E}_{7(-5)}}{{\rm SU}(2)\times {\rm SO}(12)}$  & $64$ &  & \\
    & & & & \\
  & $\frac{{\rm E}_{8(-24)}}{{\rm SU}(2)\times {\rm E}_7}$  & $112$ &  & \\
   & & & & \\
  & $\frac{{\rm USp}(2,2n_H)}{{\rm USp}(2)\times {\rm USp}(2n_H)}$  & $4n_H$ & &  \\
  \hline
\end{tabular}
\caption{Homogeneous symmetric special K\"ahler (SK) and quaternionic K\"ahler (QK) scalar manifolds in $\mathcal{N}=2$ supergravities, their real dimensions $n_s$ and the number $n_v$ of vector fields.}}\label{table2}
\end{center}
\end{table}
The action of an isometry
transformation $g\in G$ on the scalar fields $\phi^r$ parametrizing
$\Mscal$ is defined by means of a \emph{coset
representative} $\Lc(\phi)\in G/H$ as follows:
\begin{equation}
g\cdot \Lc(\phi^r)=\Lc(g\star\phi^r)\cdot
h(\phi^r,g)\,,\label{gLh}
\end{equation}
where $g\star\phi^r$ denote the transformed scalar fields,
non-linear functions of the original ones $\phi^r$, and
$h(\phi^r,g)$ is a \emph{compensator} in $H$. The coset representative is defined modulo the right-action of $H$ and is fixed by the chosen
parametrization of the manifold. Of particular relevance in supergravity is the so-called \emph{solvable parametrization}, which corresponds to fixing the action of $H$ so that $\Lc$ belongs to a solvable Lie group%
\footnote{
a solvable Lie group $G_S$ can be described (locally) as a the Lie group generated by \emph{solvable Lie algebra} $\Solv$: $G_S=\exp(\Solv) $. A Lie algebra $\Solv$ is solvable iff, for some $k>0$, ${\bf D}^k \Solv=0$, where the \emph{derivative} ${\bf D}$ of a Lie algebra $\mathfrak{g}$ is defined as follows: \,${\bf D}\mathfrak{g}\equiv [\mathfrak{g},\mathfrak{g}]$, \;${\bf D}^n\mathfrak{g}\equiv [{\bf D}^{n-1}\mathfrak{g},{\bf D}^{n-1}\mathfrak{g}]$. In a suitable basis of a given representation, elements of a solvable Lie group or a solvable Lie algebra are all described by upper (or lower) triangular matrices
}
$G_S=\exp(\Solv)$, generated by a solvable Lie algebra $\Solv$ and defined, in the symmetric case, by the Iwasawa decomposition of $G$ with respect to $H$.
The scalar fields are then parameters of the solvable Lie algebra $\Solv$:
\begin{align}
\Lc(\phi^r)&= e^{\phi^r T_r}\in \exp(\Solv)\,,
\end{align}
where $\{T_r\}$ is a basis of $\Solv$ ($r=1,\dots,\,n_s$). All homogeneous scalar manifolds occurring in supergravity theories admit this parametrization, which is useful when the four-dimensional supergravity originates from the Kaluza-Klein reduction of a higher-dimensional one on some internal compact manifold. The
solvable coordinates directly describe dimensionally reduced fields and moreover this parametrization makes the shift symmetries of the metric manifest.\par
The Lie algebra $\mathfrak{g}$ of $G$ can be decomposed into the Lie algebra $\mathfrak{H}$ generating $H$, and a coset space $\mathfrak{K}$:
\begin{equation}
\mathfrak{g}=\mathfrak{H}\oplus \mathfrak{K}\,,\label{ghkdec}
\end{equation}
where in general we have:
\begin{equation}
[\mathfrak{H},\,\mathfrak{H}]\subset \mathfrak{H}\;;\qquad
[\mathfrak{H},\,\mathfrak{K}]\subset \mathfrak{K}\;;\qquad
[\mathfrak{K},\,\mathfrak{K}]\subset \mathfrak{H}\oplus\mathfrak{K}\;,\label{hkh}
\end{equation}
that is the space $\mathfrak{K}$ supports a representation $\mathcal{K}$ of $H$ with respect to its adjoint action. An alternative choice of parametrization corresponds to defining the coset representative as an element of $\exp(\mathfrak{K})$:
\begin{align}
\Lc(\phi^r)&= e^{\phi^r K_r} \;\in\, \exp(\mathfrak{K})\,,
\end{align}
where $\{K_r\}$ is a basis of $\mathfrak{K}$. As opposed to the solvable parametrization, the coset representative is no-longer a group element, since $\mathfrak{K}$ does not close an algebra, see last of eqs.\ (\ref{hkh}). The main advantage of this parametrization is that the action of $H$ on the scalar fields is \emph{linear}:
\begin{align}
\forall h\in H\;:\quad
h\,\Lc(\phi^r)=h\,e^{\phi^r K_r}\,h^{-1}\,h=e^{\phi^r\,h\,K_r\,h^{-1}}\,h=\Lc(\phi^{\prime r})\,h\,,
\end{align}
where $\phi^{\prime r}=(h^{-1})_s{}^r\,\phi^s$, and $h_s{}^r$ describes $h$ in the representation $\mathcal{K}$. This is not the case for the solvable parametrization since $[\mathfrak{H},\,\Solv]\nsubseteq \Solv$.\par
In all parametrizations, the origin $\Or$ is defined as the point in which the coset representative equals the identity element of $G$ and thus the $H$-invariance of $\Or$ is manifest: $\Lc(\Or)=\Id$.\par
If the manifold, besides being homogeneous, is also \emph{symmetric}, the space $\mathfrak{K}$ can be defined so that:
\begin{equation}
[\mathfrak{K},\,\mathfrak{K}]\subset \mathfrak{H}\,.
\end{equation}
In this case the eq.\ (\ref{ghkdec}) defines the Cartan decomposition of $\mathfrak{g}$ into \emph{compact} and \emph{non-compact} generators, in $\mathfrak{H}$ and $\mathfrak{K}$, respectively. This means that, in a given matrix representation of $\mathfrak{g}$, a basis of the carrier vector space can be chosen so that the elements of $\mathfrak{H}$ and of $\mathfrak{K}$ are represented by anti-hermitian and hermitian matrices, respectively. \par
The geometry of $\Mscal$ is described by vielbein and an $H$-connection constructed out of the left-invariant one-form
\begin{equation}
\Omega\=\Lc^{-1}\,d\Lc\,\in\,\galg\,,\label{omegapro}
\end{equation}
satisfying the Maurer-Cartan equation:
\begin{equation}
d\Omega+\Omega\wedge \Omega=0\,.\label{MCeq}
\end{equation}
The Vielbein and $H$-connection are defined by decomposing $\Omega$ according to (\ref{ghkdec})
\begin{equation}
\Omega(\phi)=\PP(\phi)+\W(\phi)\,; \quad\quad \W\in\halg\,,\quad \PP\in\kalg\,.\label{Vom}
\end{equation}
Let us see how these quantities transform under the action of $G$. For any $g\in G$, using eq.\ (\ref{gLh}), we can write $\Lc(g\star \phi)=g\,\Lc(\phi)\,h^{-1}$, so that:
\begin{align}
\Omega(g\star \phi)&=h\,\Lc(\phi)^{-1}\,g^{-1} d(g\,\Lc(\phi)\,h^{-1})=
h\,\Lc(\phi)^{-1}\,d\Lc(\phi)\;h^{-1}+h\;dh^{-1}\,.
\end{align}
From (\ref{Vom}) we find:
\begin{align}
&\PP(g\star \phi)+\W(g\star \phi)=h\,\PP(\phi)\,h^{-1}+h\,\W(\phi)h^{-1}+h\;dh^{-1}\,.
\end{align}
Since $h\;dh^{-1}$ is the left-invariant 1-form on $\mathfrak{H}$, it has value in this algebra. Projecting the above equation over $\mathfrak{K}$ and $\mathfrak{H}$, we find:
\begin{align}
\PP(g\star \phi)&=h\,\PP(\phi)\,h^{-1}\,,\label{Ptra}\\
\W(g\star \phi)&=h\,\W(\phi)\,h^{-1}+h\;dh^{-1}\,.\label{omtra}
\end{align}
We see that $\W$ transforms as an $H$-connection while the matrix-valued one-form $\PP$ transforms linearly under $H$. The vielbein of the scalar manifold are defined by expanding $\PP$ in a basis $\{K_{\underline{s}}\}$ of $\mathfrak{K}$ (underlined indices $\underline{s},\underline{r},\underline{t},\dots$ are rigid tangent-space indices, as opposed to the curved coordinate indices $s,r,t,\dots$):
\begin{equation}
\PP(\phi)=V^{\underline{s}}( \phi)\,K_{\underline{s}}\,.
\end{equation}
From (\ref{Ptra}) it follows that the vielbein 1-forms $V^{\underline{s}}( \phi)=V_s{}^{\underline{s}}( \phi)d\phi^s$ transform under the action of $G$ as follows:
\begin{equation}
V^{\underline{s}}(g\star \phi)\=V^{\underline{t}}( \phi)\,(h^{-1})_{\underline{t}}{}^{\underline{s}}\=h^{\underline{s}}{}_{\underline{t}}V^{\underline{t}}( \phi)\,.\label{Vtra}
\end{equation}
For symmetric spaces, from (\ref{MCeq}) it follows that $\W$ and $\PP$ satisfy the following conditions
\begin{align}
\mathscr{D}\PP&~\equiv~ d\PP+\W\wedge \PP+\PP\wedge \W\=0\,,\label{DP}\\
R(\W)&~\equiv~ d\W+\W\wedge \W\=-\PP\wedge \PP\,,\label{RW}
\end{align}
where we have defined the $H$-covariant derivative $\mathscr{D}\PP$ of $\PP$ and the $\mathfrak{H}$-valued curvature $R(\W)$ of the manifold.
The latter can be written in components:
\begin{equation}
R(\W)=\frac{1}{2}\,R_{rs}\,d\phi^r\wedge d\phi^s \quad\Rightarrow\quad
R_{rs}=-[\PP_r,\,\PP_s]\in \mathfrak{H}\,.\label{Rcompo}
\end{equation}
We define the metric at the origin $\Or$ as the $H$-invariant matrix:
\begin{equation}
\eta_{\underline{s}\underline{t}}\equiv k\,{\rm Tr}(K_{\underline{s}}\,K_{\underline{t}})>0\,,
\end{equation}
where $k$ is a positive number depending on the representation, so that the metric in a generic point reads:
\begin{equation}
 ds^2(\phi)\equiv\Gm_{st}(\phi)d\phi^s\,d\phi^t\equiv V_s{}^{\underline{s}}( \phi)V_t{}^{\underline{t}}( \phi)\eta_{\underline{s}\underline{t}}\,d\phi^s\,d\phi^t=k\,{\rm Tr}(\PP_s\,\PP_t)\,.
\end{equation}
As it follows from eqs.\ (\ref{Ptra}), (\ref{Vtra}), the above metric is manifestly invariant under global $G$-transformations acting on $\Lc$ to the left (as well as local $H$-transformations acting on $\Lc$ to the right):
\begin{equation}
ds^2(g\star \phi)=ds^2(\phi)\;.
\end{equation}
The $\sigma$-model Lagrangian can be written in the form:
\begin{equation}
\Lscal=\frac{e}{2}\, \Gm(\phi)_{st}\partial_\mu\phi^s\,\partial^\mu\phi^t
=\frac{e}{2}\,k\,\Tr\big(\PP_\mu(\phi)\,\PP^\mu(\phi)\big)\,,
\qquad \PP_\mu=\PP_s\frac{\partial\phi^s}{\partial x^\mu}\,,\;\quad\label{lagrscal}
\end{equation}
and, just as the metric $ds^2$, is manifestly invariant under global $G$ and local $H$-transformations acting on $\Lc$ as in (\ref{gLh}).\par
The bosonic part of the equations of motion for the scalar fields can be derived from the Lagrangian (\ref{boslagr}) and read:
\begin{align}
\mathscr{D}_\mu (\partial^\mu
\phi^s)&=\frac{1}{4}\,\Gm^{st}\,\left[F_{\mu\nu}^\Lambda\,
\partial_t\,\I_{\Lambda\Sigma}\,F^{\Sigma\, \mu\nu}+F_{\mu\nu}^\Lambda
\partial_t\, \R_{\Lambda\Sigma}\,{}^*F^{\Sigma\,
\mu\nu}\right]\,,\label{scaleqs}
\end{align}
where $\partial_s\equiv \frac{\partial}{\partial \phi^s}$, while $\mathscr{D}_\mu$ also contains the Levi-Civita connection $\tilde{\Gamma}$ on the scalar manifold:
\begin{equation}
\mathscr{D}_\mu (\partial_\nu
\phi^s)\equiv \nabla_\mu(\partial_\nu
\phi^s)+\tilde{\Gamma}^s_{t_1 t_2}\partial_\mu \phi^{t_1}\,\partial_\nu\phi^{t_2}\,,
\end{equation}
$\nabla_\mu$ being the covariant derivative containing the Levi-Civita
connection on space-time.\par
Let us end this paragraph by introducing, in the coset geometry, the Killing vectors describing the infinitesimal action of isometries on the scalar fields. Let us denote by $t_\alpha$ the infinitesimal generators of $G$, defining a basis of its Lie algebra $ \mathfrak{g}$ and satisfying the corresponding commutation relations
\begin{equation}
[t_\alpha,\,t_\beta]={\bf f}_{\alpha\beta}{}^\gamma\,t_\gamma\,,\label{talg}
\end{equation}
${\bf f}_{\alpha\beta}{}^\gamma$ being the structure constants of $\mathfrak{g}$. Under an infinitesimal $G$-transformation generated by $\epsilon^\alpha\,t_\alpha$ ($\epsilon^\alpha\ll 1$):
\begin{equation}
g\approx \Id+\epsilon^\alpha\,t_\alpha\,,
\end{equation}
the scalars transform as:
\begin{equation}
\phi^s\rightarrow \phi^s+\epsilon^\alpha\,k^s_\alpha(\phi)\,,
\end{equation}
$k^s_\alpha(\phi)$ being the Killing vector associated with $t_\alpha$. The action of $g$ on the scalars is defined by eq.\ (\ref{gLh}), neglecting terms of order $O(\epsilon^2)$:
\begin{equation}
(\Id+\epsilon^\alpha\,t_\alpha)\,\Lc(\phi)=\Lc(\phi+\epsilon^\alpha\,k_\alpha)(\Id-\frac{1}{2}\,\epsilon^\alpha W_\alpha ^I\,J_I)\,,
\end{equation}
where $(\Id-\frac{1}{2}\,\epsilon^\alpha W_\alpha ^I\,J_I)$ denotes, expanded to linear order in $\epsilon$, the compensating transformation $h(\phi,g)$, $\{J_I\}$ being a basis of $\mathfrak{H}$. Equating the terms proportional to $\epsilon^\alpha$, multiplying to the left by $\Lc^{-1}$ and using the expansion (\ref{Vom}) of the left-invariant 1-form, we end up with the following equation:
\begin{equation}
\Lc^{-1}t_\alpha\Lc\=k_\alpha^s\,(\mathcal{P}_s+\W_s)-\frac{1}{2}\,W_\alpha ^I\,J_I\=k_\alpha^s\,V_s{}^{\underline{s}}\,K_{\underline{s}}+\frac{1}{2}\,(k_\alpha^s\omega_s^I-W_\alpha ^I)\,J_I\,,\label{kespans}
\end{equation}
where we have expanded the $H$-connection along $J_I$ as follows:
\begin{equation}
 \W_s=\frac{1}{2}\,\omega^I_s\,J_I\,.
\end{equation}
Eq.\ (\ref{kespans}) allows to compute $k_\alpha$ for homogeneous scalar manifolds by projecting $\Lc^{-1}t_\alpha\Lc$ along the directions of the coset space $\mathfrak{K}$.
These Killing vectors satisfy the following algebraic relations (note the minus sign on the right hand side with respect to (\ref{talg}) :
 \begin{equation}
[k_\alpha,\,k_\beta]=-{\bf f}_{\alpha\beta}{}^\gamma\,k_\gamma\,,
\end{equation}
We can split, according to the general structure (\ref{Hgroup}), the $H$-generators $J_I$ into $H_{\rm R}$-generators $J_{{\bf a}}$ (${{\bf a}}=1,\dots,{\rm dim}(H_{\rm R})$) and $H_{\rm matt}$-generators $J_{{\bf m}}$ (${\bf m}=1,\dots,{\rm dim}(H_{\rm matt})$), and rewrite (\ref{kespans}) in the form:
\begin{equation}
\Lc^{-1}t_\alpha \Lc\=
k_\alpha^s\,V_s{}^{\underline{s}}\,K_{\underline{s}}-\frac{1}{2}\,\Ps_\alpha^{{\bf a}}\,J_{{\bf a}}-\frac{1}{2}\,\Ps_\alpha^{{\bf m}}\,J_{{\bf m}}\,.\label{kespans2}
\end{equation}
The quantities
\begin{equation}
\Ps_\alpha^{{\bf a}}=- (k_\alpha^s\omega_s^{{\bf a}}-W_\alpha ^{{\bf a}})\,,
\end{equation}
generalize the so called \emph{momentum maps} in $\N=2$ theories, which provide a Poissonian realization of the isometries $t_\alpha$.
One can verify the general property:
\begin{equation}
 k_\alpha^s\,R^{{\bf a}}_{st}=\mathscr{D}_t \Ps_\alpha^{{\bf a}}\,,\label{KRP}
\end{equation}
where $\mathscr{D}_s$ denotes the $H$-covariant derivative and we have expanded the curvature $R[\W]$ defined in (\ref{RW}) along $J_I$:
\begin{equation}
 R[\W]=\frac{1}{2}\,R^I_{st}\,d\phi^s\wedge d\phi^t\,J_I\,.
\end{equation}
These objects are important in the gauging procedure since they enter the definition of the the gauged connections for the fermion fields as well as gravitino-shift matrix $\mathbb{S}_{AB}$ (see Sect.\ \ref{sec:3}). For all those isometries which do not produce compensating transformations in $H_{\rm R}$, $W_\alpha^{{\bf a}}=0$ and $ \Ps_\alpha^{{\bf a}}$ are easily computed to be $$ \Ps_\alpha^{{\bf a}}=- k_\alpha^s\omega_s^{{\bf a}}\,.$$ This is the case, in the solvable parametrization, for all the isometries in $\Solv$, which include translations in the axionic fields.\par
In $\N=2$ models with non-homogeneous scalar geometries, though we cannot apply the above construction of $k_\alpha,\,\Ps_\alpha^{{\bf a}}$, the momentum maps are constructed from the Killing vectors as solutions to the differential equations (\ref{KRP}).
In general, in these theories, with each isometry $t_\alpha$ of the scalar manifold, we can associate the quantities $\Ps_\alpha^{{\bf a}},\,\Ps_\alpha^{{\bf m}}$ which are related to the corresponding Killing vectors $k_\alpha$ through general
relations (see \cite{Andrianopoli:1996cm} for a comprehensive account of $\N=2$ theories).

\subsection{Vector sector}
We can associate with the electric field strengths $F_{\mu\nu}^{\Lambda}$ their magnetic duals $\Gd_{\Lambda\,\mu\nu}$ defined as:
\begin{align}
\Gd_{\Lambda\,\mu\nu}&\equiv -\epsilon_{\mu\nu\rho\sigma}
\frac{\partial \Lagr_4}{\partial
F^\Lambda_{\rho\sigma}}=\R_{\Lambda\Sigma}\,F^\Sigma_{\mu\nu}-\I_{\Lambda\Sigma}\,{}^*F^\Sigma_{\mu\nu}\,,\label{GF}
\end{align}
where we have omitted fermion currents in the expression of $\Gd_\Lambda$ since we are only focussing for the time being on the bosonic sector of the theory. In ordinary Maxwell theory (no scalar fields), $\I_{\Lambda\Sigma}=-\delta_{\Lambda\Sigma}$ and $\R_{\Lambda\Sigma}=0$, so that $\Gd_{\Lambda\,\mu\nu}$ coincides with the Hodge-dual of $F^\Lambda_{\mu\nu}$: $\Gd_{\Lambda}={}^* F^{\Lambda}$.\par
In terms of $F^\Lambda$ and $\Gd_{\Lambda}$ the bosonic part of the Maxwell equations read
\begin{equation}
\nabla^{\mu}({}^*F^\Lambda_{\mu\nu}) = 0\,;
\qquad \nabla^{\mu }({{}^*\Gd}_{\Lambda\,\mu\nu}) = 0\,,
\label{biafieq}
\end{equation}
In order to set the stage for the discussion of global symmetries, it is useful to rewrite the scalar and vector field equations in a different form.
Using (\ref{GF}) and the property that ${}^*{}^*
F^\Lambda=-F^\Lambda$, we can express ${}^* F^\Lambda$ and ${}^*
\Gd_\Lambda$ as linear functions of $F^\Lambda$ and $\Gd_\Lambda$:
\begin{align}
{}^* F^\Lambda&=
\I^{-1\,\Lambda\Sigma}\,(\R_{\Sigma\Gamma}\,F^\Gamma-\Gd_\Sigma)\;;\\
{}^*\Gd_\Lambda&=
(\R\I^{-1}\R+\I)_{\Lambda\Sigma}\,F^\Sigma-(\R\I^{-1})_\Lambda{}^\Sigma\,\Gd_\Sigma\,,\label{GF2}
\end{align}
where, for the sake of simplicity, we have omitted the space-time
indices. It is useful to arrange $F^\Lambda$ and $\Gd_\Lambda$ in a
single $2n_v$-dimensional vector $\mathbb{F}\equiv (\mathbb{F}^M)$
of two-forms:
\begin{equation}
\mathbb{F}= \left(\frac{1}{2}\,\mathbb{F}^M_{\mu\nu}\,dx^\mu\wedge dx^\nu\right) \equiv \left(\begin{matrix}F^\Lambda_{\mu\nu}\cr
\Gd_{\Lambda\mu\nu}\end{matrix}\right)\,\frac{dx^\mu\wedge dx^\nu}{2}\,,\label{bbF}
\end{equation} in terms of which the Maxwell equations read:
\begin{equation}
d\mathbb{F}=0\,,\label{Max}
\end{equation}
and eqs.\ (\ref{GF2}) are easily rewritten in the
following compact form:
\begin{eqnarray}
{}^*\mathbb{F}=-\mathbb{C}\mathcal{M}(\phi^s)\,\mathbb{F}\,,\label{FCMF}
\end{eqnarray}
where
\begin{equation}
\mathbb{C}=(\mathbb{C}^{MN})\equiv\left(\begin{matrix} \Zero & \Id
\cr -\Id & \Zero \end{matrix}\right)\,,\label{C}
\end{equation}
$\Id$, $\Zero$ being the $n_v\times n_v$ identity and zero-matrices, respectively, and
\begin{equation}
\mathcal{M}(\phi)= (\mathcal{M}(\phi)_{MN})\equiv
\left(\begin{matrix}(\R\I^{-1}\R+\I)_{\Lambda\Sigma} &
-(\R\I^{-1})_\Lambda{}^\Gamma\cr -(\I^{-1}\R)^\Delta{}_\Sigma & \I^{-1\,
\Delta \Gamma}\end{matrix}\right)\,,\label{M}
\end{equation}
is a symmetric, negative-definite matrix, function of the scalar
fields. The reader can easily verify that this matrix is also symplectic, namely that:
\begin{equation}
\mathcal{M}(\phi)\mathbb{C}\mathcal{M}(\phi)=\mathbb{C}\,.
\end{equation}
This matrix contains $\I_{\Lambda\Sigma}$ and $\R_{\Lambda\Sigma}$ as components, and therefore defines the non-minimal coupling of the scalars to the vector fields.\par
After some algebra, we can also rewrite eqs.\ (\ref{scaleqs}) in a compact form as follows
\begin{align}
\mathscr{D}_\mu (\partial^\mu\phi^s)&=
\frac{1}{8}\,\Gm^{st}\,\mathbb{F}^T_{\mu\nu}\partial_t\mathcal{M}(\phi)\,\mathbb{F}^{\mu\nu}\,,\label{scaleqs2}
\end{align}

\subsection{Coupling to gravity}
We can now compute the Einstein equations:
\begin{equation}
R_{\mu\nu}-\frac{1}{2}\,g_{\mu\nu}\,R=T^{(S)}_{\mu\nu}+T^{(V)}_{\mu\nu}+T^{(F)}_{\mu\nu}\,,\label{EEQ1}
\end{equation}
where the three terms on the right hand side are the energy-momentum tensors of the scalars, vectors and fermionic fields, respectively. The first two can be cast in the following general form
\begin{align}
T^{(S)}_{\mu\nu}&= \Gm_{rs}(\phi)\,\partial_\mu \phi^r\partial_\nu
\phi^s-\frac{1}{2}\,g_{\mu\nu}\,\Gm_{rs}(\phi)\,\partial_\rho
\phi^r\partial^\rho \phi^s\,,\\
T^{(V)}_{\mu\nu}&=\left({F}^T_{\mu\rho}\,\I\,F_{\nu}{}^\rho-\frac{1}{4}\,g_{\mu\nu}\,(F^T_{\rho\sigma} \I F^{\rho\sigma})\right)\,,\label{tv}
\end{align}
where in the last equation the vector indices $\Lambda,\Sigma$ have been suppressed for the sake of notational simplicity.
It is convenient for our next discussion, to rewrite, after some algebra, the right hand side of (\ref{tv}) as follows
\begin{equation}
T^{(V)}_{\mu\nu}=\frac{1}{2}\,\mathbb{F}^T_{\mu\rho}\,\mathcal{M}(\phi)\,\mathbb{F}_{\nu}{}^\rho\,,
\end{equation}
so that eq.\ (\ref{EEQ1}) can be finally recast in the following form:
\begin{equation}
R_{\mu\nu}=\Gm_{rs}(\phi)\,\partial_\mu \phi^r\partial_\nu
\phi^s+\frac{1}{2}\,\mathbb{F}^T_{\mu\rho}\,\mathcal{M}(\phi)\,\mathbb{F}_{\nu}{}^\rho+\dots\,,\label{EEQ2}
\end{equation}
where the ellipses refer to fermionic terms.\par
The scalar fields enter the kinetic terms of the vector fields through the matrices $\I(\phi)$ and $\R(\phi)$. As a consequence of this, a symmetry transformation of the scalar part of the Lagrangian will not in general leave the vector field part invariant.

\subsection{Global symmetry group}\label{gsg}
In extended supergravity models ($\N>1$) the (identity sector of the) global symmetry group $G$ of the scalar action can be promoted to a global invariance \cite{Gaillard:1981rj} of, at least, the field equations and the Bianchi identities, provided its (non-linear) action on the scalar fields is associated with a linear transformation on the vector field strengths $F^\Lambda_{\mu\nu}$ and their magnetic duals $\Gd_{\Lambda\,\mu\nu}$:
\begin{align}
g\in G\,:\;
\begin{cases}
\,\,\,\,\,\,\phi^r &\rightarrow \quad g\star\phi^r\;\; \qquad\qquad\qquad\qquad\qquad\qquad\qquad\text{(non--linear)},\\[\jot]
\left(\begin{matrix}
F^\Lambda \cr \Gd_\Lambda
\end{matrix}\right)
&\rightarrow\quad\Rs_v[g]\cdot
\left(\begin{matrix}
F^\Lambda\cr \Gd_\Lambda
\end{matrix}\right)=
\left(\begin{matrix}
A[g]^\Lambda{}_\Sigma & B[g]^{\Lambda\Sigma}\cr
C[g]_{\Lambda\Sigma} & D[g]_\Lambda{}^\Sigma
\end{matrix}\right)
\,\left(\begin{matrix}
F^\Sigma\cr \Gd_\Sigma
\end{matrix}\right)
\,\quad\text{(linear)}.
\end{cases}\label{dual}\nne
\end{align}
The transformations (\ref{dual}) are clearly a symmetry of the scalar action and of the Maxwell equations ($d\mathbb{F}=0$) if $F^\Lambda$ and $\Gd_\Lambda$ were independent, since the latter are clearly invariant with respect to any linear transformation on $\mathbb{F}^M$. The definition $\Gd_\Lambda$ in (\ref{GF}) as a function of $F^\Lambda,\,{}^*F^\Lambda$ and the scalar fields, which is equivalently expressed by the twisted self-duality condition (\ref{FCMF}), however poses constraints on the $2n_v\times 2n_v$ matrix $\Rs_v[g]=(\Rs_v[g]^M{}_N)$. In order for (\ref{dual}) to be an invariance of the vector equations of motion (\ref{Max}) and (\ref{FCMF}) the following conditions have to be met:
\begin{itemize}
\item[i)]{for each $g\in G$ (more precisely in the identity sector of $G$), the matrix $\Rs_v[g]$ should be \emph{symplectic}, namely
\begin{equation}
\Rs_v[g]^{T}\mathbb{C}\,\Rs_v[g]=\mathbb{C}\,;\label{SSym}
\end{equation}
}
\item[ii)]{the symplectic, scalar dependent, matrix $\mathcal{M}(\phi)$ should transform as follows:
\begin{equation}
\mathcal{M}(g\star \phi)=\Rs_v[g]^{-T}\mathcal{M}(\phi)\,\Rs_v[g]^{-1}\,,\label{traM}
\end{equation}
where we have used the short-hand notation $\Rs_v[g]^{-T}\equiv (\Rs_v[g]^{-1})^T$.
}
\end{itemize}
The reader can indeed verify that conditions i) and ii) are sufficient to guarantee invariance of (\ref{FCMF}) under (\ref{dual}).
The symplectic transformation $\Rs_v[g]$, associated with each element $g$ of $G$, mixes electric and magnetic field strengths, acting therefore as a generalized electric--magnetic duality and defines a \emph{symplectic representation} $\Rs_v$ of $G$:
\begin{equation}
\forall g\in G\,\,\,\stackrel{\Rs_v}{\longrightarrow}\,\,\,\,\,\Rs_v[g]\in {\rm Sp}(2n_v,\,\mathbb{R})\,.
\end{equation}
The field strengths and their magnetic duals transform therefore, under the duality action (\ref{dual}) of $G$ in a $2n_v$-dimensional symplectic representation.\par
We denote by $\Rs_{v*}=\Rs_v^{-T}$ the representation dual to $\Rs_v$, acting on covariant symplectic vectors, so that, for any ${\bf g}\in G$:
\begin{align}
\Rs_{v*}[{\bf g}]&=(\Rs_{v*}[{\bf g}]_M{}^N)=\Rs_v[{\bf g}]^{-T}=-\mathbb{C}\Rs_v[{\bf g}]\mathbb{C}\,\,\,\Rightarrow \nonumber\\&\Rightarrow\,\,\, \Rs_{v*}[{\bf g}]_M{}^N=\mathbb{C}_{MP}\,\Rs_v[{\bf g}]^P{}_Q\,\mathbb{C}^{NQ}\,,\end{align}
where we have used the property that $\Rs_v$ is a symplectic representation%
\footnote{
the symplectic indices {\small $M,\,N,\dots$} are raised (and lowered) with the symplectic matrix $\mathbb{C}^{MN}$ ($\mathbb{C}_{MN}$) using north-west south-east conventions: $X^{M}=\mathbb{C}^{MN}\,X_{N}$ (and $X_M=\mathbb{C}_{NM}\,X^{N}$)
}.\par
From (\ref{SSym}) and (\ref{traM}), it is straightforward to verify the manifest $G$-invariance of the scalar field equations and the Einstein equations written in the forms (\ref{scaleqs2}) and (\ref{EEQ2}).\par
Conditions i) and ii) are verified in extended supergravities as a consequence of supersymmetry. In these theories indeed supersymmetry is large enough as to connect certain scalar fields to vector fields and, as a consequence of this, symmetry transformations on the former imply transformations on the latter
(more precisely transformations on the vector field strengths $F^\Lambda$ and their duals $\Gd_\Lambda$). The existence of a symplectic representation $\Rs_v$ of $G$, together with the definition of the matrix $\mathcal{M}$ and its transformation property (\ref{traM}), are built-in in the mathematical structure of the scalar manifold. More precisely they follow from the definition on $\Mscal$ of a \emph{flat symplectic structure}. Supersymmetry totally fixes $\mathcal{M}(\phi)$ and thus the coupling of the scalar fields to the vectors, aside from a freedom in the choice of the basis of the symplectic representation (\emph{symplectic frame}) which amounts to a change in the definition of $\mathcal{M}(\phi)$ by a constant symplectic transformation $E$:
\begin{equation}
\mathcal{M}(\phi)\rightarrow \mathcal{M}'(\phi)=E\mathcal{M}(\phi)E^T\,.\label{MEtra}
\end{equation}
Clearly if $E\in \mathscr{R}_{v*}[G]\subset {\rm Sp}(2n_v,\mathbb{R})$, its effect on $\mathcal{M}(\phi)$ can be offset be a redefinition of the scalar fields, by virtue of eq.\ (\ref{traM}). On the other hand if $E$ a were block-diagonal matrix, namely an element of ${\rm GL}(n_v,\mathbb{R})\subset {\rm Sp}(2n_v,\mathbb{R})$, it could be reabsorbed in a local redefinition of the field strengths. Inequivalent symplectic frames are then connected by symplectic matrices $E$ defined modulo redefinitions of the scalar and vector fields, namely by matrices in the coset \cite{deWit:2002vt}:
\begin{equation}
E\,\in \,{\rm GL}(n_v,\mathbb{R})\backslash {\rm Sp}(2n_v,\mathbb{R})/ \mathscr{R}_{v*}[G]\,,\label{generalE}
\end{equation}
where the quotient is defined with respect to the left-action of ${\rm GL}(n_v,\mathbb{R})$ (local vector redefinitions) and to the right-action of $ \mathscr{R}_{v*}[G]$ (isometry action on the scalar fields).\par
A change in the symplectic frame amounts to choosing a different embedding $\Rs_v$ of $G$ inside ${\rm Sp}(2n_v,\,\mathbb{R})$, which is not unique.
This affects the form of the action, in particular the coupling of the scalar fields to the vectors. However, at the ungauged level,
it only amounts to a redefinition of the vector field strengths and their duals which has no physical implication.
In the presence of a gauging, namely if vectors are minimally coupled to the other fields, the symplectic frame becomes physically relevant and may lead to different vacuum-structures of the scalar potential.\par
We
emphasize here that the existence of this symplectic structure on the
scalar manifold is a general feature of all extended supergravites,
including those $\N=2$ models in which the scalar manifold is not
even homogeneous (i.e.\ the isometry group, if it exists, does not
act transitively on the manifold itself). In the $\N=2$ case, only the scalar fields belonging to the vector multiplets are non-minimally
coupled to the vector fields, namely enter the matrices
$\I(\phi),\,\R(\phi)$, and they span a \emph{special K\"ahler}
manifold. On this manifold a flat symplectic bundle is defined%
\footnote{
a special K\"ahler manifold is in general
characterized by the product of a ${\rm U}(1)$-bundle, associated
with its K\"ahler structure (with respect to which the manifold is
Hodge K\"ahler), and a flat symplectic bundle. See for instance
\cite{Andrianopoli:1996cm} for an in depth account of this issue
},
which fixes the scalar dependence of the matrices $\I(\phi),\,\R(\phi)$, aside from an initial choice of the symplectic frame, and the matrix
$ \mathcal{M}(\phi)$ defined in (\ref{M}) satisfies the property
(\ref{traM}).\par If the scalar manifold is homogeneous, we can consider at any point the coset representative $\Lc(\phi)\in G$ in the symplectic, $2n_v$-dimensional representation $\Rs_v $:
\begin{equation}
\Lc(\phi)\,\,\,\stackrel{\Rs_v}{\longrightarrow}\,\,\,\,\,\Rs_v[\Lc(\phi)]\in {\rm Sp}(2n_v,\,\mathbb{R})\,.
\end{equation}
In general the representation $\Rs_v[H]$ of the isotropy group $H$ may not be orthogonal, that is $\Rs_v[H]\nsubseteq  {\rm SO}(2n_v)$.
In this case we can always change the basis of the representation%
\footnote{
we label the new basis by underlined indices
}
by means of a matrix $\mathcal{S}$%
\begin{equation}
\mathcal{S}=(\mathcal{S}^N{}_{\underline{M}})
\,\in {\rm Sp}(2n_v,\,\mathbb{R})/{\rm U}(n)
\end{equation}
such that, in the rotated representation $\underline{\mathscr{R}}_v\equiv \mathcal{S}^{-1}\Rs_v\,\mathcal{S}$:
\begin{equation}
\underline{\mathscr{R}}_v[H]\equiv \mathcal{S}^{-1}\Rs_v[H]\,\mathcal{S}\subset {\rm SO}(2n_v)
\quad\Leftrightarrow\quad
\underline{\mathscr{R}}_v[h]^T\underline{\mathscr{R}}_v[h]=\Id\;,\quad
\forall h\in H\,.\label{hort}
\end{equation}
For any point $\phi$ on the scalar manifold define now the \emph{hybrid coset-representative matrix} $\LL(\phi)=(\LL(\phi)^M{}_{\underline{N}})$ as follows:
\begin{equation}
\LL(\phi)\equiv \Rs_v[\Lc(\phi)]\mathcal{S}
\quad\Leftrightarrow\quad
\LL(\phi)^M{}_{\underline{N}}\equiv \Rs_v[\Lc(\phi)]^M{}_N\mathcal{S}^N{}_{\underline{N}}\,.\label{hybrid}
\end{equation}
We also define the matrix
\begin{equation}
\LL(\phi)_M{}^{\underline{N}}~\equiv~ \mathbb{C}_{MP}\,\mathbb{C}^{\underline{NQ}}\;\LL(\phi)^P{}_{\underline{Q}}\;.
\end{equation}
Notice that, as a consequence of the fact that the two indices of $\LL$ refer to two different symplectic bases, $\LL$ itself is not a matrix representation of the coset representative $\Lc$.
From (\ref{gLh}), the property of $\mathscr{R}_v$ of being a representation and the definition (\ref{hybrid}) we have:
\begin{equation}
\forall {\bf g}\in G \;:\quad \mathscr{R}_v[{\bf g}]\,\LL(\phi)=\LL({\bf g}\star\phi)\,\underline{\mathscr{R}}_v[h]\,,\label{gLh2}
\end{equation}
where $h\equiv h(\phi,{\bf g})$ is the compensating transformation. The hybrid index structure of $\LL$ poses no consistency problem since, by (\ref{gLh2}), the coset representative is acted on to the left and to the right by two different groups: $G$ and $H$, respectively. Therefore, in our notations, underlined symplectic indices {\footnotesize $\underline{M},\,\underline{N},\dots$} are acted on by $H$ while non-underlined ones by $G$.\par
The
$ \mathcal{M}(\phi)$ is then expressed in terms of the coset representative as follows:
\begin{equation}
\mathcal{M}(\phi)_{MN}=\mathbb{C}_{MP}\mathbb{L}(\phi)^P{}_{\underline{L}}\mathbb{L}(\phi)^R{}_{\underline{L}}\,\mathbb{C}_{RN}
\;\;\Leftrightarrow\;\;
\mathcal{M}(\phi)=\mathbb{C}\mathbb{L}(\phi)\,\mathbb{L}(\phi)^T\,\mathbb{C}\,,\label{Mcos}
\end{equation}
where summation over the index {\footnotesize $\underline{L}$} is understood. The reader can easily verify that the definition of the matrix $\mathcal{M}(\phi)$ given above is indeed consistent, in that it is $H$-invariant, and thus only depends on the point $\phi$, and transforms according to (\ref{traM}):
\begin{align}
\forall g\in G \;:\quad \mathcal{M}(g\star\phi)&=\mathbb{C}\LL(g\star\phi)\,\LL(g\star\phi)^T\mathbb{C}=\nonumber\\&=
\mathbb{C}\Rs_v[g]\,\LL(\phi)(\underline{\Rs}_v[h]^{-1}\,\underline{\Rs}_v[h]^{-T})\LL(\phi)^T\Rs_v[g]^T\mathbb{C}=\nonumber\\&
=\Rs_v[g]^{-T}\mathbb{C}\LL(\phi)\,\LL(\phi)^T
\mathbb{C}\Rs_v[g]^{-1}=\nonumber\\
&=\Rs_v[g]^{-T}\mathcal{M}(\phi)\Rs_v[g]^{-1}\,,
\end{align}
where we have used eq.\ (\ref{gLh2}), the orthogonality property (\ref{hort}) of $\underline{\Rs}_v[h]$ and the symplectic property of $\Rs_v[g]$.
From the definition (\ref{Mcos}) of $\mathcal{M}$ in terms of the coset representative, it follows that for symmetric scalar manifolds the scalar Lagrangian (\ref{lagrscal}) can also be written in the equivalent form:
\begin{equation}
\Lscal=\frac{e}{2}\, \Gm_{st}(\phi)\partial_\mu\phi^s\,\partial^\mu\phi^t
=\frac{e}{8}\,k\,\Tr\big(\mathcal{M}^{-1}\partial_\mu\mathcal{M}\,\mathcal{M}^{-1}\partial^\mu\mathcal{M}\big)\,,\label{lagrscalM}
\end{equation}
where $k$ depends on the representation $\Rs_v$ of $G$.
\par
The transformation properties of the matrices $\I_{\Lambda\Sigma}$ and $\R_{\Lambda\Sigma}$ under $G$ can be inferred from (\ref{traM}) and can be conveniently described by defining the complex symmetric matrix \begin{equation}
\mathcalboondox{N}_{\Lambda\Sigma}\equiv \R_{\Lambda\Sigma}+i\,\I_{\Lambda\Sigma}\,.
\end{equation}
Under the action of a generic element $g\in G$, \,$\mathcalboondox{N}$ transforms as follows:
\begin{equation}
\mathcalboondox{N}(g\star\phi)=(C[g]+D[g]\,\mathcalboondox{N}(\phi))(A[g]+B[g]\,\mathcalboondox{N}(\phi))^{-1}\,,\label{Ntra}
\end{equation}
where $A[g],\,B[g],\,C[g]\,,D[g]$ are the $n_v\times n_v$ blocks of the matrix $\Rs_v[g]$ defined in (\ref{dual}).\par

\subparagraph{Parity.} We have specified above that only the elements of $G$ which belong to the identity sector, namely which are continuously connected to the identity, are associated with symplectic transformations. There may exist isometries $g\in G$ which do not belong to the identity sector and are associated with \emph{anti-symplectic} matrices ${\bf A}[g]$:
\begin{equation}
\mathcal{M}(g\star \phi)={\bf A}[g]^{-T}\,\mathcal{M}(\phi)\,{\bf A}[g]\;;
\quad\;
{\bf A}[g]^T\mathbb{C}{\bf A}[g]=-\mathbb{C}\,.
\end{equation}
Anti-symplectic matrices do not close a group but can be expressed as the product of a symplectic matrix ${\bf S}$ times a fixed anti-symplectic one ${\bf P}$, that is ${\bf A}=\S\,{\bf P}$. In a suitable symplectic frame, the matrix ${\bf P}$ can be written in the following form:
\begin{equation}
{\bf P}=\left(\begin{matrix}
\Id & \Zero \cr \Zero & -\Id \end{matrix}\right)\,.\label{Pmatrix}
\end{equation}
Due to their being implemented by anti-symplectic duality transformations (\ref{dual}), these isometries leave eq.\ (\ref{FCMF}) invariant up to a sign which can be offset by a \emph{parity transformation}, since under parity one has $\,*\,\rightarrow\,-*\,$\,.\, Indeed one can show that these transformations are a symmetry of the theory provided they are combined with parity.
Notice that this poses no problem with the generalized theta-term since, as parity reverses the sign of $\epsilon^{\mu\nu\rho\sigma}F^\Lambda_{\mu\nu} F^\Sigma_{\rho\sigma}$, under ${\bf P}$ we have:
\begin{equation}
\I_{\Lambda\Sigma}\rightarrow\I_{\Lambda\Sigma}\;;\qquad
\R_{\Lambda\Sigma}\rightarrow -\R_{\Lambda\Sigma}\,,
\end{equation}
see equation (\ref{Ntra}), so that the corresponding term $\epsilon^{\mu\nu\rho\sigma}F^\Lambda_{\mu\nu} F^\Sigma_{\rho\sigma}\R_{\Lambda\Sigma}$ in the Lagrangian is invariant.
The global symmetry group of the theory is therefore described by a group
\begin{equation}
G=G_0\times \mathbb{Z}_2=\{G_0,\,G_0\cdot p\}\,,
\end{equation}
where $G_0$ is the \emph{proper duality} group defined by the identity sector of $G$ and $p$ is the element of $G$ which corresponds, in a suitable symplectic frame, to the anti-symplectic matrix ${\bf P}$\,:\; ${\bf P}={\bf A}[p]$.

\subparagraph{Example.}
Let us discuss the simple example of the lower-half complex plane
\begin{equation}
G/H={\rm SL}(2,\mathbb{R})/{\rm SO}(2)\,.
\end{equation}
This manifold is parametrized by a complex coordinate $z$, with ${\rm Im}(z)<0$.
As symplectic representation of $G={\rm SL}(2,\mathbb{R})$ we can choose the fundamental representation and the following basis of generators of $\mathfrak{g}=\mathfrak{sl}(2,\mathbb{R})$:
\begin{align}
\mathfrak{sl}(2,\mathbb{R})=\{\sigma^1,\,i\,\sigma^2,\sigma^3\}=\left\{\left(\begin{matrix}0 & 1 \cr 1 & 0\end{matrix}\right),\,\left(\begin{matrix}0 & 1 \cr -1 & 0\end{matrix}\right),\,\left(\begin{matrix}1 & 0 \cr 0 & -1\end{matrix}\right)\right\}\,.
\end{align}
The subalgebra $\Solv$ of upper-triangular generators
\begin{align}
\Solv=\{\sigma^3,\,\sigma^+\}\,\,,\,\,\,\,\sigma^+\equiv \left(\begin{matrix}0 & 1 \cr 0 & 0\end{matrix}\right)\,.
\end{align}
defines the solvable parametrization $\phi^s=(\varphi,\,\chi)$, in which the coset representative $\LL$ has the following form:
\begin{align}
\LL(\varphi,\,\chi)\equiv e^{\chi \sigma^+}\,e^{\frac{\varphi}{2}\sigma^3}=\left(
\begin{array}{ll}
 1 & \chi  \\
 0 & 1
\end{array}
\right)\left(
\begin{array}{ll}
 e^{\varphi /2} & 0 \\
 0 & e^{-\varphi /2}
\end{array}
\right)\,\in\,\, e^{\Solv}\,.
\end{align}
The relation between the solvable coordinates and $z$ is
\begin{equation}
z\=z_1+i\,z_2\=\chi-i\,e^{\varphi}\,.
\end{equation}
The metric reads:
\begin{equation}
ds^2=\frac{{d\varphi }^2}{2}+\frac{1}{2}{d\chi }^2 e^{-2 \varphi }=\frac{1}{2z_2^2}\,dz d\bar{z}\;;
\end{equation}
and the matrix $\mathcal{M}(\phi)_{MN}$ reads:
\begin{align}
\mathcal{M}(z,\,\bar{z})_{MN}=\mathbb{C}_{MP}\,\mathbb{L}(\phi)^P{}_{\underline{L}}\,\mathbb{L}(\phi)^R{}_{\underline{L}}\,
\mathbb{C}_{RN}=
\frac{1}{z_2}\left(
\begin{array}{cc}
 1 & -{z_1} \\
 -{z_1} & |z|^2
\end{array}
\right)\,.
\end{align}
The generic isometry which is continuously connected to the identity is a holomorphic transformation
of the form
\begin{equation}
z\rightarrow z'=\frac{a z +b}{ c z +d}\,,\qquad ad-bc=1\,,
\end{equation}
corresponding to the ${\rm SL}(2,\mathbb{R})$ transformation $\S=\left(\begin{matrix}a & b\cr c & d\end{matrix}\right)$ with ${\rm det}(\S)=1$.
The reader can easily verify that:
\begin{equation}
\mathcal{M}(z',\,\bar{z}')=\S^{-T}\mathcal{M}(z,\,\bar{z})\S^{-1}\,.
\end{equation}
We also have the following isometry:
\begin{equation}
z\rightarrow -\bar{z}\,,\label{pisom}
\end{equation}
which is not in the identity sector of the isometry group, and corresponds to the anti-symplectic transformation ${\bf P}={\rm diag}(1,-1)$ in that:
 \begin{equation}
\mathcal{M}(-\bar{z},\,-z)={\bf P}^{-T}\mathcal{M}(z,\,\bar{z}){\bf P}^{-1}\,.
\end{equation}
This corresponds to a parity transformation whose effect is to change the sign of the pseudo-scalar $\chi$ while leaving the scalar $\varphi$ inert:
\begin{equation}
\mbox{parity}:\;\;\chi\rightarrow -\chi\,\,,\,\,\,\,\varphi\rightarrow \varphi\,.
\end{equation}
Notice that the correspondence between the linear transformation ${\bf P}$ and the isometry (\ref{pisom}) exists since ${\bf P}$ is an \emph{outer-automorphism} of the isometry algebra $\mathfrak{g}=\mathfrak{sl}(2,\mathbb{R})$, namely:
\begin{equation}
{\bf P}^{-1}\mathfrak{sl}(2,\mathbb{R}){\bf P}\=\mathfrak{sl}(2,\mathbb{R})\,,
\end{equation}
while ${\bf P}$ is \emph{not} in ${\rm SL}(2,\mathbb{R})$ and the above transformation cannot be offset by any conjugation by ${\rm SL}(2,\mathbb{R})$ elements. Analogous outer-automorphisms implementing parity can be found in other extended supergravities, including the maximal one in which $G=\Exc\times \mathbb{Z}_2$\, \cite{Ferrara:2013zga}.

\subparagraph{Solitonic solutions, electric-magnetic charges and duality.}
Ungauged supergravities only contain fields which are neutral with respect to the ${\rm U}(1)^{n_v}$ gauge-symmetry of the vector fields.
These theories however feature \emph{solitonic solutions}, namely configurations of neutral fields which carry ${\rm U}(1)^{n_v}$ electric-magnetic charges. These solutions are typically black holes in four dimensions or black branes in higher and have been extensively studied in the literature.
On a charged dyonic solution of this kind, we define the electric and magnetic charges as the integrals%
\footnote{
the electric and magnetic charges $(e,m)$ are expressed in the rationalized-Heaviside-Lorentz (RHL) system of units
}:
\begin{align}
e_\Lambda\equiv\int_{S^2} \Gd_{\Lambda}=\frac{1}{2}\,\int_{S^2} \Gd_{\Lambda\,\mu\nu}\,dx^\mu\wedge dx^\nu \,\,,\nonumber\\
m^\Lambda\equiv\int_{S^2} F^{\Lambda}=\frac{1}{2}\,\int_{S^2} F^{\Lambda}{}_{\mu\nu}\,dx^\mu\wedge dx^\nu \,,
\end{align}
where $S^2$ is a spatial two-sphere. They define a symplectic vector $\Gamma^M$:
\begin{align}
\Gamma=(\Gamma^M)=\left(\begin{matrix}m^\Lambda\cr e_\Lambda\end{matrix}\right)=\int_{S^2} \mathbb{F}^M\,.
\end{align}
These are the \emph{quantized charges}, namely they satisfy the Dirac-Schwinger-Zwanziger quantization condition for dyonic particles \cite{Dirac:1931kp,Schwinger:1966nj,Zwanziger:1968rs}:
\begin{equation}
\Gamma_2^T\mathbb{C}\Gamma_1=m_2^{\Lambda}\,e_{1\Lambda}-m_1^{\Lambda}\,e_{2\Lambda}= 2\pi\,\hbar\,c\,n\,\,\,;\,\,\,\,n\in \mathbb{Z}\,.\label{DZS}
\end{equation}
At the quantum level, the dyonic charges therefore belong to a symplectic lattice and this breaks the duality group $G$ to a suitable discrete subgroup $G(\mathbb{Z})$ which leaves this symplectic lattice invariant:
\begin{equation}
G(\mathbb{Z})\equiv G\cap {\rm Sp}(2n_v,\mathbb{Z})\,.
\end{equation}
This discrete symmetry group of surviving quantum corrections (or a suitable extension thereof) was conjectured in \cite{Hull:1994ys} to encode all known string/M-theory dualities.

\subsection{Symplectic frames and Lagrangians}\label{sframes}
As pointed out earlier, the duality action $\Rs_v[G]$ of $G$ depends on which elements, in the basis of the ${\bf 2\,n_v}$ representation, are chosen to be the $n_v$ electric vector fields (appearing in the Lagrangian) and which their magnetic duals namely on the choice of the \emph{symplectic frame} which determines the embedding of the group $G$ inside $\Sp(2n_v,\,\mathbb{R})$. Different choices of the symplectic frame may yield inequivalent Lagrangians (that is Lagrangians that are not related by local field redefinitions) with different global symmetries. Indeed, the global symmetry group of the Lagrangian%
\footnote{
here we only consider \emph{local} transformations on the fields
}
is defined as the subgroup $G_{el}\subset G$, whose duality action is linear on the electric field strengths
\begin{equation}
g\in G_{el}\;:
\quad \Rs_v[g]=
\left(\begin{matrix}
A^\Lambda{}_\Sigma & \Zero \cr
C_{\Lambda\Sigma} & D_\Lambda{}^\Sigma
\end{matrix}\right)\,,\label{ge}
\end{equation}
where $D=A^{-T}$ by the symplectic condition, so that
\begin{align}
g\in G_{el}\;:\quad
&F^\Lambda \rightarrow\,F^{\prime\Lambda}=A^\Lambda{}_\Sigma\,F^\Sigma\;,\nne
&\Gd_\Lambda \rightarrow\,\Gd'_{\Lambda}=C_{\Lambda\Sigma}\,F^\Sigma+
D_\Lambda{}^\Sigma\,\Gd_\Sigma
\,.\label{Gel}
\end{align}
Indeed, as the reader can verify using eq.\ (\ref{Ntra}), under the above transformation the matrices $\I,\,\R$ transform as follows:
\begin{equation}
\I_{\Lambda\Sigma}\rightarrow D_\Lambda{}^\Pi D_\Sigma{}^\Delta\,\I_{\Pi\Delta}
\,;\quad\;
\R_{\Lambda\Sigma}\rightarrow D_\Lambda{}^\Pi D_\Sigma{}^\Delta\,\R_{\Pi\Delta}+C_{\Lambda\Pi}\,D_{\Sigma}{}^\Pi\,,
\end{equation}
and the consequent variation of the Lagrangian reads
\begin{equation}
\LB=
\frac{1}{8}\,C_{\Lambda\Pi}\,A^\Pi{}_{\Sigma}\epsilon^{\mu\nu\rho\sigma}\,F^\Lambda_{\mu\nu}F^\Sigma_{\rho\sigma}\,,\label{deltaLC}
\end{equation}
which is a \emph{total derivative} since $C_{\Lambda\Pi}\,A^\Pi{}_{\Sigma}$ is constant. These transformations are called \emph{Peccei-Quinn transformations } and follow from shifts in certain axionic scalar fields. They are a symmetry of the classical action, while invariance of the perturbative path-integral requires the variation (\ref{deltaLC}), integrated over space-time, to be proportional through an integer to $2\pi \hbar$. This constrains the symmetries to close to a discrete subgroup $G(\mathbb{Z})$ of $G$ whose duality action is implemented by integer-valued matrices ${\Rs_v}[g]$. Such restriction of $G$ to $G(\mathbb{Z})$ in the quantum theory was discussed earlier as a consequence of the Dirac-Schwinger-Zwanziger quantization condition for dyonic particles (\ref{DZS}).\par
From (\ref{Gel}) we see that, while the vector field strengths $F^\Lambda_{\mu\nu}$ and their duals $\Gd_{\Lambda\,\mu\nu}$ transform together under $G$ in the ($2n_v$--dimensional) symplectic representation $\Rs_v$, the vector field strengths alone transform linearly under the action of $G_{el}$ in a smaller representation ${\bf n_v}$, defined by the $A$-block in (\ref{ge}).\par\medskip
Different symplectic frames of a same ungauged theory may originate from different compactifications.
A distinction here is in order. In $\N\geq 3$ theories, scalar fields always enter the same multiplets as the vector fields. Supersymmetry then implies their non-minimal coupling to the latter and that the scalar manifold is endowed with a symplectic structure associating with each isometry a constant symplectic matrix. In $\N=2$ theories, scalar fields may sit in vector multiplets or hypermultiplets. The former span a \emph{special K\"ahler manifold}, the latter a \emph{quaternionic K\"ahler} one, so that the scalar manifold is always factorized in the product of the two:
\begin{equation}
\Mscal^{\scalebox{0.5}{$\,(\N=2)$}}\=\MsSK\times \MsQK\,.\label{SKQK}
\end{equation}
The scalar fields in the hypermultiplets are not connected to vector fields through supersymmetry and consequently they do not enter the matrices $\I(\phi)$ and $\R(\phi)$. As a consequence of this the isometries of the Quaternionic-K\"ahler manifolds spanned by these scalars are associated with trivial duality transformations
\begin{equation}
g\,\in\,\text{isom. of}\;\MsQK
\;\;\;\,\,\,\Rightarrow\quad
\Rs_v[g]=\Id\;,\label{qisom}
\end{equation}
while only $\MsSK$ features a flat symplectic structure which defines the embedding of its isometry group inside ${\rm Sp}(2n_v,\mathbb{R})$ and the couplings of the vector multiplet-scalars to the vector fields through the matrix $\mathcal{M}(\phi)$. It is important to remark that such structure on a special K\"ahler manifold exists even if the manifold itself is not homogeneous. This means that one can still define the symplectic matrix $\LL(\phi)$ and, in terms of the components $\I_{\Lambda\Sigma}$ and $\R_{\Lambda\Sigma}$, also the matrix $\mathcal{M}(\phi)$ as in (\ref{Mcos}), although $\LL(\phi)$ has no longer the interpretation of a coset representative for non-homogeneous manifolds.\par
It is convenient for later purposes to rewrite the transformation properties of the bosonic fields the group $G$, discussed in this section, in the following infinitesimal form:
\begin{align}
G\;:\quad
\begin{cases}
\delta\,\LL = \Lambda^\alpha\,t_\alpha\,\LL\;,\nn\\
\delta \mathbb{F}^{{M}}_{\mu\nu} =
-\Lambda^{\alpha}\,(t_{\alpha})_{{{N}}}{}^{{{M}}}\;\mathbb{F}_{\mu\nu}^{{{N}}}\,,
\end{cases}
\end{align}
in terms of the infinitesimal generators $t_\alpha$ of $G$ earlier introduced, satisfying the relation (\ref{talg}).
The matrices $(t_{\alpha})_{M}{}^{N}$ define the infinitesimal duality action of $G$ and are symplectic generators
\begin{equation}
(t_{\alpha})_{M}{}^{N}\,\CC_{NP} = (t_{\alpha})_{P}{}^{N}\,\CC_{NM}\, \qquad\;  {\scriptstyle M},\,{\scriptstyle N},\dotsc=1,\dotsc,\,2n_v\;.
\end{equation}
This is equivalently stated as the property of the tensor $t_{\alpha\,MN}\equiv (t_{\alpha})_{M}{}^{P}\,\mathbb{C}_{PN}$ of being symmetric in {\footnotesize $M\,N$}:
\begin{equation}
(t_\alpha)_{MN}=(t_\alpha)_{NM}\,.
\end{equation}

\subsection{The fermionic sector} \label{fsector} Fermions in supergravity transform covariantly with respect to the isotropy group $H$ of the scalar manifold, which has the general form (\ref{Hgroup}), while they do not transform under $G$, as opposed to the bosonic fields. Bosons and fermions have therefore definite transformation properties with respect to different groups of internal symmetry. The matrix $\mathbb{L}$, defining the coset representative for homogeneous scalar manifolds, transforms under the action of $G$ to the left and of $H$ to the right, according to (\ref{gLh})
\begin{equation}
G\,\,\rightarrow \,\,\,\LL\,\,\,\leftarrow\,\, H\,,
\end{equation}
and thus has the right index structure to ``mediates'' in the Lagrangian between bosons and fermions. This means that we can construct $G$-invariant terms by contracting $\LL$ to the left by bosons (scalars, vectors and their derivatives), and to the right by fermions
\begin{equation}
(\mbox{Bosons}) \star \mathbb{L}(\phi)\star (\mbox{Fermions})\,,\label{BLF}
\end{equation}
the two $\star$ symbols denote some contraction of indices: $G$-invariant to the left and $H$-invariant to the right.
The ``Boson'' part of (\ref{BLF}) may also contain $\LL$ and its derivatives.
These are the kind of terms occurring in the field equations. If under a transformation $g\in G$, symbolically:
\begin{equation}
\mbox{Bosons}\;\rightarrow\;\mbox{Bosons}'=\mbox{Bosons}\star g^{-1}\,,
\end{equation}
and the \emph{fermions are made to transform under the compensating transformation} $h(\phi,g)$ in (\ref{gLh}):
\begin{equation}
\mbox{Fermions}\;\rightarrow\;\mbox{Fermions}'=h(\phi,g)\star \mbox{Fermions}\,.\label{Hfermi}
\end{equation}
Using (\ref{gLh}) we see that (\ref{BLF}) remains invariant:
\begin{equation}
(\mbox{Bosons})' \star \mathbb{L}(g\star\phi)\star (\mbox{Fermions}')=(\mbox{Bosons}) \star \mathbb{L}(\phi)\star (\mbox{Fermions})\,.
\end{equation}
The Lagrangian is manifestly invariant under local $H$-transformations since the covariant derivatives on the fermion fields contain the $H$-connection%
\footnote{
we define $\W_\mu\equiv \W_s\,\partial_\mu\phi^s$
}
$\W_\mu$:
\begin{equation}
\mathscr{D}_\mu\xi=\nabla_\mu\xi+\W_\mu\star \xi\,,\label{Dxi}
\end{equation}
where, as usual, the $\star$ symbol denotes the action of the $\mathfrak{H}$-valued connection $\W_\mu$ on $\xi$ in the corresponding $H$-representation. The reader can verify that (\ref{Dxi}) is indeed covariant under local $H$-transformations (\ref{Hfermi}), provided $\W$ is transformed according to (\ref{omtra}). As opposed to the gauge groups we are going to introduce by the gauging procedure, which involve minimal couplings to the vector fields of the theory, the local $H$-symmetry group of the ungauged theory is not gauged by the vector fields, but by a \emph{composite connection} $\W_\mu$, which is a function of the scalar fields and their derivatives. The minimal coupling $\W_\mu\star \xi$ is an example of the boson-fermion interaction term (\ref{BLF}).\par
It is useful to write the coupling (\ref{BLF}) in the following form:
\begin{equation}
{\bf f}(\phi,\mbox{Bosons}) \star (\mbox{Fermions})\,,\label{BLF2}
\end{equation}
where we have introduced the $H$-covariant \emph{composite field}:
\begin{equation}
{\bf f}(\phi,\mbox{Bosons})\equiv (\mbox{Bosons}) \star \mathbb{L}(\phi)\,,
\end{equation}
obtained by \emph{dressing} the bosonic fields and their derivatives with the coset-representative so as to obtain an $H$-covariant quantity with the correct $H$-index structure to contract with fermionic currents. Indeed under a $G$-transformation
\begin{equation}
{\bf f}(g\star\phi,\mbox{Bosons}')\equiv {\bf f}(\phi,\mbox{Bosons})\star h(\phi,g)^{-1}\,,
\end{equation}
The manifest $H$-invariance of the supergravity theory requires the supersymmetry transformation properties of the femionic fields to be $H$-covariant. Indeed such transformation rules, which in rigid supersymmetric theories (i.e.\ theories which are invariant only under global supersymmetry) can be schematically described as follows%
\footnote{
this is a schematic representation in which we have suppressed the Lorentz indices and gamma-matrices
}:
\begin{equation}
\delta\mbox{Fermion}=\sum_{{\tiny \mbox{Bosons}}}\partial\mbox{Boson}\cdot\epsilon\,,
\end{equation}
and in supergravity theories have the following general $H$-covariant form%
\footnote{
the gravitino field has an additional term $\mathscr{D}\epsilon$ which is its variation as the gauge field of local supersymmetry
}
\begin{equation}
\delta\mbox{Fermion}=\sum_{{\tiny \mbox{Bosons}}}{\bf f}(\phi,\mbox{Bosons})\cdot\epsilon\,,
\end{equation}
where the space-time derivatives of the bosonic fields are dressed with the scalars in the definition of ${\bf f}(\phi,\mbox{Bosons})$.
Examples of composite fields ${\bf f}(\phi,\mbox{Bosons})$ are the vielbein of the scalar manifold (pulled back on space-time) $\mathcal{P}_\mu\equiv \mathcal{P}_s\,\partial_\mu\phi^s$, the $H$-connection $\W_\mu$ in (\ref{Dxi}), the dressed vector field-strengths
\begin{equation}
{\bf F}(\phi,\partial A)^{\;\underline{M}}_{\mu\nu}\equiv -(\LL(\phi)^{-1})^{\underline{M}}{}_N\,\mathbb{F}^N_{\mu\nu}\,,
\end{equation}
or the \emph{$\Tb$-tensor}, to be introduced later, in which the bosonic field to be dressed by the coset representative is the \emph{embedding tensor} $\Theta$ defining the choice of the gauge algebra.

\section{Gauging Supergravities} \label{sec:3}
We have reviewed the field content and the Lagrangian of ungauged supergravity, as well as the action of the global symmetry group $G$. Now we want to discuss how to construct a gauged theory from an ungauged one.\par
In the following, we will employ a covariant formalism in which the possible gaugings will be encoded into an object called embedding tensor, that can be characterized group-theoretically \cite{Cordaro:1998tx,Nicolai:2000sc,deWit:2002vt}.

\subsection{The gauging procedure step-by-step}\label{gaugingsteps}
As anticipated in the Introduction, the gauging procedure consists in promoting a suitable global symmetry subgroup $G_g\subset G_{el}$ of the Lagrangian to a local symmetry gauged by the vector fields of the theory. This requirement gives us a preliminary condition
\begin{equation}
\dim(G_g) ~\le~ n_v\;.\label{preliminary}
\end{equation}
As explained in Sect.\ \ref{sframes}, \emph{different symplectic frames correspond to ungauged Lagrangians with different global symmetry groups $G_{el}$ and thus to different choices for the possible gauge groups.}\par
The first condition for the global symmetry subgroup $G_g$ to become a viable gauge group, is that there should exist a subset $\{A^{\hat{\Lambda}}\}$ of the vector fields%
\footnote{
we describe by hatted-indices those pertaining to the symplectic frame in which the Lagrangian is defined
}
which transform under the co-adjoint representation of the duality action of $G_g$. These fields will become the \emph{gauge vectors} associated with the \emph{generators} $X_{\hat\Lambda}$ of the subgroup $G_g$. \par
We shall name \emph{electric frame} the symplectic frame defined by our ungauged Lagrangian and labeled by hatted indices.\par
Note that, once the gauge group is chosen within $G_{el}$, its action on the various fields is fixed, being it defined by the action of $G_g$ as a global symmetry group of the ungauged theory (duality action on the vector field strengths, non-linear action on the scalar fields and indirect action through $H$-compensators on the fermionic fields): fields are thus automatically associated with representations of $G_g$.\par
After the initial choice of $G_g$ in $G_{el}$, the first part of the procedure is quite standard in the construction of non-abelian gauge theories: we introduce a gauge-connection, gauge-curvature (i.e.\ non-abelian field strengths) and covariant derivatives. We will also need to introduce an extra topological term needed for the gauging of the Peccei-Quinn transformations (\ref{deltaLC}). This will lead us to construct a gauged Lagrangian $\Lgaug^{(0)}$ with manifest local $G_g$-invariance. Consistency of the construction will imply constraints on the possible choices of $G_g$ inside $G$. The minimal couplings will however break supersymmetry.\par
The second part of the gauging procedure consists in further deforming the Lagrangian $\Lgaug^{(0)}$ in order to restore the original supersymmetry of the ungauged theory and, at the same time, preserving local $G_g$-invariance.
\smallskip

\subparagraph{\textbf{Step 1.} Choice of the gauge algebra.}
We start by introducing the gauge connection:
\begin{equation}
\Omega_{g}=\Omega_{g\,\mu}dx^\mu\;;
\quad \Omega_{g\,\mu}\equiv
g\,A^{\hat{\Lambda}}_\mu\,X_{\hat{\Lambda}}\,,\label{gconnection}
\end{equation}
$g$ being the coupling constant.
The gauge-algebra relations can be written
\begin{equation}
\left[X_{\hat{\Lambda}},\,X_{\hat{\Sigma}}\right]\=f_{{\hat{\Lambda}}{\hat{\Sigma}}}{}^{\hat{\Gamma}}\,X_{\hat{\Gamma}}\,,\label{gaugealg}
\end{equation}
and are characterized by the structure constants
$f_{{\hat{\Lambda}}{\hat{\Sigma}}}{}^{\hat{\Gamma}}$. This closure condition should be regarded as a constraint on $X_{\hat{\Lambda}}$, since the structure constants are not generic but fixed in terms of the action of the gauge generators on the vector fields as global symmetry generators of the original ungauged theory.
To understand this, let us recall that $G_g$ is a subgroup of $G_{el}$ and thus its electric-magnetic duality action, as a global symmetry group, will have the form (\ref{ge}). The duality action on the vector field strengths and their duals of the infinitesimal generators $X_{\hat{\Lambda}}$ will then by represented by a symplectic matrix of the form (\ref{ge})
\begin{equation}
\left(X_{\hat{\Lambda}}\right)^{\hat{M}}{}_{\hat{N}}\=
\left(\begin{matrix}
X_{\hat{\Lambda}}{}^{{\hat{\Lambda}}}{}_{{\hat{\Sigma}}}  &  \Zero \cr
X_{{\hat{\Lambda}}\,{\hat{\Gamma}}{\hat{\Sigma}}} &  X_{{\hat{\Lambda}}\,{\hat{\Gamma}}}{}^{\hat{\Delta}}
\end{matrix}\right)
\,,\label{xsymp}
\end{equation}
where $X_{\hat{\Lambda}}{}^{{\hat{\Lambda}}}{}_{{\hat{\Sigma}}}$ and $X_{{\hat{\Lambda}}\,{\hat{\Gamma}}}{}^{\hat{\Delta}}$ are the infinitesimal generators of the $A$ and $D$-blocks in (\ref{ge}) respectively, while $X_{{\hat{\Lambda}}\,\hat{\Gamma}\hat{\Sigma}}$ describes the infinitesimal $C$-block.
It is worth emphasizing here that we do not identify the generator $X_{{\hat{\Lambda}}}$ with the symplectic matrix defining its electric-magnetic duality action. As pointed our in Sect.\ \ref{sframes}, there are isometries in $\N=2$ models which do not have duality action, see eq.\ (\ref{qisom}), namely for which the matrix in (\ref{xsymp}) is null.\par
The variation of the field strengths under an infinitesimal transformation $\xi^{{\hat{\Lambda}}}\,X_{{\hat{\Lambda}}}$, whose duality action is described by (\ref{xsymp}), is:
\begin{equation}
\delta \mathbb{F}^{\hat{M}}=\xi^{{\hat{\Lambda}}}\,(X_{{\hat{\Lambda}}}){}^{\hat{M}}{}_{\hat{N}}\,\mathbb{F}^{\hat{N}}
\;\;\Rightarrow\;\;
\begin{cases}\delta F^{\hat{\Lambda}}=
\xi^{{\hat{\Gamma}}}X_{{\hat{\Gamma}}}{}^{\hat{\Lambda}}{}_{\hat{\Sigma}}\,F^{\hat{\Sigma}}\,,\cr \delta\Gd_{\hat{\Lambda}}=
\xi^{{\hat{\Gamma}}}X_{{\hat{\Gamma}}\,{\hat{\Lambda}}{\hat{\Sigma}}} F^{\hat{\Sigma}}+\xi^{{\hat{\Gamma}}}X_{{\hat{\Gamma}}{\hat{\Lambda}}}{}^{\hat{\Sigma}}\,\Gd_{\hat{\Sigma}}\,.
\end{cases}
\label{deltas}
\end{equation}
The imposed symplectic condition on the matrix $X_{\hat{\Lambda}}$ implies
the properties:
\begin{align}
X_{{\hat{\Lambda}} \hat{M}}{}^{\hat{P}}\,\mathbb{C}_{\hat{N} \hat{P}}=X_{{\hat{\Lambda}} \hat{N}}{}^{\hat{P}}\,\mathbb{C}_{\hat{M} \hat{P}}
\;\quad\Leftrightarrow\quad\;
\begin{cases}
X_{\hat{\Lambda}}{}^{{\hat{\Sigma}}}{}_{{\hat{\Gamma}}}\=-X_{{\hat{\Lambda}}{\hat{\Gamma}}}{}^{\hat{\Sigma}}\,,\cr
X_{{\hat{\Lambda}}\,{\hat{\Gamma}}{\hat{\Sigma}}}\=X_{{\hat{\Lambda}}\,{\hat{\Sigma}}{\hat{\Gamma}}}\,.
\end{cases}
\label{sympconde}
\end{align}
The condition that $A^{\hat{\Lambda}}_\mu$ transform in the co-adjoint representation of the gauge group:
\begin{equation}
\delta F^{\hat{\Lambda}}=
\xi^{{\hat{\Gamma}}}\,f_{{\hat{\Gamma}}{\hat{\Sigma}}}{}^{\hat{\Lambda}}F^{\hat{\Sigma}}\,,
\end{equation}
and the transformation properties (\ref{deltas}), leads us to identify the structure constants of the gauge group in (\ref{gaugealg}) with
the diagonal blocks of the symplectic matrices $X_{\hat{\Lambda}}$:
\begin{equation}
f_{{\hat{\Gamma}}{\hat{\Sigma}}}{}^{\hat{\Lambda}}=-X_{{\hat{\Gamma}}{\hat{\Sigma}}}{}^{\hat{\Lambda}}\,,\label{idenfx}
\end{equation}
so that the closure condition reads
\begin{equation}
\left[X_{\hat{\Lambda}},\,X_{\hat{\Sigma}}\right]\=-X_{{\hat{\Lambda}}{\hat{\Sigma}}}{}^{\hat{\Gamma}}\,X_{\hat{\Gamma}}\,,\label{gaugealg2}
\end{equation}
and is a quadratic constraint on the tensor $X_{{\hat{\Lambda}}}{}^{\hat{M}}{}_{\hat{N}}$. The identification (\ref{idenfx}) also implies
\begin{align}
X_{({\hat{\Gamma}}{\hat{\Sigma}})}{}^{\hat{\Lambda}}=0\,.\label{lin1}
\end{align}
\medskip
The closure condition (\ref{gaugealg2}) can thus be interpreted in two equivalent ways:
\begin{enumerate}[$\circ$,itemsep=1ex]
\item{the vector fields $A^{\hat{\Lambda}}_\mu$ transform in the co-adjoint representation of $G_g$ under its action as global symmetry, namely
     \begin{equation}
     {\bf n_v}=\text{co-adj}(G_g)\,;
     \end{equation}
     }
\item{the gauge generators $X_{{\hat{\Lambda}}}$ are invariant under the action of $G_g$ itself:
\begin{equation}
\delta_{{\hat{\Lambda}}}X_{{\hat{\Sigma}}}\equiv [X_{\hat{\Lambda}},\,X_{{\hat{\Sigma}}}]+X_{{\hat{\Lambda}}{\hat{\Sigma}}}{}^{\hat{\Gamma}}\,X_{\hat{\Gamma}}=0\,.
\end{equation}}
\end{enumerate}
\medskip

\subparagraph{\textbf{Step 2.} Introducing gauge curvatures and covariant derivatives.}
Having defined the gauge connection (\ref{gconnection}) we also define its transformation property under a local $G_g$-transformation $ {\bf g}(x)\in G_g$:
\begin{equation}
\Omega_g \;\rightarrow\; \Omega_g'={\bf g}\,\Omega_g\,{\bf g}^{-1}+d {\bf g}\,{\bf g}^{-1}=g\,A^{\prime \hat{\Lambda}}\,X_{\hat{\Lambda}}\,.\label{Omtrasg}
\end{equation}
Under an infinitesimal transformation \;${\bf g}(x)\equiv \Id+g\,\zeta^{\hat{\Lambda}}(x)\,X_{\hat{\Lambda}}$,\, eq.\ (\ref{Omtrasg}) implies the following transformation property of the gauge vectors:
\begin{equation}
\delta A^{\hat{\Lambda}}_\mu\=\cD_\mu\zeta^{\hat{\Lambda}}~\equiv~ \partial_\mu\zeta^{\hat{\Lambda}}+g\,A_\mu^{\hat{\Sigma}}X_{\hat{\Sigma}\hat{\Gamma}}{}^{\hat{\Lambda}}\,\zeta^{\hat{\Gamma}} \,,
\end{equation}
where we have introduced the $G_g$-covariant derivative of the gauge parameter $\cD_\mu\zeta^{\hat{\Lambda}}$.\par
As usual in the construction of non-abelian gauge-theories, we define the gauge curvature%
\footnote{
here we use the following convention for the definition of the components of a form: $\omega_{(p)}=\frac{1}{p!}\,\omega_{\mu_1\dots\mu_p}\,dx^{\mu_1}\wedge \dots dx^{\mu_p}$
}
\begin{equation}
g\,\mathcal{F}=g\,F^{\hat{\Lambda}}\,X_{\hat{\Lambda}}=\frac{g}{2}\,F^{\hat{\Lambda}}_{\mu\nu}\,dx^\mu\wedge dx^\nu\,X_{\hat{\Lambda}}\equiv d\Omega_g-\Omega_g\wedge \Omega_g\,,\label{calF}
\end{equation}
which, in components, reads:
 \begin{equation}
F_{\mu\nu}^{{\hat{\Lambda}}} \= \partial_\mu A^{\hat{\Lambda}}_\nu-\partial_\nu A^{\hat{\Lambda}}_\mu -
g\,f_{{\hat{\Gamma}}{\hat{\Sigma}}}{}^{\hat{\Lambda}}\,A^{\hat{\Gamma}}_\mu\,A^{\hat{\Sigma}}_\nu\= \partial_\mu A^{\hat{\Lambda}}_\nu-\partial_\nu A^{\hat{\Lambda}}_\mu +
g\,X_{{\hat{\Gamma}}{\hat{\Sigma}}}{}^{\hat{\Lambda}}\,A^{\hat{\Gamma}}_\mu\,A^{\hat{\Sigma}}_\nu\;.\label{defF}
\end{equation}
The gauge curvature transforms covariantly under a transformation $ {\bf g}(x)\in G_g$:
\begin{equation}
\mathcal{F} \;\rightarrow\; \mathcal{F}'={\bf g}\,\mathcal{F}\,{\bf g}^{-1}\,,\label{gaugecovF}
\end{equation}
and satisfies the Bianchi identity:
\begin{equation}
\cD\mathcal{F}\equiv d\mathcal{F}-\Omega_g\wedge \mathcal{F}+\mathcal{F}\wedge \Omega_g=0
\;\;\Leftrightarrow\;\;
\cD F^{{\hat{\Lambda}}}\equiv dF^{{\hat{\Lambda}}}+g\,X_{{\hat{\Sigma}}{\hat{\Gamma}}}{}^{{\hat{\Lambda}}}A^{{\hat{\Sigma}}}\wedge F^{{\hat{\Lambda}}}=0\,,
\end{equation}
where we have denoted by $\cD F^{{\hat{\Lambda}}}$ the $G_g$--covariant derivative acting on $F^{{\hat{\Lambda}}}$.
In the original ungauged Lagrangian we then replace the abelian field strengths by the new $G_g$-covariant ones:
\begin{equation}
\partial_\mu A^{\hat{\Lambda}}_\nu-\partial_\nu A^{\hat{\Lambda}}_\mu\,\,\rightarrow\,\,\,\,\partial_\mu A^{\hat{\Lambda}}_\nu-\partial_\nu A^{\hat{\Lambda}}_\mu +
g\,X_{{\hat{\Gamma}}{\hat{\Sigma}}}{}^{\hat{\Lambda}}\,A^{\hat{\Gamma}}_\mu\,A^{\hat{\Sigma}}_\nu\,.\label{replaceF}
\end{equation}
After having given the gauge fields a $G_g$-covariant description in the Lagrangian through the non-abelian field strengths, we now move to the other fields.
The next step in order to achieve local invariance of the Lagrangian under $G_g$ consists in replacing ordinary derivatives by covariant ones
\begin{equation}
\dmu\;\;\longrightarrow\;\;
\cD_\mu\=\dmu-g\,A^{\hat{\Lambda}}\,X_{\hat{\Lambda}} \,.\label{covder}
\end{equation}
As it can be easily ascertained, the covariant derivatives satisfy the identity which is well known from gauge theories:
\begin{equation}
\cD^2=-g\,\mathcal{F}=-g\,F^{\hat{\Lambda}}\,X_{\hat{\Lambda}}
\quad\Leftrightarrow\quad
[\cD_\mu,\,\cD_\nu]=-g\,F^{\hat{\Lambda}}_{\mu\nu}\,X_{\hat{\Lambda}}\,.\label{D2F}
\end{equation}
Aside from the vectors and the metric, the remaining bosonic fields are the scalars $\phi^s$, whose derivatives are covariantized using the Killing vectors $k_{\hat{\Lambda}}$ associated with
the action of the gauge generator $X_{\hat{\Lambda}}$ as an isometry:
\begin{equation}
\dmu\;\;\longrightarrow\;\;
\cD_\mu\phi^s\=\dmu\phi^s-g\,A^{\hat{\Lambda}}\,k^s_{\hat{\Lambda}}(\phi)\,,\label{covderphi}
\end{equation}
The replacement (\ref{covder}), and in particular (\ref{covderphi}), amounts to the \emph{introduction of minimal couplings} for the vector fields.\par\smallskip
Care is needed for the fermion fields which, as we have discussed above, do not transform directly under $G$, but under the corresponding compensating transformations in $H$. This was taken into account by writing the $H$-connection $\W$ in the fermion $H$-covariant derivatives. Now we need to promote such derivatives to $G_g$-covariant ones, by minimally coupling the fermions to the gauge fields. This is effected by modifying the $H$-connection.\par
For homogeneous scalar manifolds redefine the left-invariant 1-form $\Omega$ (pulled-back on space-time), defined on them in (\ref{omegapro}), by a \emph{gauged} one obtained by covariantizing the derivative on the coset representative:
\begin{equation}
\Omega_\mu= \Lc^{-1}\partial_\mu \Lc
\;\;\;\longrightarrow\;\;\;
\hat{\Omega}_\mu\equiv \Lc ^{-1}\cD \Lc =
\Lc^{-1}\left(\partial_\mu-g\,A^{\hat{\Lambda}}_\mu\,X_{\hat{\Lambda}}\right) \Lc =\hat{\mathcal{P}}_\mu+\hat{\W}_\mu
\label{hatOm}
\end{equation}
where, as usual, the space-time dependence of the coset representative is defined by the scalar fields $\phi^s(x)$:\; $\partial_\mu \Lc\equiv \partial_s \Lc\, \partial_\mu\phi^s$.\par
The \emph{gauged} vielbein and connection are related to the ungauged ones as follows:
\begin{align}
\hat{\mathcal{P}}_\mu&={\mathcal{P}}_\mu-g\,A^{\hat{\Lambda}}_\mu\,{\mathcal{P}}_{\hat{\Lambda}}\;;
\;\;\quad\;\;
\hat{\W}_\mu=
{\W}_\mu-g\,
A^{\hat{\Lambda}}_\mu\,{\W}_{\hat{\Lambda}}\,.\label{gaugedPW}
\end{align}
The matrices ${\mathcal{P}}_{\hat{\Lambda}},\,{\W}_{\hat{\Lambda}}$ begin the projections onto $\mathfrak{K}$ and $\mathfrak{H}$, respectively, of $\Lc^{-1}X_{\hat{\Lambda}}\Lc$:
\begin{align}
{\mathcal{P}}_{\hat{\Lambda}}&\equiv \left.\Lc^{-1}X_{\hat{\Lambda}}\Lc\right\vert_{\mathfrak{K}}\;;
\quad\;
{\W}_{\hat{\Lambda}}\equiv \left.\Lc^{-1}X_{\hat{\Lambda}}\Lc\right\vert_{\mathfrak{H}}\,.\label{wPproj}
\end{align}
Using eq.\ (\ref{kespans2}) we can express the above quantities as follows:
\begin{align}
{\mathcal{P}}_{\hat{\Lambda}}&= k_{\hat{\Lambda}}^s\,V_s{}^{\underline{s}}\,K_{{\underline{s}}}\;;
\quad
{\W}_{\hat{\Lambda}}= -\frac{1}{2}\,\Ps_{\hat{\Lambda}}^{{\bf a}}\,J_{{\bf a}}-\frac{1}{2}\,\Ps_{\hat{\Lambda}}^{{\bf m}}\,J_{{\bf m}}\,,\label{gaugedPW2}
\end{align}
where $\Ps_{\hat{\Lambda}}^{{\bf a}}$ were defined in Sect.\ \ref{ghsect}.\par\smallskip
For non-homogeneous scalar manifolds we cannot use the construction (\ref{hatOm}) based on the coset representative. Nevertheless
we can still define $\Ps_{\hat{\Lambda}}^{{\bf m}},\,\Ps_{\hat{\Lambda}}^{{\bf a}}$ in terms of the Killing vectors, see discussion below eq.\ (\ref{KRP}).
From these quantities one then defines gauged vielbein $\hat{P}_\mu$ and $H$-connection $\hat{\W}_\mu$ using (\ref{gaugedPW}) and (\ref{gaugedPW2}), where now $K_{{\underline{s}}}$ should be intended as a basis of the tangent space to the manifold at the origin (and not as isometry generators) and $\{J_{{\bf a}},\,J_{{\bf m}}\}$ a basis of the holonomy group.\par
Notice that, as a consequence of eqs.\ (\ref{gaugedPW2}) and (\ref{gaugedPW}), the gauged vielbein 1-forms (pulled-back on space-time) can be written as the ungauged ones in which the derivatives on the scalar fields are replaced by the covariant ones (\ref{covderphi}). This is readily seen by applying the general formula (\ref{kespans}) for homogeneous manifolds to the isometry $X_{\hat{\Lambda}}$ in (\ref{hatOm}), and projecting both sides of this equation on the coset space $\mathfrak{K}$:
\begin{equation}
\hat{\mathcal{P}}_\mu=\mathcal{P}_s\,\cD_\mu z^s\,.\label{hatPPDz}
\end{equation}
Consequently the replacement (\ref{covderphi}) is effected by replacing everywhere in the Lagrangian $\PP_\mu$ by $\hat{\PP}_\mu$.\par
Consider now a local $G_g$-transformation ${\bf g}(x)$ whose effect on the scalars is described by eq.\ (\ref{gLh}):\; ${\bf g}\Lc(\phi)=\Lc({\bf g}\star\phi)\,h(\phi,{\bf g})$.\; From (\ref{hatOm}) and from the fact that $\cD$ is the $G$-covariant derivative, the reader can easily verify that:
\begin{equation}
\hat{\Omega}_\mu(g\star \phi)=h\,\hat{\Omega}_\mu(\phi)\,h^{-1}+hdh^{-1}
\;\;\Rightarrow\;\;
\begin{cases}
\hat{\PP}(g\star \phi)=h\,\hat{\PP}(\phi)\,h^{-1}\,,\cr
\hat{\W}(g\star \phi)=h\,\hat{\W}(\phi)\,h^{-1}+hdh^{-1}\,,
\end{cases}
\label{PWhattra}
\end{equation}
where $ h=h(\phi,{\bf g})$.
By deriving (\ref{hatOm}) we find the \emph{gauged} Maurer-Cartan equations:
\begin{equation}
d\hat{\Omega}+\hat{\Omega}\wedge \hat{\Omega}=-g\,\Lc^{-1}\mathcal{F}\Lc\;,
\end{equation}
where we have used (\ref{D2F}). Projecting the above equation onto $\mathfrak{K}$ and $\mathfrak{H}$ we find the gauged version of eqs.\ (\ref{DP}), (\ref{RW}):
\begin{align}
\mathscr{D}\hat{\PP}&\equiv d\hat{\PP}+\hat{\W}\wedge \hat{P}+\hat{P}\wedge \hat{\W}=-g\,F^{\hat{\Lambda}}\,{\mathcal{P}}_{\hat{\Lambda}}\,,\label{DP2}\\
\hat{R}(\hat{\W})&\equiv d\hat{\W}+\hat{\W}\wedge \hat{\W}=-\hat{\PP}\wedge \hat{\PP}-g\,F^{\hat{\Lambda}}\,{\W}_{\hat{\Lambda}}\,.\label{RW2}
\end{align}
The above equations are manifestly $G_g$-invariant. Using (\ref{hatPPDz}) one can easily verify that the gauged curvature 2-form (with value in $\mathfrak{H}$) can be written in terms of the curvature components $R_{rs}$ of the manifold, given in eq.\ (\ref{Rcompo}), as follows:
\begin{equation}
\hat{R}(\hat{\W})=\frac{1}{2}\,R_{rs}\,\mathscr{D}\phi^r\wedge \mathscr{D}\phi^s-g\,F^{\hat{\Lambda}}\,{\W}_{\hat{\Lambda}}\,.
\end{equation}
The gauge-covariant derivatives, when acting on a generic fermion field $\xi$, is defined using $\hat{\W}_\mu$, so that (\ref{Dxi}) is replaced by
\begin{equation}
\cD_\mu\xi=\nabla_\mu\xi+\hat{\W}_\mu\star \xi\,.\label{Dxi2}
\end{equation}
Summarizing, local invariance of the action under $G_g$ requires replacing everywhere in the Lagrangian the abelian field strengths by the non abelian ones, eq.\ (\ref{replaceF}) and the ungauged vielbein $\mathcal{P}_\mu$ and $H$-connection $\W_\mu$ by the gauged ones:
\begin{equation}
\mathcal{P}_\mu\;\rightarrow\;\hat{\mathcal{P}}_\mu\;;
\;\quad\;
\W_\mu\;\rightarrow\;\hat{\W}_\mu\,.\label{replacePW}
\end{equation}
Clearly supersymmetry of the gauged action would require as a necessary, though not sufficient, condition to perform the above replacements also in the supersymmetry transformation laws of the fields.
\par

\subparagraph{\textbf{Step 3.} Introducing topological terms.}
If the symplectic duality action (\ref{xsymp}) of $X_{\hat{\Lambda}}$ has a non-vanishing off-diagonal block $X_{{\hat{\Lambda}}{\hat{\Gamma}}{\hat{\Sigma}}}$, that is if the gauge transformations include Peccei-Quinn shifts, then an infinitesimal (local) gauge transformation $\xi^{\hat{\Lambda}}(x)\,X_{{\hat{\Lambda}}}$ would produce a variation of the Lagrangian of the form (\ref{deltaLC}):
\begin{equation}
\delta\LB=
\frac{g}{8}\,\xi^{\hat{\Lambda}}(x)X_{{\hat{\Lambda}}{\hat{\Gamma}}{\hat{\Sigma}}}\epsilon^{\mu\nu\rho\sigma}\,
F^{\hat{\Gamma}}_{\mu\nu}F^{\hat{\Sigma}}_{\rho\sigma}\,.\label{deltaLX}
\end{equation}
Being $\xi^{\hat{\Lambda}}(x)$ a local parameter, the above term is no longer a total derivative and thus the transformation is not a symmetry of the action.
In \cite{deWit:1984px} it was proven that the variation (\ref{deltaLX}) can be canceled by adding to the Lagrangian a topological term of the form
\begin{equation}
\Lagr_{\rm top.}
=\frac{1}{3}\,g\,\epsilon^{\mu\nu\rho\sigma}\,X_{{\hat{\Lambda}}{\hat{\Gamma}}{\hat{\Sigma}}}\;A^{\hat{\Lambda}}_\mu\,
A^{\hat{\Sigma}}_\nu\,
\left(\partial_\rho A^{\hat{\Gamma}}_\sigma+\frac{3}{8}\,g\,X_{{\hat{\Delta}}{\hat{\Pi}}}{}^{\hat{\Gamma}}\,A^{\hat{\Delta}}_\rho\,A^{\hat{\Pi}}_\sigma\right)
\,,\label{top}
\end{equation}
provided the following condition holds
\begin{equation}
X_{({\hat{\Lambda}}{\hat{\Gamma}}{\hat{\Sigma}})}\=0\;.\label{xsymmetr}
\end{equation}
We will see in the following that condition (\ref{xsymmetr}), together with the closure constraint (\ref{gaugealg2}), is part of a set of constraints on the gauge algebra which is implied by supersymmetry. Indeed, even if the Lagrangian $\Lagr^{(0)}_{g}$ constructed so far is locally $G_g$-invariant, the presence of minimal couplings explicitly breaks both supersymmetry and the duality global symmetry $G$.
\par

\paragraph{Gauge algebra and embedding tensor.}
We have seen that the gauging procedure corresponds to promoting some suitable subgroup $G_g\subset G_{el}$ to a local symmetry. This subgroup is defined selecting a subset of generators within the global symmetry algebra $\mathfrak{g}$ of $G$. Now, all the information about the gauge algebra can be encoded in a $G_{el\,}$-covariant object $\theta$, which expresses the gauge generators as linear combinations of the global symmetry generators $t_\alpha$ of the subgroup $G_{el}\subset G$
\begin{equation}
X_{\hat{\Lambda}}=\theta_{\hat{\Lambda}}{}^\sigma\,t_\sigma\;; \quad\quad
\theta_{\hat{\Lambda}}{}^\sigma \in {\bf n_v}\times\Adj(G_{el})\;,
\label{gentheta}
\end{equation}
with \,${\hat{\Lambda}}=1,\,\dotsc,\,n_v$\; and with \,$\sigma=1,\dotsc,\,\dim(G_{el})$. \;The advantage of this description is that the $G_{el\,}$-invariance of the original ungauged Lagrangian $\Lagr$ is restored at the level of the gauged Lagrangian $\Lagr_{\rm gauged}$, to be constructed below, provided $\theta_{\hat{\Lambda}}{}^\sigma$ is
transformed under $G_{el}$ as well. However, the full global symmetry group $G$ of the field equations and Bianchi identities is still broken, since the parameters $\theta_{\hat{\Lambda}}{}^\sigma$ can be viewed as a number $n_{el}=\dim(G_{el})$ of electric charges, whose presence manifestly break electric-magnetic duality invariance. In other words we are working in a specific symplectic frame defined by the ungauged Lagrangian we started from. \par
\medskip
We shall give later on a definition of the gauging procedure which is completely freed from the choice of the symplectic frame. For the time being, it is useful to give a description of the gauge algebra (and of the consistency constraints on it) which does not depend on the original symplectic frame, namely which is manifestly $G$-covariant. This is done by encoding all information on the initial symplectic frame in a symplectic matrix $E\equiv(E_M{}^N)$ and writing the gauge generators, through this matrix, in terms of new generators
\begin{equation}
X_M=(X_\Lambda,\,X^\Lambda)
\end{equation}
which are at least twice as many as the $X_{\hat{\Lambda}}$:
\begin{eqnarray}
\left(\begin{matrix}
X_{\hat{\Lambda}} \cr 0
\end{matrix}\right)
=E\,\left(\begin{matrix}
X_\Lambda\cr X^\Lambda
\end{matrix}\right)
\;.\label{EXL}
\end{eqnarray}
This description is therefore redundant and this is the price we have to pay in order to have a manifestly symplectic covariant formalism. We can then rewrite the gauge connection in a symplectic fashion
\begin{align}
A^{\hat{\Lambda}}\,X_{\hat{\Lambda}} = A^{\hat{\Lambda}}\,E_{\hat{\Lambda}}{}^\Lambda\,X_\Lambda  +A^{\hat{\Lambda}}\,E_{{\hat{\Lambda}}\,\Lambda}\,X^\Lambda
=\AL\,X_\Lambda+\ALd\,X^\Lambda=A^M_\mu\,X_M\;,\label{syminvmc}
\end{align}
where we have introduced the vector fields $\AL$ and the corresponding dual ones $\ALd$, that can be regarded as components of a symplectic vector
\begin{eqnarray}
\AM\equiv(\AL,\,\ALd)\,.
\end{eqnarray}
 These are clearly not independent, since they are all expressed in terms of the only electric vector fields $A^{\hat{\Lambda}}$ of our theory (those entering the vector kinetic terms):
\begin{eqnarray}
\AL=E_{\hat{\Lambda}}{}^\Lambda\,A^{\hat{\Lambda}}_\mu\;,\qquad
\ALd=E_{{\hat{\Lambda}}\,\Lambda}\,A^{\hat{\Lambda}}_\mu\;.
\end{eqnarray}
In what follows, it is useful to adopt this symplectic covariant description in terms of $2n_v$ vector fields $\AM$ and $2n_v$ generators $X_M$, bearing in mind the above definitions through the matrix $E$, which connects our initial symplectic frame to any other.\par
\smallskip
The components of the symplectic vector $X_M$ are generators in the isometry algebra $\mathfrak{g}$ and thus can be expanded in a basis $t_\alpha$ of generators of $G$:
\begin{equation}
X_M=\Theta_M{}^\alpha\,t_\alpha\,,\qquad \alpha=1,\dotsc,\,\dim(G)
\,.\label{Thdef}
\end{equation}
The coefficients of this expansion $\Theta_M{}^\alpha$ represent an extension of the definition of $\theta$ to a $G$-covariant tensor:
\begin{equation}
\theta_\Lambda{}^\sigma \,\;\dashrightarrow\;\;
\Th \equiv (\theta^{\Lambda\,\alpha},\;\theta_\Lambda{}^\alpha)\,;
\qquad \Th\,\in\,\Rs_{v*}\times\Adj(G)
\,,\label{embtens}
\end{equation}
which describes the explicit embedding of the gauge group $G_g$ into the global symmetry group $G$, and combines the full set of deformation parameters of the original ungauged Lagrangian. The advantage of this description is that it allows to recast all the consistency conditions on the choice of the gauge group into $G$-covariant (and thus independent of the symplectic frame) constraints on $\Theta$.\par
We should however bear in mind that, just as the redundant set of vectors $A^M_\mu$, also the components of $\Theta_M{}^\alpha$ are not independent since, by eq.\ (\ref{EXL}),
\begin{equation}
\theta_{\hat{\Lambda}}{}^\alpha=E_{\hat{\Lambda}}{}^M\,\Theta_M{}^\alpha\;,
\;\quad\;
0=E^{\hat{\Lambda}\,M}\,\Theta_M{}^\alpha
\,,\label{elET}
\end{equation}
so that
\begin{equation}
\dim(G_g)\=\rank(\theta)=\rank(\Theta)\,.
\end{equation}
The above relations (\ref{elET}) imply for $\Theta_M{}^\alpha$ the following symplectic-covariant condition:
\begin{equation}
\Theta_\Lambda{}^\alpha\,\Theta^{\Lambda\,\beta}-\Theta_\Lambda{}^\beta\,\Theta^{\Lambda\,\alpha}=0
\quad\Leftrightarrow\quad
\mathbb{C}^{MN}\Theta_M{}^\alpha\Theta_N{}^\beta=0\quad
\,.\label{locality}
\end{equation}
Vice versa, one can show that if $\Theta_M{}^\alpha$ satisfies the above conditions, there exists a symplectic matrix $E$ which can rotate it to an electric frame, namely such that eqs.\ (\ref{elET}) are satisfied for some $\theta_{\hat{\Lambda}}{}^\alpha$.\, Equations (\ref{locality}) define the so-called \emph{locality constraint} on the embedding tensor $\Theta_M{}^\alpha$ and they clearly imply:
\begin{equation}
\dim(G_g)=\rank(\Theta)\le n_v\;,
\end{equation}
which is the preliminary consistency condition (\ref{preliminary}).\par
The electric-magnetic duality action of $X_M$, in the generic symplectic frame defined by the matrix $E$, is described by the tensor:
\begin{equation}
X_{MN}{}^P~\equiv~
\Theta_M{}^\alpha\,t_{\alpha\,N}{}^P\=
E^{-1}{}_M{}^{\hat{M}}E^{-1}{}_N{}^{\hat{N}}\,X_{\hat{M}\hat{N}}{}^{\hat{P}}\,E_{\hat{P}}{}^P\,.\label{XEhatX}
\end{equation}
For each value of the index {\footnotesize $M$}, the tensor $X_{MN}{}^P$ should generate symplectic transformations. This implies that:
\begin{equation}
X_{MNP}\equiv X_{MN}{}^Q\mathbb{C}_{QP}=X_{MPN}\,,
\end{equation}
which is equivalent to eqs.\ (\ref{sympconde}). The remaining linear constraints (\ref{lin1}), (\ref{xsymmetr}) on the gauge algebra can be recast in terms of $X_{MN}{}^P$ in the following symplectic-covariant form:
\begin{equation}
X_{(MNP)}=0
\;\quad\Leftrightarrow\quad\;
\begin{cases}
2\,X_{(\Lambda\Sigma)}{}^\Gamma=X^\Gamma{}_{\Lambda\Sigma}\,,\cr
2\,X^{(\Lambda\Sigma)}{}_\Gamma=X_\Gamma{}^{\Lambda\Sigma}\,,\cr X_{(\Lambda\Sigma\Gamma)}=0\,.
\end{cases}
\label{lconstr}
\end{equation}
Notice that the second of equations (\ref{lconstr}) implies that in the electric frame, in which $X^{\hat{\Lambda}}=0$, also the $B$-block (i.e.\ the upper-right one) of the infinitesimal gauge generators $\mathscr{R_v}[X_{\hat{\Lambda}}]$ vanishes, being $X_{\hat{\Gamma}}{}^{\hat{\Lambda}\hat{\Sigma}}=0$, so that the gauge transformations are indeed in $G_{el}$.\par
Finally, the closure constraints (\ref{gaugealg2}) can be written, in the generic frame, in the following form:
\begin{equation}
[X_M,\,X_N]=-X_{MN}{}^P\,X_P
\quad\Leftrightarrow\quad
\Theta_M{}^\alpha\Theta_N{}^\beta{\rm f}_{\alpha\beta}{}^\gamma+\Theta_M{}^\alpha\,t_{\alpha\,N}{}^P\Theta_P{}^\gamma=0\,.
\end{equation}
The above condition can be rephrased, in a $G$-covariant fashion, as the condition that the \emph{embedding tensor $\Theta_M{}^\alpha$ be invariant under the action of the gauge group it defines}:
\begin{align}
\delta_M\Theta_N{}^\alpha=0\,.\\
\nn
\end{align}
Summarizing we have found that consistency of the gauging requires the following set of linear and quadratic algebraic, $G$-covariant constraints to be satisfied by the embedding tensor:
\begin{enumerate}[$\circ$,itemsep=1ex]
\item{\emph{Linear constraint:}
   \begin{align}
   X_{(MNP)}&=0\,,\label{linear2}
   \end{align}
    }
\item{\emph{Quadratic constraints:}
   \begin{align}
   &\mathbb{C}^{MN}\Theta_M{}^\alpha\Theta_N{}^\beta=0\,,\label{quadratic1}\\
   &[X_M,\,X_N]=-X_{MN}{}^P\,X_P\,.\label{quadratic2}
   \end{align}
    }
\end{enumerate}
The linear constraint (\ref{linear2}) amount to a projection of the embedding tensor on a specific $G$-representation $\Rs_\Theta$ in the decomposition of the product $\Rs_{v*}\times {\rm Adj}(G)$ with respect to $G$
\begin{equation}
\Rs_{v*}\times {\rm Adj}(G)\;\;\stackrel{G}{\longrightarrow}\;\;
\Rs_\Theta +\, \dots
\end{equation}
and thus can be formally written as follows:
\begin{equation}
\mathbb{P}_\Theta\cdot \Theta=\Theta\,,
\end{equation}
where $\mathbb{P}_\Theta$ denotes the projection on the representation $\Rs_\Theta$. For this reason (\ref{linear2}) is also named \emph{representation constraint}.\par
The first quadratic constraint (\ref{quadratic1}) guarantees that a symplectic matrix $E$ exists which rotates the embedding tensor $\Theta_M{}^\alpha$ to an electric frame in which the \emph{magnetic components} $\Theta^{\hat{\Lambda}\,\alpha}$ vanish. The second one (\ref{quadratic2}) is the condition that the gauge algebra close within the global symmetry one $\mathfrak{g}$ and implies that $\Theta$ is a singlet with respect to $G_g$. In a general theory, the three constraints \eqref{linear2}, \eqref{quadratic1} and \eqref{quadratic2} should be imposed independently. As we shall prove below, in theories (such as the maximal one) in which all scalar fields enter the same supermultiplets as the vector ones, the locality constraint \eqref{quadratic1} follows from the other two. In maximal supergravity, however, the closure constraint \eqref{quadratic2} follows from \eqref{linear2} and \eqref{quadratic1} and thus, once the linear constraint is imposed, the two quadratic ones are equivalent.\par\smallskip
The second part of the gauging procedure, which we are going to discuss below, has to do with restoring supersymmetry after minimal couplings have been introduced and the $G_g$-invariant Lagrangian $\Lgaug^{(0)}$ have been constructed.
As we shall see, the supersymmetric completion of $\Lgaug^{(0)}$ requires no more constraints on $G_g$ (i.e.\ on $\Theta$) than the linear (\ref{linear2}) and quadratic ones (\ref{quadratic1}), (\ref{quadratic2}) discussed above.\par\medskip
As a final remark let us prove that the locality constraint (\ref{quadratic1}) is independent of the
others only in theories featuring scalar isometries with no duality action, namely in which the symplectic duality representation $\Scr{R}_v$ of the isometry algebra $\mathfrak{g}$ is not faithful. This is
the case of the quaternionic isometries in $\N = 2$ theories, see eq.\ (\ref{qisom}) of Sect.\ \ref{sframes}.
Let us split the generators $t_\alpha$ of $G$ into $t_{\ell}$, which have a non-trivial duality action, and $t_{m}$, which do not:
\begin{equation}
(t_{\ell})_M{}^N\neq 0\;; \;\quad\; (t_{m})_M{}^N= 0\,.
\end{equation}
From equation (\ref{quadratic2}) we derive, upon symmetrization of the {\footnotesize $M,\,N$} indices, the following condition:
\begin{equation}
X_{(MN)}{}^P\,X_{P}=X_{(MN)}{}^P\,\Theta_{P}{}^\alpha\,t_\alpha=0\,,\label{quad2n}
\end{equation}
where $t_\alpha$ on the right hand side are \emph{not} evaluated in the $\Rs_v$ representation and thus are all non-vanishing.
Using the linear constraint (\ref{linear2}) we can then rewrite $X_{(MN)}{}^P$ as follows:
\begin{equation}
X_{(MN)}{}^P=-\frac{1}{2}\,\mathbb{C}^{PQ}\,X_{QMN}=-\frac{1}{2}\,\mathbb{C}^{PQ}\,\Theta_Q{}^\ell t_{\ell\,MN}\,,
\end{equation}
so that (\ref{quad2n}) reads
\begin{equation}
\mathbb{C}^{QP}\,\Theta_Q{}^\ell\Theta_{P}{}^\alpha\,t_\alpha\,t_{\ell\,MN}\,=0
\,.\label{quad2nbis}
\end{equation}
Being $t_\alpha$ and $t_{\ell\,MN}$ independent for any $\alpha$ and $\ell$, conditions (\ref{linear2}) and (\ref{quadratic2}) only imply \emph{part of} the locality constraint (\ref{quadratic1}):
\begin{equation}
\mathbb{C}^{QP}\,\Theta_Q{}^\ell\Theta_{P}{}^\alpha=0\,,\label{quad2n2}
\end{equation}
while the remaining constraints (\ref{quadratic1})
\begin{equation}
\mathbb{C}^{QP}\,\Theta_Q{}^m\Theta_{P}{}^{n}=0\,,\label{quad2n3}
\end{equation}
need to be imposed independently. 
Therefore in theories in which all scalar fields sit in the same supermultiplets as the vector ones, as it is the case of $\N>2$ or $\N=2$ with no hypermultiplets, the locality condition \eqref{quadratic1} is not independent but follows from the other constraints.

\subsection{The gauged Lagrangian}
The three steps described above allow us to the construction of a Lagrangian $\Lgaug^{(0)}$ which is locally $G_g$-invariant starting from the ungauged one. Now we have to check if this deformation is compatible with local supersymmetry. As it stands, the Lagrangian $\Lgaug^{(0)}$ is no longer invariant under supersymmetry, due to the extra contributions that arise from variation of the vector fields in the covariant derivatives.\par
Consider, for instance, the supersymmetry variation of the (gauged) Rarita-Schwinger term in the Lagrangian
\begin{equation}
\Lagr_{\textsc{rs}}=i\,e\,\bar{\psi}^A_\mu\gamma^{\mu\nu\rho}\cD_\nu\psi_{A\,\rho}~+~\text{h.c.}\;,
\end{equation}
where $\cD_\nu$ is the gauged covariant derivative defined in eq.\ (\ref{Dxi2}). Under supersymmetry variation of $\psi_\mu$:
\begin{equation}
\delta\psi_\mu=\cD_\mu \epsilon+\dots\,,
\end{equation}
$\epsilon$ being the local supersymmetry parameter%
\footnote{
the ellipses refer to terms containing the vector field strengths
}.
The variation of $\Lagr_{\textsc{rs}}$ produces a term
\begin{align}
\delta\Lagr_{\textsc{rs}}&\=\dots+2i\,e\,\bar{\psi}^A_\mu\gamma^{\mu\nu\rho}\cD_\nu \cD_{\rho}\epsilon_A~+~\text{h.c.}\=\nne
&\=-i\,g\,e\,\bar{\psi}^A_\mu\gamma^{\mu\nu\rho}F_{\nu\rho}^{\hat{\Lambda}}\,(\W_{\hat{\Lambda}}\epsilon)_A~+~\text{h.c.}
\;,\label{RSvar}
\end{align}
where we have used the property (\ref{D2F}) of the gauge covariant derivative. Similarly we can consider the supersymmetry variation of the spin-$1/2$ fields:
\begin{equation}
\delta \lambda^{\mathcal{I}}=i\,\hat{\mathcal{P}}_\mu^{\mathcal{I}\,A}\,\gamma^\mu\epsilon_A+\dots\,,
\end{equation}
where the dots denote terms containing the vector fields and $\hat{\mathcal{P}}_\mu^{\mathcal{I}\,A}$ is a specific component of the $\mathfrak{K}$-valued matrix $\hat{\mathcal{P}}_\mu$.
The resulting variation of the corresponding kinetic Lagrangian contains terms of the following form:
\begin{align}
\delta\left(-ie\,\bar{\lambda}_{\mathcal{I}}\gamma^\mu \cD_\mu\lambda^{\mathcal{I}}~+~\text{h.c.}\right)&=\dots-2i\,e\,\bar{\lambda}_{\mathcal{I}}\gamma^{\mu\nu} \cD_\mu\hat{P}_\nu^{\mathcal{I}\,A}\,\epsilon_A~+~\text{h.c.}=\nonumber\\&=\dots+ig\,e\,\bar{\lambda}_{\mathcal{I}}\gamma^{\mu\nu} F_{\mu\nu}^{\hat{\Lambda}}\,{\mathcal{P}}_{\hat{\Lambda}}^{\mathcal{I}\,A}\,\epsilon_A~+~\text{h.c.}
\label{lambdatra}
\end{align}
We see that the supersymmetry variation of the minimal couplings in the fermion kinetic terms have produced $O(g)$-terms which contain the tensor
\begin{equation}
F_{\mu\nu}^{\hat{\Lambda}}\,\Lc^{-1}X_{\hat{\Lambda}} \Lc=\mathbb{F}_{\mu\nu}^{M}\,\Lc^{-1}X_M \Lc
\label{genvar}
\end{equation}
projected on $\mathfrak{H}$ and contracted with the $\bar{\psi}\epsilon$ current in (\ref{RSvar}), or restricted to $\mathfrak{K}$ and contracted with the $\bar{\lambda}\epsilon$ current in the second case (\ref{lambdatra}). On the right hand side of (\ref{genvar}) the summation over the gauge generators has been written in the symplectic invariant form defined in eq.\ (\ref{syminvmc}):\; $\mathbb{F}^M\, X_M\equiv F^{\hat{\Lambda}}\,E_{\hat{\Lambda}}{}^M\,X_M$\,.\, These are instances of the various terms occurring in the supersymmetry variation $\delta\Lgaug^{(0)}$.
Just as (\ref{RSvar}) and (\ref{lambdatra}), these terms are proportional to an $H$-tensor defined as follows%
\footnote{
in the formulas below we use the coset representative in which the first index (acted on by $G$) is in the generic symplectic frame defined by the matrix $E$ and which is then related to the same matrix in the electric frame (labeled by hatted indices) as follows:
\begin{equation}
\Lc(\phi)_{\hat{M}}{}^{\underline{N}}~=~E_{\hat{M}}{}^P\,\Lc(\phi)_{P}{}^{\underline{N}}
\quad\Rightarrow\quad
\mathcal{M}(\phi)_{\hat{M}\hat{N}}=E_{\hat{M}}{}^PE_{\hat{N}}{}^Q\mathcal{M}(\phi)_{PQ}\,,
\end{equation}
last equation being (\ref{MEtra})
}:
\begin{align}
\Tb(\Theta,\phi)_{\underline{M}}&~\equiv~ \frac{1}{2}\,\LL(\phi)^{-1}{}_{\underline{M}}{}^N\,\Lc(\phi)^{-1}X_N\,\Lc(\phi)\=
\frac{1}{2}\,\LL(\phi)^{-1}{}_{\underline{M}}{}^N\,\Theta_N{}^\beta\,\Lc(\phi)_\beta{}^\alpha\,t_\alpha\=\nonumber\\
&\=\Tb(\Theta,\phi)_{\underline{M}}{}^\alpha\,t_\alpha
\,,\label{TT}
\end{align}
where
\begin{equation}
\Tb(\Theta,\phi)_{\underline{M}}{}^\alpha\equiv \frac{1}{2}\,\LL(\phi)^{-1}{}_{\underline{M}}{}^N\,\Theta_N{}^\beta \Lc(\phi)_\beta{}^\alpha=
\frac{1}{2}\,(\Lc^{-1}(\phi)\star \Theta)_{\underline{M}}{}^\alpha\,,
\end{equation}
where $\star$ denotes the action of $\Lc^{-1}$ as an element of $G$ on $\Theta_M{}^\alpha$ in the corresponding $\Rs_\Theta$-representation.
The tensor $\Tb(\phi,\,\Theta)=\Lc^{-1}(\phi)\star \Theta$ is called the \emph{$\Tb$-tensor} and was first introduced in \cite{deWit:1982ig}.\par
If $\Theta$ and $\phi$ are simultaneously transformed with $G$, the $\Tb$-tensor transforms under the corresponding $H$-compensator:
\begin{align}
\forall {\bf g}\in G\;:\quad
&\Tb({\bf g}\star\phi,\,{\bf g}\star\Theta)=\frac{1}{2}\, \Lc^{-1}({\bf g}\star\phi)\star ({\bf g}\star\Theta)=\nonumber\\
&=\frac{1}{2}\,(h({\bf g},\phi)\star \Lc^{-1}(\phi)\star {\bf g}^{-1})\star ({\bf g}\star\Theta)=h({\bf g},\phi)\star\Tb(\phi,\,\Theta)\,.
\end{align}
This quantity $\Tb$ naturally belongs to a representation of the group $H$ and is an example of \emph{composite field}
discussed at the end of Sect.\ \ref{fsector}.\par
If, on the other hand, we fix $\phi$ and only transform $\Theta$, $\Tb$ transforms in the same $G$-representation $\Rs_\Theta$ as $\Theta$, being $\Tb$ defined (aside for the factor $1/2$) by acting on the embedding tensor with the $G$-element $\LL^{-1}$.
As a consequence of this, $\Tb$ satisfies the same constraints (\ref{linear2}), (\ref{quadratic1}) and (\ref{quadratic2}) as $\Theta$:
\begin{align}
\Tb_{\underline{NM}}{}^{\underline{N}}\=\Tb_{(\underline{MNP})}\=0\,,\nonumber\\
\mathbb{C}^{\underline{MN}}\,\Tb_{\underline{M}}{}^\alpha\,\Tb_{\underline{N}}{}^\beta\=0\,,\nonumber\\
[\Tb_{\underline{M}},\,\Tb_{\underline{N}}]+\Tb_{\underline{MN}}{}^{\underline{P}}\,
\Tb_{\underline{P}}\=0\,,\label{Tids}
\end{align}
where we have defined $\Tb_{\underline{MN}}{}^{\underline{P}}\equiv \Tb_{\underline{M}}{}^\alpha\,t_{\alpha \underline{N}}{}^{\underline{P}}$. Equations (\ref{Tids}) have been originally derived within maximal supergravity in \cite{deWit:1982ig}, and dubbed \emph{$\Tb$-identities}%
\footnote{
recall that in maximal supergravity the locality constraint follows from the linear and the closure ones
}.\par
Notice that, using eqs.\ (\ref{wPproj}) and (\ref{gaugedPW2}) we can rewrite the $\Tb$-tensor in the following form:
\begin{equation}
\Tb_{\underline{M}}\=\frac{1}{2}\,\LL^{-1}{}_{\underline{M}}{}^N\,\Theta_N{}^\alpha\left(k_{\alpha}^s\,V_s{}^{\underline{s}}\,K_{{\underline{s}}}-\frac{1}{2}\,\Ps_{\alpha}^{{\bf a}}\,J_{{\bf a}}-\frac{1}{2}\,\Ps_{\alpha}^{{\bf m}}\,J_{{\bf m}}\right)\,, \label{Tgen}
\end{equation}
which can be extended to $\N=2$ theories with non-homogeneous scalar manifolds, see discussion at the end of this section. \par\smallskip
To cancel the supersymmetry variations of $\Lgaug^{(0)}$ and to construct a gauged Lagrangian $\Lgaug$ preserving the original supersymmetries, one can apply the general Noether method (see \cite{VanNieuwenhuizen:1981ae} for a general review) which consists in adding new terms to $\Lgaug^{(0)}$ and to the supersymmetry transformation laws, iteratively in the gauge coupling constant.
In our case the procedure converges by adding terms of order one ($\Delta\Lgaug^{(1)}$) and two ($\Delta\Lgaug^{(2)}$) in $g$, so that
\begin{equation}
\Lgaug=\Lgaug^{(0)}+\Delta\Lgaug^{(1)}+\Delta\Lgaug^{(2)}\,.
\end{equation}
The additional $O(g)$-terms are of \emph{Yukawa type} and have the general form:
\begin{equation}
e^{-1}\Delta\Lgaug^{(1)}=g\left(2\bar{\psi}^A_\mu\;\gamma^{\mu\nu}\;\psi_\nu^B\;\mathbb{S}_{AB}
~+~i\,\bar{\lambda}^{\mathcal{I}}\;\gamma^\mu\;\psi_{\mu\,A}\;\mathbb{N}_{\mathcal{I}}{}^A
~+~\bar{\lambda}^{\mathcal{I}}\,\lambda^{\mathcal{J}}\;\mathbb{M}_{\mathcal{IJ}}\right)
~+~\text{h.c.}
\;,\label{fmassterms}
\end{equation}
characterized by the scalar-dependent matrices $\mathbb{S}_{AB}$ and $\mathbb{N}^{\mathcal{I} A}$ called \emph{fermion shift matrices}, and a matrix $\mathbb{M}^{\mathcal{IJ}}$ that can be rewritten in terms of the previous mixed mass tensor $\mathbb{N}^{\mathcal{I} A}$ (see the subsequent sections).\par
The $O(g^2)$-terms consist of a scalar potential:
\begin{equation}
e^{-1}\Delta\Lgaug^{(2)}=-g^2\,V(\phi) \,.\label{spot}
\end{equation}
At the same time the fermionic supersymmetry transformations need to be suitably modified. To this end, we shall \emph{add order--$g$ terms to the fermion supersymmetry transformation rules} of the gravitino ($\psi_{\mu A}$) and of the other fermions ($\chi^\mathcal{I}$)
\begin{align}
\delta_\epsilon\psi_{\mu A}&\=\cD_\mu\epsilon_A
+i\,g\;\mathbb{S}_{AB}\;\gamma_\mu\,\epsilon^B+\dotsc\,,\nn\\
\delta_\epsilon\lambda_{\mathcal{I}}&\=g\,\mathbb{N}_{\mathcal{I}}{}^{A}\,\epsilon_A+\dotsc
\label{fermshifts}
\end{align}
depending on the same matrices $\mathbb{S}_{AB},\,\mathbb{N}_{\mathcal{I}}{}^{A}$ entering the mass terms.
The fermion shift-matrices are composite fields belonging to some appropriate representations $\Rs_S,\,\Rs_N$ of the $H$ group, such that (\ref{fmassterms}) is $H$-invariant.\par
These additional terms in the Lagrangian and supersymmetry transformation laws are enough to cancel the original $O(g)$ variations in $\delta\Lgaug^{(0)}$ --- like (\ref{RSvar}) and (\ref{lambdatra}), together with new $O(g)$ terms depending on $\mathbb{S}$ and $\mathbb{N}$ in the supersymmetry variation of $\Lgaug^{(0)}$ --- provided the shift-tensors $\mathbb{S}_{AB},\,\mathbb{N}^{\mathcal{I} A}$ are identified with suitable $H$-covariant components of the $\Tb$-tensor:
\begin{equation}
\Rs_\Theta\;\stackrel{H}{\longrightarrow}\;\Rs_N+\Rs_S+\Rs_{{\tiny \mbox{other}}}\,,
\end{equation}
and that additional $H$-representations $\Rs_{{\tiny \mbox{other}}}$ in the $\Tb$-tensor do not enter the supersymmetry variations of the Lagrangian. This can be formulated as a $G$-covariant restriction on the representation $\Rs_\Theta$ of the $\Tb$-tensor or, equivalently, of embedding tensor, which can be shown to be no more than the representation constraint (\ref{linear2}) discussed earlier.\par
The identification with components of the $\Tb$-tensor defines the expression fermion shift-tensors as $H$-covariant composite fields in terms of the embedding tensor and the scalar fields:
\begin{equation}
\mathbb{S}_{AB}=\mathbb{S}_{AB}(\phi,\Theta)=\left.\Tb(\phi,\Theta)\right\vert_{\Rs_S}\,;
\quad\;
\mathbb{N}_{\mathcal{I}}{}^A=\mathbb{N}_{\mathcal{I}}{}^A(\phi,\Theta)=\left.\Tb(\phi,\Theta)\right\vert_{\Rs_N}\,.
\end{equation}
Finally, in order to cancel the $O(g^2)$-contributions resulting from the variations (\ref{fermshifts}) in (\ref{fmassterms}), we need to add an \emph{order-$g^2$ scalar potential} $V(\phi)$ whose expression is totally determined by supersymmetry as a bilinear in the shift matrices by the condition
\begin{equation}
\delta_B{}^A\,V(\phi)\;=\; g^2\,\left(\mathbb{N}_{\mathcal{I}}{}^{A}\,\mathbb{N}^{\mathcal{I}}{}_{B}-12\;\mathbb{S}^{AC}\,\mathbb{S}_{BC}\right)\,,\label{WID}
\end{equation}
where we have defined \,$\mathbb{N}^{\mathcal{I}}{}_{A}\equiv (\mathbb{N}_{\mathcal{I}}{}^{A})^*$ \,and\, $\mathbb{S}^{AB}\equiv (\mathbb{S}_{AB})^*$.
The above condition is called \emph{potential Ward identity} \cite{Ferrara:1985gj,Cecotti:1984wn} (for a comprehensive discussion of the supersymmetry constraints on the fermion shifts see \cite{D'Auria:2001kv}). This identity defines the scalar potential as a quadratic function of the embedding tensor and non-linear function of the scalar fields. As a constraint on the fermion shifts, once these have been identified with components of the $\Tb$-tensor, it follows from the $\Tb$-identities (\ref{Tids}) or, equivalently, from the quadratic constraints (\ref{quadratic1}), (\ref{quadratic2}) on $\Theta$.
The derivation of quadratic supersymmetry constraints on the fermion shifts in maximal supergravity from algebraic constraints (i.e.\ scalar field independent) on the embedding tensor, was originally accomplished in \cite{Cordaro:1998tx}, though in a specific symplectic frame, and in maximal $D=3$ theory in \cite{Nicolai:2000sc}. In \cite{deWit:2002vt} the four-dimensional result was extended to a generic symplectic frame of the $\N=8$ model, i.e.\ using the $G$-covariant constraint (\ref{linear2}),(\ref{quadratic1}), (\ref{quadratic2}) on the embedding tensor%
\footnote{
in a generic gauged model, supersymmetry further require the fermion shifts to be related by differential ``gradient flow'' relations \cite{D'Auria:2001kv} which can e shown to follow from the identification of the shifts with components of the $\Tb$-tensor and the geometry of the scalar manifold
}.\par\smallskip
Let us comment on the case of $\N=2$ theories with a non-homogeneous scalar manifold (\ref{SKQK}). In this case we cannot define a coset representative. However, as mentioned earlier, one can still define a symplectic matrix $\LL^M{}_{\underline{N}}$ depending on the complex scalar fields in the vector multiplets (which has no longer the interpretation of a coset representative). We can then define the $\Tb$-tensor in these theories as in (\ref{Tgen})
where $\{K_{{\underline{s}}}\}$ should be intended as a basis of the tangent space to the origin (and not as isometry generators), while $\{J_I\}=\{J_{{\bf a}},\,J_{{\bf m}}\}$ are holonomy group generators%
\footnote{
the $H_{\rm R}={\rm U}(2)$-generators $\{J_{{\bf a}}\}$ naturally split into a ${\rm U}(1)$-generator $J_0$ of the K\"ahler transformations on $\MsSK$ and ${\rm SU}(2)$-generators $J_x$ ($x=1,2,3$) in the holonomy group of the quaternionic K\"ahler manifold $\Ms_{\textsc{qk}}$
}.
Recall that $\{\Ps_{\alpha}^{{\bf a}},\,\Ps_{\alpha}^{{\bf m}}\}$ enter the definition of the gauged composite connection (\ref{gaugedPW2}) on the scalar manifold and, as mentioned earlier, are related to the Killing vectors by general properties of the spacial K\"ahler and quaternionic K\"ahler geometries \cite{Andrianopoli:1996cm}.\par\smallskip
It is a characteristic of supergravity theories that -- in contrast to globally supersymmetric ones -- by virtue of the negative contribution due to the gravitino shift-matrix, the scalar potential is in general not positive definite, but may, in particular, feature AdS vacua. These are maximally symmetric solutions whose negative cosmological constant is given by the value of the potential at the corresponding extremum:\, $\Lambda=V_0<0$.\; Such vacua are interesting in the light of the AdS/CFT holography conjecture \cite{Maldacena:1997re}, according to which stable AdS solutions describe conformal critical points of a suitable gauge theory defined on the boundary of the space.
In this perspective, domain wall solutions to the gauged supergravity interpolating between AdS critical points of the potential describe renormalization group (RG) flow (from an ultra-violet to an infra-red fixed point) of the dual gauge theory and give important insights into its non-perturbative properties. The spatial evolution of such holographic flows is determined by the scalar potential $V(\phi)$ of the gauged theory.\par
In some
cases the effective scalar potential $V(\phi)$, at the classical
level, is non--negative and defines vacua with vanishing
cosmological constant in which supersymmetry is spontaneously
broken and part of the moduli are fixed. Models of this type are
generalizations of the so called ``no--scale'' models
\cite{Cremmer:1983bf}, \cite{Ellis:1984bm}, \cite{Barbieri:1985wq} which were subject to intense study during the
eighties.

\subsection{Dualities and flux compactifications}\label{dfcomp}
Let us summarize what we have learned so far.
\begin{enumerate}[$\circ$,itemsep=1ex]
\item{The most general local internal symmetry group $G_g$ which can be introduced in an extended supergravity is defined by an embedding tensor $\Theta$, covariant with respect to the on-shell global symmetry group $G$ of the ungauged model and defining the embedding of $G_g$ inside $G$. Since a scalar potential $V(\phi)$ can only be introduced through the gauging procedure, $\Theta$ also defines the most general choice for \,$V=V(\phi,\Theta)$. }
\item{Consistency of the gauging at the level of the bosonic action requires $\Theta$ to satisfy a number of (linear and quadratic) $G$-covariant constraints. The latter, besides completely determining the gauged bosonic action, also allow for its consistent (unique) supersymmetric extension.}
\item{Once we find a solution $\Theta_M{}^\alpha$ to these algebraic constraints, a suitable symplectic matrix $E$, which exists by virtue of (\ref{quadratic1}), will define the corresponding electric frame, in which its magnetic components vanish.}
\end{enumerate}
Although we have freed our choice of the gauge group from the original symplectic frame, the resulting gauged theory is still defined in an electric frame and thus depends on the matrix $E$: whatever solution $\Theta$ to the constraints is chosen for the gauging, the kinetic terms of the gauged Lagrangian are always written in terms of the only \emph{electric} vector fields $A^{\hat{\Lambda}}_\mu$, namely of the vectors effectively
involved in the minimal couplings, see eq.\ (\ref{syminvmc}). We shall discuss in the next section a more general formulation of the gauging which no longer depends on the matrix $E$.\par

\paragraph{Dual gauged supergravities.} All the deformations of the ungauged model required by the gauging procedure depend on $\Theta$ in a manifestly $G$-covariant way. This means that, if we transform all the fields $\Phi$ (bosons and fermions) of the model under $G$ (the fermions transforming under corresponding compensating transformations in $H$) and at the same time transform $\Theta$ and the matrix $E$, the field equations and Bianchi identities -- collectively denoted by $\Es(E,\,\Phi,\,\Theta)=0$ -- are left invariant:
\begin{equation}
\forall g\in G\;:\;\;
\Es(E,\,\Phi,\,\Theta)=0
\;\;\Leftrightarrow\;\;
\Es(E',\,g\star\Phi,\,g\star\Theta)=0
\quad\;\; (\text{with } \, E'=E\,\Rs_{v}[g]^T)\,.
\end{equation}
Since the embedding tensor $\Theta$ is a \emph{spurionic}, namely non-dynamical, object, the above on-shell invariance should not be regarded as a symmetry of a single theory, but rather as an equivalence (or proper duality) between two different theories, one defined by $\Theta$ and the other by $g\star \Theta$.\, Gauged supergravities are therefore classified in \emph{orbits} with respect to the action of $G$ (or better $G(\mathbb{Z})$) on $\Theta$. This property has an important bearing on the study of flux compactifications mentioned in the Introduction. Indeed, in all instances of flux compactifications, the internal fluxes manifest themselves in the lower-dimensional effective gauged supergravity as components of the embedding tensor defining the gauging \cite{Angelantonj:2003rq,D'Auria:2003jk,deWit:2003hq}:
\begin{equation}
\Theta= \mbox{Internal Fluxes}\,.
\end{equation}
This allows us to formulate a precise correspondence between general fluxes (form, geometric and non-geometric) and the gauging of the resulting supergravity. Moreover, using this identification, the quadratic constraints (\ref{quadratic1}), (\ref{quadratic2}) precisely reproduce the consistency conditions on the internal fluxes deriving from the Bianchi identities and field equations in the higher dimensional theory such as, in the presence of RR fluxes, the tadpole cancelation condition \cite{Grana:2005jc,Angelantonj:2003rq,deWit:2003hq}. \par
Consider the limit in which the lower-dimensional gauged theory provides a reliable description of the low-energy string or M-theory dynamics on a flux background. This limit is defined by the condition that the flux-induced masses in the effective action be much smaller than the scale of the Kaluza-Klein masses (of order $1/R$, where $R$ is the size of the internal manifold)%
\footnote{
for string theory compactifications we should also require this latter scale to be negligible compared to the mass-scale of the string excitations (order $1/\sqrt{\alpha'}$)
}:
\begin{equation}
\mbox{Flux-induced masses}\,\,\ll\,\,\frac{1}{R}\,.\label{sugracond}
\end{equation}
In this case, fields and fluxes in the lower-dimensional supergravity arrange in representations with respect to the characteristic symmetry group $G_{int}$ the internal manifold would have in the absence of fluxes. In the case of compactifications on $T^n$, such characteristic group is ${\rm GL}(n,\,\mathbb{R})$, acting transitively on the internal metric moduli.\par
\begin{figure}[H]
\centerline{\includegraphics[width=0.6\textwidth]{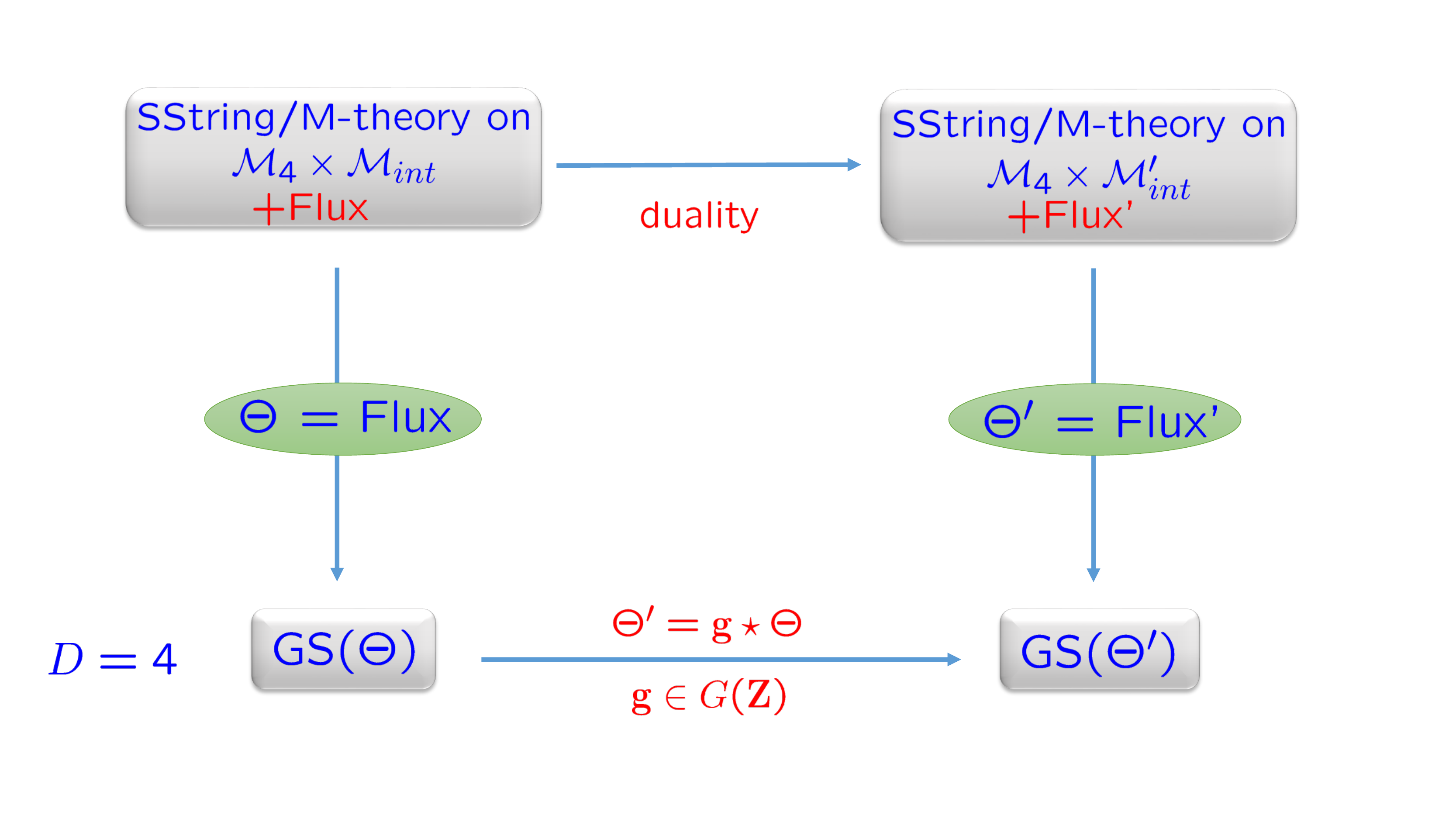}}
 \caption{\scriptsize Dualities between known flux compactifications (``GS'' stands for ``gauged supergravity'').}\label{fig1}
\end{figure}

In general, in the absence of fluxes, $G_{int}$ is a global symmetry group of the action: $G_{int}\subset G_{el}$.\; By branching $\mathscr{R}_\Theta$ with respect to $G_{int}$, we can identify within $\Theta$ the components corresponding to the various internal fluxes. The effect of any such background quantities in the compactification is reproduced by simply switching on the corresponding components of $\Theta$. The gauging procedure does the rest and the resulting gauged model is thus uniquely determined.
Since, as mentioned earlier at the end of Sect.\ \ref{gsg}, a suitable subgroup $G(\mathbb{Z})$ of $G$ was conjectured to encode all known string/M-theory dualities, the embedding tensor formulation of the gauging procedure provides an ideal theoretical laboratory where to systematically study the effects of these dualities on fluxes. Some elements of $G(\mathbb{Z})$ will map gauged supergravity descriptions of known compactifications into one another, see Fig.\ \ref{fig1}.\par
\begin{figure}[H]
\centerline{\includegraphics[width=0.6\textwidth]{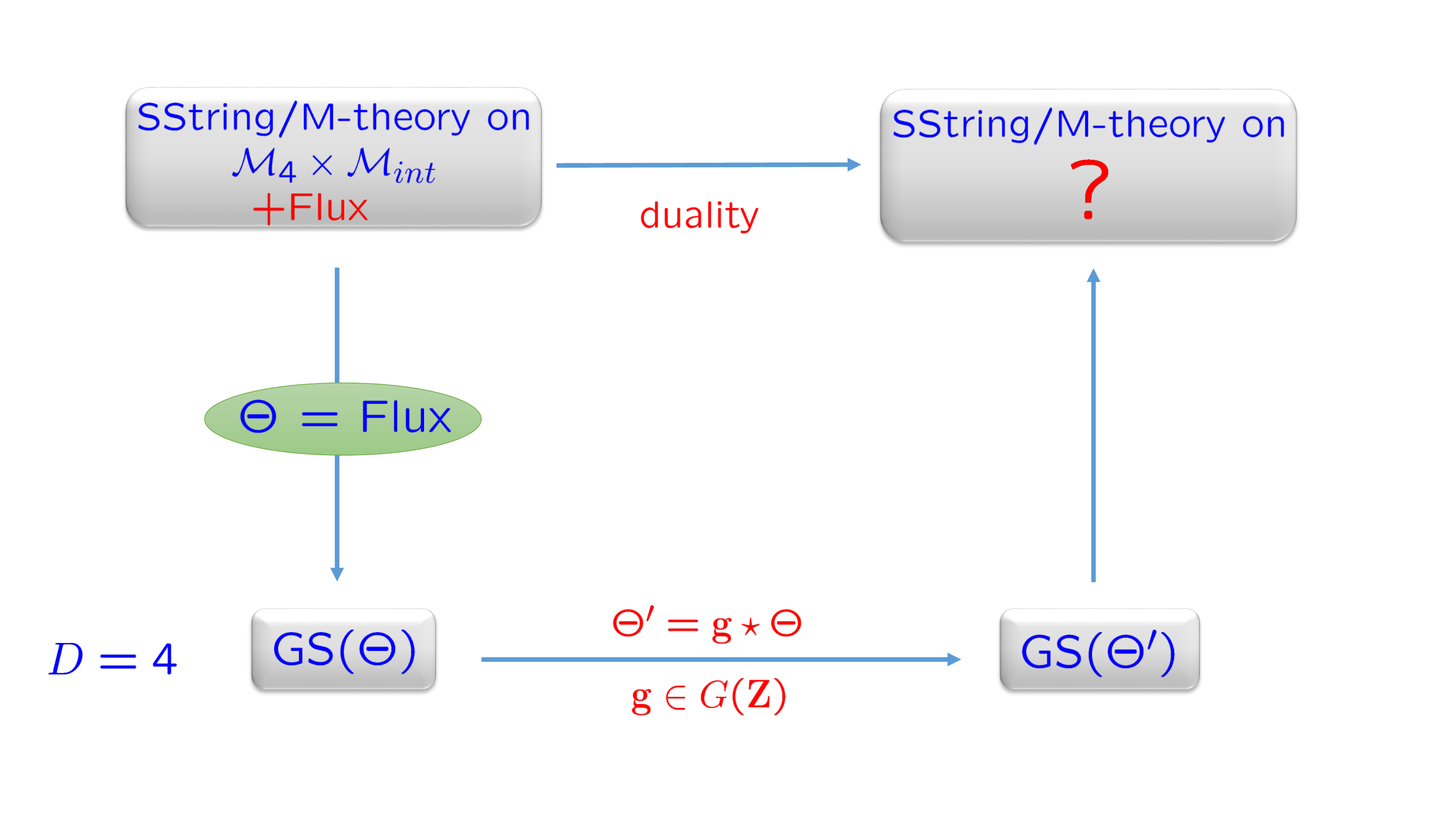}}
 \caption{\scriptsize Dualities connecting known flux compactifications to unknown ones.}
\label{fig2}
\end{figure}

\noindent
Other elements of $G(\mathbb{Z})$ will map gauged supergravities, originating from known compactifications, into theories whose string or M-theory origin is unknown, see Fig.\ \ref{fig2}.
\noindent
In this case we can use the duality between the corresponding low-energy descriptions to make sense of new compactifications as ``dual'' to known ones.\par
The so-called \emph{non-geometric} fluxes naturally fit in the above description as dual to certain compactifications with NS-NS H-flux. If we consider superstring theory compactified to four-simensions on a six-torus $T^6$ without fluxes, the resulting (classical) ungauged supergravity features a characteristic ${\rm O}(6,6)$ global symmetry group, which contains the T-duality group ${\rm O}(6,6;\mathbb{Z})$ and which acts transitively on the moduli originating from the metric and Kalb-Ramond $B$-field in ten dimensions. The $G$-representation $\mathscr{R}_\Theta$ of the embedding tensor, defining the most general gauging, contains the representation ${\bf 220}$ of ${\rm O}(6,6)$
\begin{equation}
\mathscr{R}_\Theta\,\stackrel{{\rm O}(6,6)}{\longrightarrow}\;
{\bf 220}~+~\dots
\end{equation}
which in turn branches with respect to the characteristic group $G_{int}={\rm GL}(6,\mathbb{R})$ of the torus as follows:
\begin{equation}
{\bf 220}\;\stackrel{{\rm GL}(6,\mathbb{R})}{\longrightarrow}\;
{\bf 20}_{-3}~+~({\bf 84+6})_{-1}~+~({\bf 84'+6'})_{+1}~+~{\bf 20}_{+3}\,.\label{220}
\end{equation}
The component ${\bf 20}_{-3}$ can be identified with the H-flux $H_{\alpha\beta\gamma}$ (that is the flux of the field strength of the Kalb-Ramond field $B$) along a 3-cycle of the torus. Switching on only the ${\bf 20}_{-3}$ representation in $\Theta$, the gauging procedure correctly reproduces the couplings originating from a toroidal dimensional reduction with H-flux. What (\ref{220}) tells us is that the action of the T-duality group ${\rm O}(6,6;\mathbb{Z})$ will generate, from an H-flux in the ${\bf 20}_{-3}$, all the other representations:
\begin{align}
{\bf 84+6}_{-1} \;&:\;\; \tau_{\alpha\beta}{}^\gamma\,,\nonumber\\
{\bf 84'+6'}_{+1} \;&:\;\; Q_{\alpha}{}^{\beta\gamma}\,,\nonumber\\
{\bf 20}_{+3} \;&:\;\; R^{\alpha\beta\gamma}\,.
\end{align}
The first tensor $\tau_{\alpha\beta}{}^\gamma$ is an instance of \emph{geometric flux}, being a background quantity which characterizes the geometry of the internal manifold. It describes a compactification on a space which is no longer a torus, but is locally described by a group manifold \cite{Scherk:1979zr} with structure constants $\tau_{\alpha\beta}{}^\gamma$. The constraint (\ref{quadratic2}) indeed implies for $\tau_{\alpha\beta}{}^\gamma$ the Jacobi identity: $\tau_{[\alpha\beta}{}^\gamma\tau_{\sigma]\gamma}{}^{\delta}=0$.\; This new internal manifold is called \emph{twisted torus} \cite{Kaloper:1999yr} (see also \cite{Grana:2005jc} and references therein).\par
The T-duality picture is completed by the remaining two representations, described by the tensors $Q_{\alpha}{}^{\beta\gamma},\,R^{\alpha\beta\gamma}$. Their interpretation as originating from a string theory compactification is more problematic, since in their presence the internal space cannot be given a global or even local description as a differentiable manifold. For this reason they are called \emph{non-geometric} fluxes \cite{Mathai:2004qq,Hull:2004in,Shelton:2005cf} (see also \cite{Grana:2005jc} and references therein).
The $H,\,\tau,\,Q,\,R$-fluxes can all be given a unified description as quantities defining the geometry of more general internal manifolds, having the T-duality group as structure group. Such manifolds are defined in the context of \emph{generalized geometry} \cite{Hitchin:2010qz,Gualtieri:2003dx} (see also \cite{Grana:2005jc} and references therein), by doubling the tangent space to the internal manifold in order to accommodate a representation of ${\rm O}(6,6)$ and introducing on it additional geometric structures, or of \emph{double geometry}/\emph{double field theory} \cite{Hull:2007jy,Dabholkar:2005ve,Hull:2009mi,Hohm:2010pp}, in which the internal manifold itself is enlarged, and parametrized by twice as many coordinates as the original one.\par

\begin{figure}[!h]
\centerline{\includegraphics[width=0.6\textwidth]{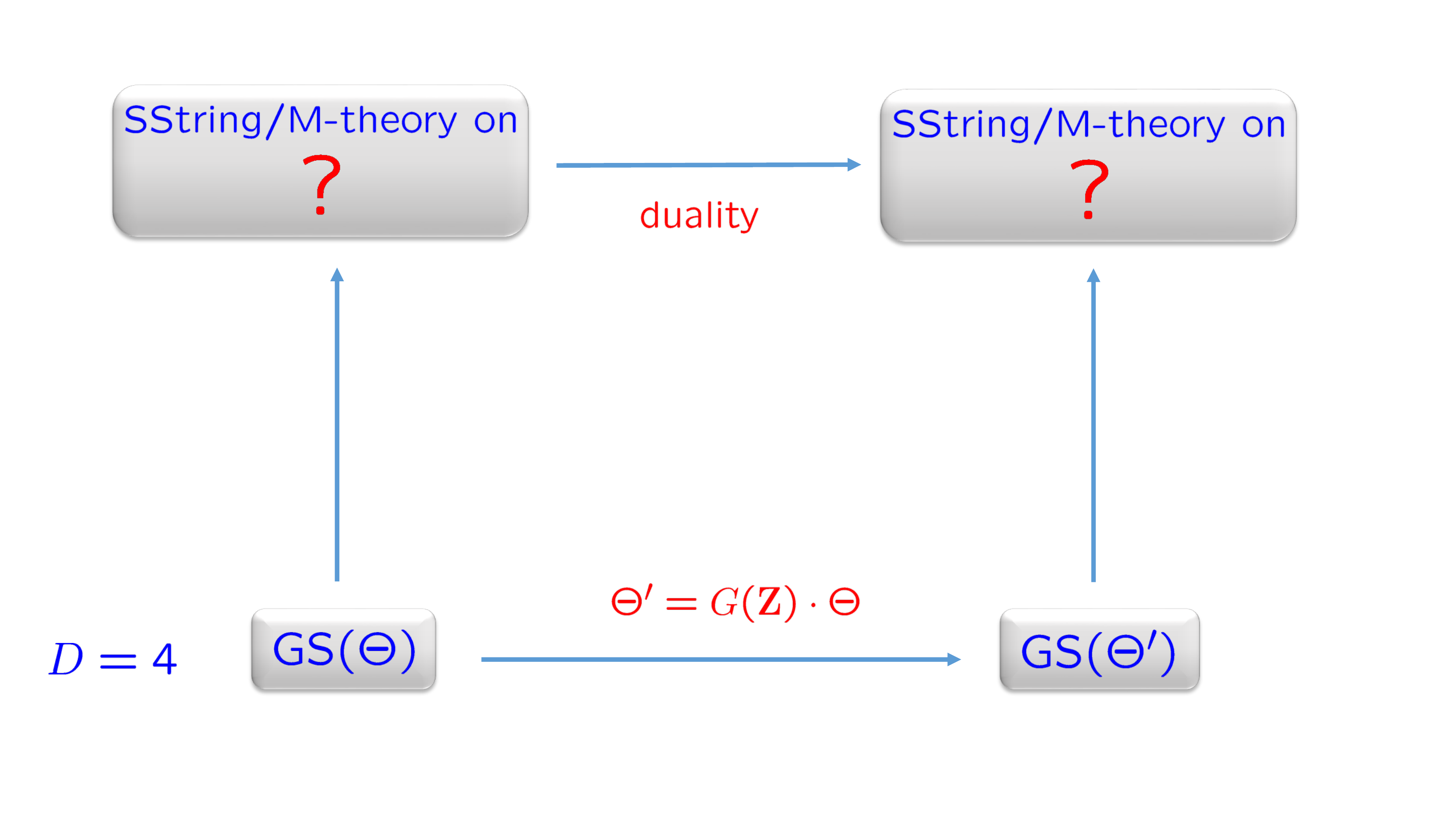}}
 \caption{\scriptsize Intrinsically non-geometric theories.}\label{fig3}
\end{figure}

Finally there are gauged supergravities which are not $G(\mathbb{Z})$-dual to models with a known string or M-theory origin, Fig.\ \ref{fig3}.\;
\noindent
Finding an ultra-violet completion of these theories, which are sometimes called \emph{intrinsically non-geometric}, in the context of string/M-theory is an open challenge of theoretical high-energy physics. Progress in this direction has been achieved in the context of extended generalized geometry \cite{Hull:2007zu,Pacheco:2008ps} or exceptional field theory \cite{Hohm:2013pua,Hohm:2013uia,Hohm:2014qga}.\par
If the hierarchy condition (\ref{sugracond}) is not met, the gauged supergravity cannot be intended as a description of the low-energy string/M-theory dynamics, but just as a \emph{consistent truncation} of it, as in the case of the spontaneous compactification of $D=11$ supergravity on $\AdS_4\times S^7$. In this case, the back-reaction of the fluxes on the internal geometry will manifest in extra geometric fluxes, to be identified with additional components of $\Theta$.\par

\paragraph{Vacua and dualities.}
The scalar potential
\begin{equation}
V(\phi,\Theta)\;=\;\frac{g^2}{\mathcal{N}} \,\left(\mathbb{N}_{\mathcal{I}}{}^{A}\,\mathbb{N}^{\mathcal{I}}{}_{B}-12\;\mathbb{S}^{AC}\,\mathbb{S}_{BC}\right)\,,\label{Pot}
\end{equation}
being expressed as an $H$-invariant combination of composite fields (the fermion shifts),
is invariant under the simultaneous action of $G$ on $\Theta$ and $\phi^s$:
\begin{equation}
\forall g\in G \;:\quad
V(g\star\phi,\,g\star\Theta)=V(\phi,\,\Theta)\,.
\end{equation}
This means that, if $V(\phi,\,\Theta)$ has an extremum in $\phi_0$
\begin{equation}
\left.\frac{\partial}{\partial\phi^s}V(\phi,\,\Theta)\right\vert_{\phi_0}=0\,,
\end{equation}
$V(\phi,g\star\Theta)$ has an extremum at $\phi'_0=g\star \phi_0$ with the same properties (value of the potential at the extremum and its derivatives):
\begin{equation}
\left.\frac{\partial}{\partial\phi^s}V(\phi,g\star\Theta)\right\vert_{g\star\phi_0}=0\,,
\qquad g\in G\;.
\end{equation}
If the scalar manifold is homogeneous, we can map any point $\phi_0$ to the origin $\Or$, where all scalars vanish, by the inverse of the coset representative $\Lc(\phi_0)^{-1}\in G$. We can then map a generic vacuum $\phi_0$ of a given theory (defined by an embedding tensor $\Theta$) to the origin of the theory defined by \,$\Theta'=\Lc(\phi_0)^{-1}\star \Theta$.\;
As a consequence of this, when looking for vacua with given properties (residual (super)symmetry, cosmological constant, mass spectrum etc.), with no loss of generality we can compute all quantities defining the gauged theory -- fermion shifts and mass matrices -- at the origin: \begin{equation}
\mathbb{N}(\Or,\,\Theta)\,,\;\;\mathbb{S}(\Or,\,\Theta)\,,\;\;\mathbb{M}(\Or,\,\Theta)\,,
\end{equation}
and translate the properties of the vacuum in conditions on $\Theta$. In this way, we can search for the vacua by scanning though all possible gaugings \cite{Dibitetto:2011gm,DallAgata:2011aa,inverso2010sitter}.

\subsection{Gauging \,$\N=8$\,,\; $D=4$}

\paragraph{Ungauged action.}
The four dimensional maximal supergravity is characterized by
having $\N=8$ supersymmetry (that is $32$ supercharges), which is
the maximal amount of supersymmetry allowed by a consistent theory of gravity.\par\smallskip
We shall restrict ourselves to the (ungauged) $\N=8$ theory with no antisymmetric tensor field -- which would eventually be dualized to scalars. The theory, firstly constructed in \cite{Cremmer:1978ds,Cremmer:1979up}, describes a single massless graviton supermultiplet consisting of the graviton $g_{\mu\nu}$, $8$ spin-$3/2$ gravitini $\psi^A_\mu$ ($A=1,\dots, 8$) transforming in the
fundamental representation of the R--symmetry group $\SU(8)$,
$28$ vector fields $A^\Lambda_\mu$ (with $\Lambda=0,\dotsc,\,27$), $56$ spin-$1/2$ dilatini $\chi_{ABC}$ in the ${\bf
56}$ of $\SU(8)$ and $70$ real scalar fields $\phi^r$:
\begin{align}
\big[\;\;
1\;\times\;\underbracket[0.15pt][3pt]{g_{\mu\nu}}_{\mathclap{j=2}}\;,\quad
8\;\times\;\underbracket[0.15pt][3pt]{\psi^A_\mu}_{\mathclap{j=\frac32}}\;,\quad
28\;\times\;\underbracket[0.15pt][3pt]{A^\Lambda_\mu}_{\mathclap{j=1}}\;,\quad
56\;\times\;\underbracket[0.15pt][3pt]{\chi_{ABC}}_{\mathclap{j=\frac12}}\;,\quad
70\;\times\;\underbracket[0.15pt][3pt]{\phi^r}_{\mathclap{j=0}}
\;\;\;\big]
\;.\label{N8}
\end{align}
The scalar fields are described by a non-linear $\sigma$-model on the Riemannian manifold $\Mscal$, that in the $\N=8$ model has the form
\begin{eqnarray}
\Mscal\=\frac{G}{H}\=\frac{\Exc}{\SU(8)}\,,
\end{eqnarray}
the isometry group being $G=\Exc$, and $H=\SU(8)$
being the R--symmetry group. The bosonic Lagrangian has the
usual form (\ref{boslagr}).
The global symmetry group of the maximal four-dimensional theory $G=\Exc$ has 133 generators $t_\alpha$. The (abelian) vector field strengths $F^\Lambda=dA^\Lambda$ and their magnetic duals $\Gd_\Lambda$ together transform in the $\mathscr{R}_v={\bf 56}$ fundamental representation of the $\Exc$ duality group with generators $(t_\alpha)_M{}^N$, so that
\begin{equation}
\delta \mathbb{F}^M_{\mu\nu} ~=
\left(\begin{matrix}
\delta F^\Lambda_{\mu\nu} \cr \delta \Gd_{\Lambda\,\mu\nu}
\end{matrix}\right)
=~ -\Lambda^\alpha\,(t_\alpha)_N{}^M\,\mathbb{F}^N_{\mu\nu}\;.
\end{equation}

\paragraph{Gauging.}
According to our general discussion of Sect.\ (\ref{gaugingsteps}), the most general gauge group $G_g$ which can be introduced in this theory is defined by an embedding tensor $\Theta_M{}^\alpha$ ({\footnotesize $M$}$~=1,\dots, 56$\, and \,$\alpha=1,\dots, 133$), which expresses the gauge generators $X_M$ as linear combinations of the the global symmetry group ones $t_\alpha$ (\ref{Thdef}). The embedding tensor encodes all parameters (couplings and mass deformations) of the gauged theory.
This object is solution to the $G$-covariant constraints (\ref{linear2}), (\ref{quadratic1}), (\ref{quadratic2}).\par
The embedding tensor formally belongs to the product
\begin{equation}
\Theta_M{}^\alpha ~\in~ \mathscr{R}_v\otimes\Adj(G)
\={\bf 56}\;\otimes\;{\bf 133} \=
{\bf 56}\;\oplus\;{\bf 912}\;\oplus\;{\bf 6480}\;.\label{56133}
\end{equation}
The linear constraint (\ref{linear2}) sets to zero all the representation in the above decomposition which are contained in the 3-fold symmetric product of the ${\bf 56}$ representation:
\begin{equation}
X_{(MNP)}~\in~ ({\bf 56}\;\otimes\;{\bf 56}\;\otimes\;{\bf 56})_{{\rm sym.}}
\,\rightarrow\;{\bf 56}\;\oplus\;{\bf 6480}\;\oplus\;{\bf 24320}\,.
\end{equation}
The representation constraint therefore selects the ${\bf 912}$ as the representation $\mathscr{R}_\Theta$ of the embedding tensor%
\footnote{
we can relax this constraint by extending this representation to include the ${\bf 56}$ in (\ref{56133}). Consistency however would require the gauging
of the scaling symmetry of the theory (which is never an off-shell symmetry), also called \emph{trombone symmetry} \cite{LeDiffon:2008sh,LeDiffon:2011wt}. This however leads to gauged theories which do not have an action. We shall not discuss these gaugings here
}.\par
The quadratic constraints pose further restrictions on the $\Exc$-orbits of the ${\bf 912}$ representation which $\Theta_M{}^\alpha$ should belong to. In particular the locality constraint implies that the embedding tensor can be rotated to an electric frame through a suitable symplectic matrix $E$, see eq.\ (\ref{elET}).\par
Steps 1,2 and 3 allow to construct the bosonic gauged Lagrangian in this electric frame. We shall discuss in Sect.\ \ref{sec:4} a frame-independent formulation of the gauging procedure in which, for a given solution $\Theta$ to the constraints, we no longer need to switch to the corresponding electric frame.\par The complete supersymmetric gauged Lagrangian is then obtained by adding fermion mass terms, a scalar potential and additional terms in the fermion supersymmetry transformation rules, according to the prescription given in Step 4. All these deformations depend on the fermion shift matrices $\mathbb{S}_{AB},\,\mathbb{N}_\mathcal{I}{}^A$.
In the maximal theory $\mathcal{I}=[ABC]$ labels the spin-$1/2$ fields $\chi_{ABC}$ and the two fermion shift-matrices are conventionally denoted by the symbols $A_1=(A_{AB}),\,A_2=(A^D{}_{ABC})$. The precise correspondence is%
\footnote{
in the previous sections we have used, for the supergravity fields, notations which are different from those used in the literature of maximal supergravity (e.g. in \cite{deWit:2007mt}) in order to make contact with the literature of gauged $\N<8$ theories, in particular $\N=2$ ones \cite{Andrianopoli:1996cm}. Denoting by a hat the quantities in \cite{deWit:2007mt}, the correspondence between the two notations is:
\begin{align}
&\hat{\gamma}^\mu =i\gamma^\mu\,;\quad
\hat{\gamma}_5 =\gamma_5\,,\nonumber\\
&\hat{\epsilon}_i=\frac{1}{\sqrt{2}}\,\epsilon^A\,;\quad
\hat{\epsilon}^i=\frac{1}{\sqrt{2}}\,\epsilon_A\,;\quad
(i=A)\,,\nonumber\\
&\hat{\psi}_{i\mu} =\sqrt{2}\,\psi^A_\mu\,;\quad
\hat{\psi}^i_{\mu} =\sqrt{2}\,\psi_{A\,\mu}\,;\quad
(i=A)\,,\nonumber\\
&\hat{\chi}_{ijk}=\chi^{ABC}\,;\quad
\hat{\chi}^{ijk}=\chi_{ABC}\,;\quad
([ijk]=[ABC])\,,\nonumber\\
&\hat{A}_{ij}=(\hat{A}_{ij})^*=A^{AB}\,;\quad
\hat{A}_i{}^{jkl}=(\hat{A}^i{}_{jkl})^*=A^A{}_{BCD}\,;\quad
(i=A,\,j=B,\,k=C,\,l=D)\,,\nonumber\\
&\mathcal{V}^{\Lambda\,ij}=
-\frac{i}{\sqrt{2}}\,\mathbb{L}^\Lambda{}_{AB}\,;\quad
\mathcal{V}_{\Lambda}{}^{ij}=
\frac{i}{\sqrt{2}}\,\mathbb{L}_{\Lambda AB}\,;\quad
(i=A,\,j=B)\,,
\end{align}
where in the last line the $28\times 28$ blocks of $\mathcal{V}_M{}^{\underline{N}}$ have been put in correspondence with those of $\LL^M{}_{\underline{N}}$. The factor $\sqrt{2}$ originates from a different convention with the contraction of antisymmetric couples of ${\rm SU}(8)$-indices:\; $\hat{V}_{ij}\hat{V}^{ij}=\frac{1}{2}\,V^{AB}\,V_{AB}$
}:
\begin{equation}
\mathbb{S}_{AB}=-\frac{1}{\sqrt{2}}\,A_{AB}\,;\quad
\mathbb{N}_{ABC}{}^D=-\sqrt{2}\,A^D{}_{ABC}{}\,,
\end{equation}
where
\begin{equation}
A_{AB}=A_{BA}\,;\quad\;
A_{ABC}{}^D=A_{[ABC]}{}^D\,;\quad\;
A_{DBC}{}^D=0\,.
\end{equation}
The above properties identify the ${\rm SU}(8)$ representations of the two tensors:
\begin{equation}
A_{AB}\in {\bf 36}\,;\quad\; A_{ABC}{}^D\in {\bf 420}\,.
\end{equation}
The $\Tb$-tensor, defined in (\ref{TT}) as an $\Exc$-object, transforms in $\mathscr{R}_\Theta={\bf 912}$, while as an ${\rm SU}(8)$-tensor it belongs to the following sum of representations:
\begin{equation}
\Tb~\in~ {\bf 912}\;
\stackrel{{\rm SU}(8)}{\longrightarrow}\;\,
{\bf 36}\;\oplus\;\ov{{\bf 36}}\;\oplus\;
{\bf 420}\;\oplus\;\ov{{\bf 420}}\;,
\end{equation}
which are precisely the representations of the fermion shift-matrices and their conjugates $A_{AB}\,A^{AB},\;A^A{}_{BCD},\;A_A{}^{BCD}$.\;
This guarantees that the $O(g)$-terms in the supersymemtry variation of $\Lagr_{{\rm gauged}}^{(0)}$, which depend on the $\Tb$-tensor, only contain ${\rm SU}(8)$-structures which can be canceled by the new terms containing the fermion shift-matrices. This shows that the linear condition $\Theta \in \mathscr{R}_\Theta$ is also required by supersymmetry.\par
The same holds for the quadratic constraints, in particular for (\ref{quadratic2}), which implies the $\Tb$-identities and also the Ward identity (\ref{WID}) for the potential \cite{deWit:1982ig,deWit:2007mt}:
\begin{equation}
V(\phi)\,\delta^B_A\=
\frac{g^2}{6}\,\mathbb{N}^{CDE}{}_A\mathbb{N}_{CDE}{}^B-12\,g^2\,\mathbb{S}_{AC}\mathbb{S}^{BC}
\=\frac{g^2}{3}A^B{}_{CDE} A_A{}^{CDE}-6\,g^2\, A_{AC}\,A^{BC}\,,
\end{equation}
from which we derive:
\begin{equation}
V(\phi)=g^2\,\left(
\frac{1}{24}\,|A^B{}_{CDE}|^2-\frac{3}{4}\, |A_{AB}|^2\right)\,.
\end{equation}
The scalar potential can also be given in a manifestly $G$-invariant form \cite{deWit:2007mt}\,:
\begin{equation}
V(\phi)=
-\frac{g^2}{672}\,\Big(X_{MN}{}^{R}\,X_{PQ}{}^{S}\,\M^{MP}\,\M^{NQ}\,\M_{RS}
+ 7\,X_{MN}{}^{Q}\,X_{PQ}{}^{N}\,\M^{MP}\Big)\;,
\label{potentialN8}
\end{equation}
where $\mathcal{M}^{MN}$ is the inverse of the (negative definite) matrix $\mathcal{M}_{MN}$ defined in (\ref{M}) and, as usual, $X_{MN}{}^{R}$ describe the symplectic duality action of the generators $X_M$ in the $\mathscr{R}_{v*}$-representation:\, $X_{MN}{}^{R}\equiv \mathscr{R}_{v*}[X_M]_N{}^R$.

\subsection{Brief account of old and new gaugings}
As mentioned in Sect.\ (\ref{gaugingsteps}), different symplectic frames (i.e.\ different ungauged Lagrangians) correspond to different choices for the viable gauge groups and may originate from different compactifications (see \cite{deWit:2002vt} for a study of the different symplectic frames for the ungauged maximal theory).\par
The toroidal compactification of eleven dimensional theory performed in \cite{Cremmer:1978ds}, upon dualization of all form-fields to lower order ones, yields an ungauged Lagrangian with global symmetry $G_{el}={\rm SL}(8,\mathbb{R})$. We shall refer to this symplectic frame as the ${\rm SL}(8,\mathbb{R})$-frame.
The first gauging of the maximal theory was performed in this symplectic frame by choosing $G_g={\rm SO}(8)\subset {\rm SL}(8,\mathbb{R})$ \cite{deWit:1982ig}. The scalar potential features a maximally supersymmetric anti-de Sitter vacuum which corresponds \cite{deWit:1986iy} to the spontaneous compactification of eleven dimensional supergravity on $\AdS_4\times S^7$. The range of possible gaugings in the ${\rm SL}(8,\mathbb{R})$-frame was extended to include non-compact and non semisimple groups $G_g={\rm CSO}(p,q,r)$ (with $p+q+r=8$) \cite{Hull:1984qz}. These were shown in \cite{Cordaro:1998tx} to exhaust all possible gaugings in this frame.\par
The discovery of inequivalent Lagrangian formulations of the ungauged maximal theory broadened the choice of possible gauge groups. Flat-gaugings in $D = 4$ describing Scherk-Schwarz reductions of maximal $D = 5$ supergravity \cite{Cremmer:1979uq} and yielding no-scale models, were first constructed in \cite{Andrianopoli:2002mf}. The corresponding symplectic frame is the one originating from direct dimensional reduction of the maximal five-dimensional theory on a circle and has a manifest off-shell symmetry which contains the global symmetry group of the parent model%
\footnote{
see Table \ref{tab:T-tensor-repr} at the end of Sect.\ \ref{sec:4}
}
${\rm E}_{6(6)}$: one has in fact $G_{el}={\rm O}(1,1)\times {\rm E}_{6(6)}$.\par\smallskip
Exploiting the freedom in the initial choice of the sympectic frame, it was recently possible to discover a new class of gauging generalizing the original ${\rm CSO}(p,q,r)$ ones \cite{Dall'Agata:2012sx,Dall'Agata:2012bb,Dall'Agata:2014ita}. These models are obtained by gauging, in a different frame, the same ${\rm CSO}(p,q,r)$.\par
Consider two inequivalent frames admitting $G_g={\rm CSO}(p,q,r)$ as gauge group, namely for each of which ${\rm CSO}(p,q,r)\subset G_{el}$. Let $\hat{\mathscr{R}}_v$ and ${\mathscr{R}}_v$ be the corresponding symplectic duality representations of $G$. We can safely consider one of them ($\hat{\mathscr{R}}_v$) as electric. The duality action of the gauge generators $\hat{\mathscr{R}}_{v*}$ and ${\mathscr{R}}_{v*}$ are described by two tensors $X_{\hat{M}\hat{N}}{}^{{\hat{P}}}$ and $X_{MN}{}^P$, respectively, related by a suitable matrix $E$ (\ref{XEhatX}):
\begin{equation}
X_{\hat{M}\hat{N}}{}^{{\hat{P}}}=E_{\hat{M}}{}^M\,E_{\hat{N}}{}^N\,(E^{-1})_P{}^{{\hat{P}}}\,X_{MN}{}^P\,.
\end{equation}
The matrices $\mathcal{M}(\phi)$ in the two frames are then related by (\ref{MEtra}).
The two embedding tensors describe the same gauge group provided that $\{X_{M}\}$ and $\{E\,X_M\,E^{-1}\}$ define different bases of the same gauge algebra $\mathfrak{g}_g=\mathfrak{cso}(p,q,r)$ in the Lie algebra $\mathfrak{e}_{7(7)}$ of ${\rm E}_{7(7)}$. In other words, $E$ should belong to the \emph{normalizer} of $\mathfrak{cso}(p,q,r)$ in ${\rm Sp}(2n_v,\mathbb{R})$. At the same time the effect of $E$ should not be offset by local (vector and scalar field) redefinitions, see (\ref{generalE}).
The duality action of $G_g$ in both $\hat{\mathscr{R}}_{v*}$ and ${\mathscr{R}}_{v*}$ is block-diagonal:
\begin{equation}
\hat{\mathscr{R}}_{v*}[G_g]={\mathscr{R}}_{v*}[G_g]=\left(\begin{matrix}G_g & \Zero \cr\Zero & G_g^{-T}\end{matrix}\right)\,.
\end{equation}
For semisimple gauge groups $G_g={\rm SO}(p,q)$ (with $p+q=8$), it was shown in \cite{Dall'Agata:2014ita} that the most general $E$ belongs to an ${\rm SL}(2,\mathbb{R})$-subgroup of ${\rm Sp}(56,\mathbb{R})$ and has the general form:
\begin{equation}
E=\left(
\begin{matrix}
a\,\Id & b\,\eta\cr
c\,\eta & d\,\Id
\end{matrix}
\right)
~\in~ {\rm Sp}(56,\mathbb{R})\:;\quad\;
ad-bc=1
\,,\label{Eimage}
\end{equation}
where $\eta_{\Lambda\Sigma}$ is the $\mathfrak{so}(p,q)$-Cartan Killing metric, normalized so that $\eta^2=\Id$. The most general
${\rm SL}(2,\mathbb{R})$-matrix can be written, using the Iwasawa decomposition, as follows:
\begin{equation}
 \left(\begin{matrix}a & b\cr c & d\end{matrix}\right)=\left(\begin{matrix}\lambda & 0\cr 0 & \frac{1}{\lambda}\end{matrix}\right)\left(\begin{matrix}1 & \vartheta \cr 0 & 1\end{matrix}\right)\left(\begin{matrix}\cos(\omega) & \sin(\omega) \cr -\sin(\omega) & \cos(\omega)\end{matrix}\right)\,.
\end{equation}
The leftmost block corresponds in $E$ to an unphysical rescaling of the vectors (in ${\rm GL}(28,\mathbb{R})$). The middle block realizes, in going from the unhatted frame to the hatted one, a constant shift in the generalized $\theta$-angle matrix $\R$:\, $\R\rightarrow \R+\vartheta\,\eta$. This can have effects at the quantum level, but does not affect field equations \cite{Dall'Agata:2014ita}.\par
The rightmost block has, on the other hand, important bearing on the physics of the classical theory. Let $E(\omega)$ be the symplectic image (\ref{Eimage}) of this block only, and let ${\mathscr{R}}_{v}$ be the ${\rm SL}(8,\mathbb{R})$-frame, where the ${\rm CSO}(p,q,r)$ gaugings were originally constructed and in which the matrices $\LL$ and $\mathcal{M}$ are given by well know general formulas \cite{Cremmer:1978ds,deWit:1982ig}.\, For $\omega\neq 0$, this frame is no longer electric, but is related to the electric one by $E(\omega)$. Using (\ref{elET}) we can write:
\begin{equation}
X_{\hat{\Lambda}}=\cos(\omega) X_\Lambda+ \sin(\omega)\eta_{\Lambda\Sigma}\,X^\Sigma\,;
\quad\;
0=-\sin(\omega) \eta^{\Lambda\Sigma}\,X_\Sigma+ \cos(\omega)X^\Lambda\,,
\end{equation}
where $(\eta^{\Lambda\Sigma})\equiv \eta^{-1}=\eta$.\; The above relation is easily inverted:
\begin{equation}
X_\Lambda=\cos{(\omega)}\,X_{\hat{\Lambda}}\,\,,\,\,\,\,\,\,X^\Lambda=\sin{(\omega)}\,\eta^{\Lambda\Sigma}X_{\hat{\Sigma}}\,.
\end{equation}
We can then write the symplectic invariant connection (\ref{syminvmc}) in the following way:
\begin{equation}
\Omega_{g\,\mu}=A^M_\mu\,X_M=A^\Lambda_\mu\,X_\Lambda+A_{\Lambda\,\mu}\,X^\Lambda=(\cos{\omega}\,A^\Lambda_\mu+\sin(\omega)\,A_{\Lambda\,\mu})
X_{\hat{\Lambda}}=A^{\hat{\Lambda}}_\mu\,X_{\hat{\Lambda}}\,.
\end{equation}
In other words, the gauging defined by $X_M$ amounts to gauge, in the ${\rm SL}(8,\mathbb{R})$-frame, the same ${\rm SO}(p,q)$-generators by a linear combination of the electric $A^\Lambda_\mu$ and magnetic $A_{\Lambda\,\mu}$ vector fields. The true electric vectors are all and only those entering the gauge connection, that is $A^{\hat{\Lambda}}_\mu$, and define the electric frame. We shall denote by $\Theta[\omega]$ the corresponding embedding tensor.\par
The gauged model can be constructed either directly in the ${\rm SL}(8,\mathbb{R})$-frame, using the covariant formulation to be discussed in Sect.\ \ref{sec:4}, or in the electric frame, along the lines described in Sect.\ \ref{sec:3}. The range of values of $\omega$ is restricted by the discrete symmetries of the theory. One of these is parity (see Sect.\ \ref{gsg}), whose duality representation ${\bf P}$ in the ${\rm SL}(8,\mathbb{R})$-frame has the form (\ref{Pmatrix}) \cite{Ferrara:2013zga}. The reader can verify that its effect on the $\Tb$-tensor (\ref{TT}) is:
\begin{align}
\Tb(\Theta[\omega],\phi)_{\underline{M}}&\={\bf P}\star \Tb(\Theta[-\omega],\phi_p)
\label{PTtras}
\end{align}
by using the properties
\begin{equation}
{\bf P}_{\hat{M}}{}^{\hat{N}}\,{\bf P}^{-1} X_{\hat{N}}{\bf P}=X_{\hat{M}}\,;\quad\;
{\bf P}^{-1} E(\omega){\bf P}=E(-\omega)\,;\quad\;
{\bf P}^{-1}\LL(\phi){\bf P}=\LL(\phi_p)\,,
\end{equation}
where $\phi_p$ denote the parity-transformed scalar fields. Eq.\ (\ref{PTtras}) shows that parity maps $\phi$ into $\phi_p$ and $\omega$ in $-\omega$. In other words $\omega$ is \emph{parity-odd parameter}. The overall ${\bf P}$ transformation on $\Tb$ in
(\ref{PTtras}) is ineffective, since it will cancel everywhere in the Lagrangian, being ${\bf P}$ an ${\rm O}(2n_v)$-transformation.
Similarly, we can use other discrete global symmetries of the ungauged theory, which include the ${\rm SO}(8)$-triality transformations $S_3\subset {\rm E}_{7(7)}$ for the ${\rm SO}(8)$-gauging, to further restrict the range of values of $\omega$.
One finds that \cite{Dall'Agata:2012bb,Dall'Agata:2014ita}:
\begin{align}
&\omega~\in~\left[0,\frac{\pi}{8}\right]\;,\quad
\text{${\rm SO}(8)$ and $\SO(4,4)$ gaugings}\,,\nonumber\\
&\omega~\in~\left[0,\frac{\pi}{4}\right]\;,\quad
\text{other non-compact ${\rm SO}(p,q)$ gaugings}\,.
\end{align}
These are called ``$\omega$-rotated'' ${\rm SO}(p,q)$-models, or simply ${\rm SO}(p,q)_{\omega}$-models. The ${\rm SO}(8)$ ones, in particular, came as a surprise since they contradicted the common belief that the original de Wit-Nicolai ${\rm SO}(8)$-gauged model was unique.\par
For the non-semisimple ${\rm CSO}(p,q,r)$-gaugings, the non-trivial matrix $E$ does not depend on continuous parameters but is fixed, thus yielding for each gauge group only one rotated-model \cite{Dall'Agata:2012sx,Dall'Agata:2014ita}.\par
Even more surprisingly, these new class of gauged theories feature a broader range of vacua than the original models. In this sense the $\omega\rightarrow 0$ limit can be considered a singular one, in which some of the vacua move to the boundary of the moduli space at infinity and thus disappear.\par
\begin{figure}[!h]
\centerline{\includegraphics[width=0.8\textwidth]{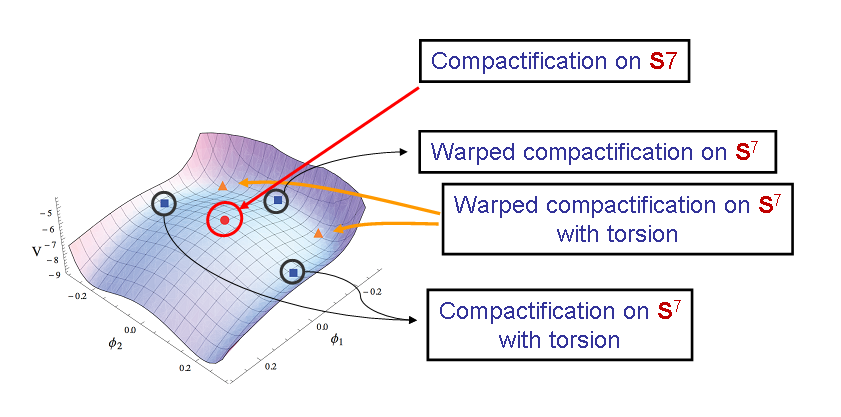}}
 \caption{\scriptsize The $G_2$-invariant vacua of the de Wit-Nicolai model, with their interpretation in terms of compactifications of the eleven-dimensional theory.}\label{fign8}
\end{figure}

Consider for instance the ${\rm SO}(8)_{\omega}$-models. They all feature an $\AdS_4$, $\N=8$ vacuum at the origin with the same cosmological constant and mass spectrum as the original  ${\rm SO}(8)$ theory. The parameter $\omega$ manifests itself in the higher order interactions of the effective theory. They also feature new vacua, which do not have counterparts in the $\omega=0$ model.
Fig.\ \ref{fign8} illustrates some of the vacua of the de Wit-Nicolai model ($\omega=0$), namely those which feature a residual symmetry group $G_2\subset{\rm SO}(8)$.\par
\begin{figure}[!h]
\centerline{\includegraphics[width=1\textwidth]{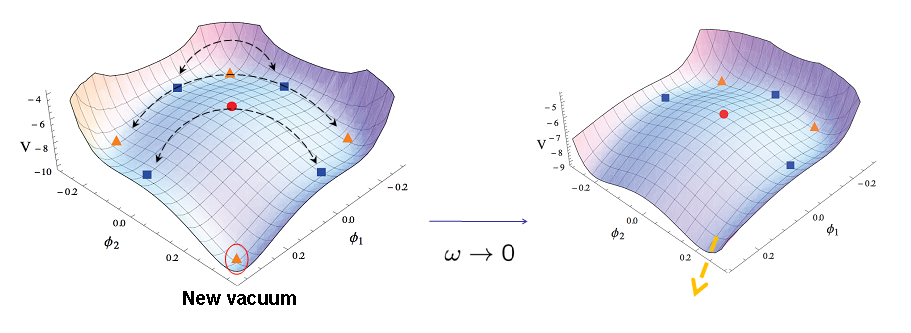}}
 \caption{\scriptsize On the left the $G_2$-invariant vacua of the ${\rm SO}(8)_\omega$ model, with $\omega=\frac{\pi}{8}$. The dashed lines
 represent identifications of vacua due to a discrete symmetry of the theory which is a combination of triality and parity. All of them have an $\omega=0$ counterpart, except the lowest one, marked by a circle, which disappears in the $\omega\rightarrow 0$ limit.}\label{fign82}
\end{figure}

Fig.\ \ref{fign82} shows the $G_2$-invariant vacua of a particular ${\rm SO}(8)_\omega$ model and the disappearance of one of the vacua in the $\omega\rightarrow 0$ limit \cite{Dall'Agata:2012bb}.
The vacua of these models have been extensively studied \cite{Borghese:2012qm,Borghese:2012zs,Borghese:2013dja,Guarino:2013gsa,Gallerati:2014xra} also in the context of renormalization group flows interpolating between (or simply originating from) AdS vacua \cite{Tarrio:2013qga,Pang:2015mra} and AdS black holes \cite{Anabalon:2013eaa,Lu:2014fpa,Wu:2015ska,Anabalon:2017yhv}.\par
Determining a string or M-theory origin of the $\omega$-rotated models is, to date, an open problem \cite{Lee:2015xga}. They seem to provide examples of what we named \emph{intrinsically non-geometric} models in Sect.\ \ref{dfcomp}. 
The only exceptions so far are the dyonic nonsemisimple gaugings ${\rm CSO}(p, q,r)$. Indeed, the ${\rm ISO}(p,7-p)$ gaugings were shown to be related to compactifications of massive Type-IIA theory \cite{Guarino:2015jca,Guarino:2015qaa,Guarino:2015vca,Cassani:2016ncu}. The $p=7$ theory features $\N=2$ \cite{Guarino:2015jca} and $\N=3$ \cite{Gallerati:2014xra,Pang:2015vna,Pang:2015rwd} AdS-vacua, all corresponding to backgrounds with topology $\AdS_4\times S^6$. The uplift of the generic ${\rm CSO}(p, q,r)$-model to Type-IIA or Type-IIB theory was eventually achieved in \cite{Inverso:2016eet}.

\section{Duality covariant gauging}\label{sec:4}
Let us discuss in this section a formulation of the gauging procedure in four-dimensions which was developed in \cite{deWit:2005ub,deWit:2007mt} and which no longer depends on the matrix $E$, so that the kinetic terms are not written in terms of the vector fields in the electric frame.\par

\paragraph{Step 1, 2 and 3 revisited.}
We start from a symplectic-invariant gauge connection of the form%
\footnote{
here, for the sake of simplicity, we reabsorb the gauge coupling constant $g$ into $\Theta$:\; $g\,\Theta\rightarrow \Theta$
}:
\begin{equation}
\Omega_{g\mu}\equiv A^M_\mu\,X_M=A^\Lambda_\mu\,X_\Lambda+A_\Lambda^\mu\,X_\Lambda=A^M_\mu\,\Theta_M{}^\alpha\,t_\alpha\,,\label{newcon}
\end{equation}
where $\Theta_M{}^\alpha$ satisfies the constraints (\ref{linear2}), (\ref{quadratic1}), (\ref{quadratic2}). The fields $A^\Lambda_\mu$ and $A_{\Lambda\,\mu}$ are now taken to be independent. This is clearly a redundant choice and, as we shall see, half of them play to role of auxiliary fields. Eq.\ (\ref{quadratic1}) still implies that at most $n_v$ linear combinations $A^{\hat{\Lambda}}_\mu$ of the $2n_v$ vectors $A^\Lambda_\mu,\,A_{\Lambda\,\mu}$ effectively enter the gauge connection (and thus the minimal couplings):
\begin{equation}
A^M_\mu\,X_M=A^{\hat{\Lambda}}_\mu\,X_{\hat{\Lambda}}\,,
\end{equation}
where $X_{\hat{\Lambda}}$ are defined in eq.\ (\ref{elET}) through the matrix $E$, whose existence is guaranteed by (\ref{quadratic1}), and where $A^{\hat{\Lambda}}_\mu\equiv E^{-1}{}_M{}^{\hat{\Lambda}}\,A^M_\mu$.\par
In the new formulation we wish to discuss, however, the vectors $A^\Lambda_\mu$ instead of $A^{\hat{\Lambda}}_\mu$ enter the kinetic terms.
The covariant derivatives are then defined in terms of (\ref{newcon}) as in Step 2 of the Section \ref{gaugingsteps}, and, as prescribed there, should replace ordinary derivative everywhere in the action.
The infinitesimal gauge variation of $A^M$ reads:
\begin{equation}
\delta A^M_\mu=\mathcal{D}_\mu\zeta^M\equiv \partial_\mu\zeta^M+\,A^N_\mu X_{NP}{}^M\,\zeta^P\,,\label{deltaA}
\end{equation}
where, as usual, $X_{MP}{}^R\equiv \mathscr{R}_{v*}[X_{M}]_{P}{}^R$.
We define for this set of electric-magnetic vector fields a symplectic covariant generalization $\mathbb{F}^M$ of the non-abelian field strengths $F^{\hat{\Lambda}}$ (\ref{defF}):
\begin{equation}
{F}^M_{\mu\nu}\equiv \partial_\mu A^M_\nu-\partial_\nu A^M_\mu+\,X_{[NP]}{}^M\,A^N_\mu A^P_\nu
\quad\Leftrightarrow\quad
{F}^M\equiv dA^M+\frac{g}{2}\,X_{NP}{}^M\,A^N\wedge A^P\,,\label{FMdef}
\end{equation}
where in the last equation we have used the form-notation for the fields strengths. The gauge algebra-valued curvature $\mathcal{F}$ is defined as in (\ref{calF}):
\begin{equation}
\mathcal{F}\equiv {F}^M\,X_M\,.\label{gcurv}
\end{equation}
The first problem one encounters in describing the vectors $A^\Lambda_\mu$ in the kinetic terms is that, in a symplectic frame which is not the electric one, such fields are not well defined, since their curvatures fail to satisfy the Bianchi identity.
This comes with no surprise, since the components $\Theta^{\Lambda\,\alpha}$ of the embedding tensor are nothing but \emph{magnetic charges}.
One can indeed verify that:
\begin{equation}
\mathcal{D}{F}^M~\equiv~d{F}^M+\,X_{NP}{}^M\,A^N\wedge {F}^P
\=\,X_{(PQ)}{}^M\,A^P\wedge\left( dA^Q+\frac{g}{3}\,X_{RS}{}^QA^R\wedge A^S\right)
~\neq~ 0
\,.\label{Bianchifail}
\end{equation}
In particular $\mathcal{D}F^\Lambda\neq 0$ since $X_{(PQ)}{}^\Lambda=-\frac{1}{2}\,\Theta^{\Lambda\alpha}\,t_{\alpha\,M}{}^P\mathbb{C}_{PN}\neq 0$, being in the non-electric frame $\Theta^{\Lambda\alpha}\neq 0$.\;
To deduce (\ref{Bianchifail}) we have used the quadratic constraint (\ref{quadratic2}) on the gauge generators $X_M$ in the $\mathscr{R}_{v*}$-representation, which reads:
\begin{equation}
X_{MP}{}^R X_{NR}{}^Q-X_{NP}{}^R X_{MR}{}^Q+X_{MN}{}^R X_{RP}{}^Q=0\,.
\end{equation}
From the above identity, after some algebra, one finds:
\begin{equation}
X_{[MP]}{}^R X_{[NR]}{}^Q+X_{[PN]}{}^R X_{[MR]}{}^Q+X_{[NM]}{}^R X_{[PR]}{}^Q=-(X_{NM}{}^R\,X_{(PR)}{}^Q)_{[MNP]}\,,\label{nojacobi}
\end{equation}
that is the \emph{generalized structure constants} $X_{[MP]}{}^R$ entering the definition (\ref{FMdef}) do not satisfy the Jacobi identity, and this feature is at the root of (\ref{Bianchifail}). Related to this is the non-gauge covariance of ${F}^M$. The reader can indeed verify that (using the form-notation):
\begin{equation}
\delta F^M=-\,X_{NP}{}^M\,\zeta^N\,F^P+\,\left(2 \,X_{(NP)}{}^M\,\zeta^N\,F^P-X_{(NP)}{}^M\,A^N\wedge \delta A^P\right)\neq -\,X_{NP}{}^M\,\zeta^N\,F^P\,,\label{deltsFnc}
\end{equation}
where $\delta A^M$ is given by (\ref{deltaA}) and where we have used the general property
\begin{equation}
\delta F^M=\mathcal{D}\delta A^M+X_{(PQ)}{}^M\,A^P\wedge \delta A^Q
\,,\label{deltaFgen}
\end{equation}
valid for generic $\delta A^M$.\; We also observe that the obstruction to the Bianchi identity (\ref{Bianchifail}), as well as the non-gauge covariant terms in (\ref{deltsFnc}), are proportional to a same tensor $X_{(MN)}{}^P$. This quantity, as a consequence of (\ref{quadratic2}) and (\ref{quad2n}), vanishes if contracted with the gauge generators $X_M$, namely with the first index of the embedding tensor: $X_{(MN)}{}^P\,\Theta_P{}^\alpha=0$.
Therefore the true electric vector fields $A^{\hat{\Lambda}}_\mu$ and the gauge connection which only depends on them, are perfectly well defined. Indeed, one can easily show using the matrix $E$ that the gauge curvature (\ref{gcurv}) only contains the field strengths $F^{\hat{\Lambda}}$ associated with $A^{\hat{\Lambda}}$ and defined in (\ref{defF}):
\begin{equation}
\mathcal{F} ~\equiv~ {F}^M\,X_M \= F^{\hat{\Lambda}}\,X_{\hat{\Lambda}}\,.
\end{equation}
On the other hand, using (\ref{Bianchifail}) and (\ref{quad2n}) we have:
\begin{equation}
 \mathcal{D}\mathcal{F}\= \mathcal{D}{F}^M\,X_M\=0\,.\label{BianchiFgauge}
\end{equation}
The gauge covariance (\ref{gaugecovF}) of $\mathcal{F}$, and thus of $F^{\hat{\Lambda}}$, is also easily verified by the same token, together with eq.\ (\ref{D2F}):\; $\mathcal{D}^2=-\mathcal{F}$.\par\smallskip
In order to construct gauge-covariant quantities describing the vector fields, we combine the vector field strengths $F^M_{\mu\nu}$ with a set of  massless antisymmetric tensor fields%
\footnote{
these fields will also be described as 2-forms $B_\alpha\equiv \frac{1}{2}\,B_{\mu\nu}\,dx^\mu\wedge dx^\nu$
}
$B_{\alpha\,\mu\nu}$ in the adjoint representation of $G$ through the matrix
\begin{equation}
Z^{M\,\alpha}\equiv \frac{1}{2}\,\mathbb{C}^{MN}\,\Theta_N{}^\alpha\,,\label{defZ}
\end{equation}
and define the following new field strengths:
\begin{align}
\mathcal{H}^M_{\mu\nu}~\equiv~ F^M_{\mu\nu}+Z^{M\,\alpha}\,B_{\alpha\,\mu\nu}\;:\;\;
\begin{cases}
\mathcal{H}^\Lambda=dA^\Lambda+\frac{1}{2}\,\Theta^{\Lambda\alpha}\,B_\alpha\,,\cr
\mathcal{H}_\Lambda=dA_\Lambda-\frac{1}{2}\,\Theta_{\Lambda}{}^{\alpha}\,B_\alpha\,.
\end{cases}
\label{HZB}
\end{align}
From the definition (\ref{defZ}) and (\ref{quadratic1}) we have:
\begin{equation}
Z^{M\,\alpha}\,\Theta_{M}{}^\beta=0
\quad\Leftrightarrow\quad
Z^{M\,\alpha}\,X_{M}=0\,.\label{Zort}
\end{equation}
The reader can verify, using the linear constraint (\ref{linear2}), that:
\begin{equation}
X_{(NP)}{}^M=-\frac{1}{2}\,\mathbb{C}^{MQ}\,X_{QN}{}^R\mathbb{C}_{RP}=-\frac{1}{2}\,
\mathbb{C}^{MQ}\,\Theta_Q{}^\alpha\,t_{\alpha\,N}{}^R\mathbb{C}_{RP}=-Z^{M\,\alpha}\,t_{\alpha\,NP}\,,\label{linear22}
\end{equation}
where, as usual, we have defined $t_{\alpha\,NP}\equiv t_{\alpha\,N}{}^R\mathbb{C}_{RP}$.\par
The reason for considering the combination (\ref{HZB}) is that the non-covariant terms in the gauge variation of $F^M_{\mu\nu}$, being proportional to $X_{(NP)}{}^M$, that is to $Z^{M\,\alpha}$, can be canceled by a corresponding variation of the tensor fields $\delta B_{\alpha\mu\nu}$:
\begin{align}
\delta \mathcal{H}^M&=\,X_{PN}{}^M\,\zeta^N\,F^P+Z^{M\alpha}\,\left(\delta B_\alpha+t_{\alpha NP}\,A^N\wedge \delta A^P\right)=\nonumber\\
&=X_{PN}{}^M\,\zeta^N\,\mathcal{H}^P+Z^{M\alpha}\,\left(\delta B_\alpha+t_{\alpha NP}\,A^N\wedge \delta A^P\right)=\nonumber\\
&=-X_{NP}{}^M\,\zeta^N\,\mathcal{H}^P+2\,X_{(NP)}{}^M\,\zeta^N\,\mathcal{H}^P+Z^{M\alpha}\,\left(\delta B_\alpha+t_{\alpha NP}\,A^N\wedge \delta A^P\right)=\nonumber\\
&=-X_{NP}{}^M\,\zeta^N\,\mathcal{H}^P+Z^{M\alpha}\,\left[\delta B_\alpha+t_{\alpha NP}\,(A^N\wedge \delta A^P-2\,\zeta^N\,\mathcal{H}^P)\right]\,,
\end{align}
where, in going from the first to the second line, we have used (\ref{Zort}), so that\, $X_{PN}{}^M\,F^P=X_{PN}{}^M\,\mathcal{H}^P$.\;
If we define:
\begin{equation}
\delta B_\alpha ~\equiv~ t_{\alpha NP}\,(2\,\zeta^N\,\mathcal{H}^P-A^N\wedge \delta A^P)\,,\label{Btra1}
\end{equation}
the term proportional to $Z^{M\,\alpha}$ vanishes and $ \mathcal{H}^M$ transforms covariantly.
The kinetic terms in the Lagrangian are then written in terms of $\mathcal{H}^\Lambda_{\mu\nu}$:
\begin{equation}
\frac{1}{e}\Lagr_{v,\,{\rm kin}}=
\frac{1}{4}\,\I_{\Lambda\Sigma}(\phi)\,\mathcal{H}^\Lambda_{\mu\nu}\,\mathcal{H}^{\Sigma\,\mu\nu}
+\frac{1}{8\,e}\,\R_{\Lambda\Sigma}(\phi)\,\eps^{\mu\nu\rho\sigma}\,\mathcal{H}^\Lambda_{\mu\nu} \,\mathcal{H}^{\Sigma}_{\rho\sigma}\,.
\label{bosoniclagr}
\end{equation}
The above transformation property (\ref{Btra1}) should however be modified since the quantity we want to transform covariantly is not quite $\mathcal{H}^M$, but rather the symplectic vector:
\begin{equation}
\Gd^M\equiv \left(\begin{matrix}\mathcal{H}^\Lambda\cr \Gd_\Lambda\end{matrix}\right)\;;\quad\;
\Gd_{\Lambda\,\mu\nu}\equiv -\epsilon_{\mu\nu\rho\sigma}
\frac{\partial \Lagr}{\partial
\mathcal{H}^\Lambda_{\rho\sigma}}\,,
\end{equation}
corresponding, in the ungauged theory, to the field-strength-vector $\mathbb{F}^M$ of eq.\ (\ref{bbF}), and which contains inside $\Gd_\Lambda$ fermion bilinears coming from Pauli terms in the Lagrangian.
Consistency of the construction will then imply that the two quantities $\mathcal{H}^M$ and $\Gd^M$, which are off-shell different since the former depends on the magnetic vector fields $A_\Lambda$ as opposed to the latter, \emph{will be identified on-shell} by the equation \begin{equation}(\mathcal{H}^M-\Gd^M)\,\Theta_M{}^\alpha=(\mathcal{H}_\Lambda-\Gd_\Lambda)\,\Theta^{\Lambda\,\alpha}=0\,.\label{HG0} \end{equation}
These equations will in particular identify the field strengths of the auxiliary fields $A_\Lambda$ in $\mathcal{H}_\Lambda$ with the duals to $\mathcal{H}^\Lambda$.
The best that we can do is to make $\Gd^M$ on-shell covariant under $G_g$, namely upon use of (\ref{HG0}).
To this end, we modify eq.\ (\ref{Btra1}) as follows:
\begin{equation}
\delta B_\alpha\equiv t_{\alpha NP}\,(2\,\zeta^N\,\Gd^P-A^N\wedge \delta A^P)\,,\label{Btra12}
\end{equation}
so that the variations of the symplectic vectors $\mathcal{H}^M$ and $\Gd^M$ read:
\begin{align}
\delta \mathcal{H}^M&=-X_{NP}{}^M\,\zeta^N\,\mathcal{H}^P~+~\text{non-covariant terms}\,,\nonumber\\
\delta \Gd^M&=-X_{NP}{}^M\,\zeta^N\,\Gd^P~+~\text{non-covariant but on-shell vanishing terms}\,.
\end{align}
\par
Consistent definition of $B_\alpha$ requires the theory to be gauge-invariant with respect to transformations parmetrized by 1-forms: $\Xi_\alpha=\Xi_{\alpha\mu}\,dx^\mu$.
Such transformations should in turn be $G_g$-invariant and leave $\mathcal{H}^M$ unaltered:
\begin{equation}
A^M\rightarrow A^M+\delta_\Xi A^M\;;\quad\;
B_\alpha\rightarrow B_\alpha+\delta_\Xi B_\alpha
\quad\Rightarrow\quad
\delta_\Xi\mathcal{H}^M=0\,.
\end{equation}
Let us use (\ref{deltaFgen}) then to write
\begin{equation}
\delta_\Xi\mathcal{H}^M=\mathcal{D}\delta_\Xi A^M+Z^{M\,\alpha}\,\left(\delta_\Xi B_{\alpha}+t_{\alpha NP}\,A^N\wedge \delta_\Xi A^P\right)\,.
\end{equation}
If we set
\begin{equation}
\delta_\Xi A^M=-Z^{M\alpha}\,\Xi_\alpha\,,\label{deltaxi1}
\end{equation}
the invariance of $\mathcal{H}^M$ implies:
\begin{equation}
\delta_\Xi B_{\alpha}=\mathcal{D}\Xi_\alpha-t_{\alpha NP}\,A^N\wedge \delta_\Xi A^P\,,\label{deltaxi2}
\end{equation}
where
\begin{equation}
\mathcal{D}\Xi_\alpha\equiv d\Xi_\alpha+\Theta_M{}^\beta\,{\rm f}_{\beta\alpha}{}^\gamma A^M\wedge\Xi_\gamma\,.
\end{equation}
Let us now introduce field strengths for the 2-forms:
\begin{equation}
\mathcal{H}^{(3)}_\alpha\equiv \mathcal{D}B_\alpha-t_{\alpha PQ}A^P\wedge\left( dA^Q+\frac{1}{3}\,X_{RS }{}^Q\,A^R\wedge A^S\right)\,.
\end{equation}
Writing the forms in components,
\begin{equation}
\mathcal{H}^{(3)}_\alpha=\frac{1}{3!}\,\mathcal{H}_{\alpha\,\mu\nu\rho}\,dx^\mu\wedge dx^\nu\wedge dx^\rho\;;\quad\;
\mathcal{D}B_\alpha=\frac{1}{2}\,\mathcal{D}_{\mu}B_{\alpha\,\nu\rho}\,dx^\mu\wedge dx^\nu\wedge dx^\rho\,,
\end{equation}
we have:
\begin{equation}
\mathcal{H}_{\alpha\,\mu\nu\rho}=3\,\mathcal{D}_{[\mu}B_{\alpha\,\nu\rho]}-6\,t_{\alpha PQ}\left(A^P_{[\mu}\partial_\nu A_{\rho]}^Q+\frac{1}{3}\,X_{RS }{}^Q\,A^P_{[\mu}A^R_\nu A^S_{\rho]}\right)\,.
\end{equation}
The reader can verify that the following Bianchi identities hold:
\begin{align}
\mathcal{D}\mathcal{H}^M&= Z^{M\alpha}\,\mathcal{H}^{(3)}_\alpha\,,\label{Bid1n}\\
\mathcal{D}\mathcal{H}^{(3)}_\alpha &= X_{NP}{}^M\,\mathcal{H}^N\wedge \mathcal{H}^P\,.
\end{align}
Just as in Step 3. of Section \ref{gaugingsteps}, gauge invariance of the bosonic action requires the introduction of topological terms, so that the final gauged bosonic Lagrangian reads:
\begin{eqnarray}
\Lagr_{\text{bos}} &=& -\frac{e}{2}\,R+\frac{e}{2}\,\Gm_{st}(\phi)\,\mathcal{D}_\mu\phi^s\,\mathcal{D}^\mu\phi^t+
\nonumber\\
&&+\frac{e}{4} \, {\cal
I}_{\Lambda\Sigma}\,\mathcal{H}_{\mu\nu}{}^{\Lambda}
\mathcal{H}^{\mu\nu\,\Sigma} +\frac{1}{8} {\cal
R}_{\Lambda\Sigma}\;\varepsilon^{\mu\nu\rho\sigma}
\mathcal{H}_{\mu\nu}{}^{\Lambda}
\mathcal{H}_{\rho\sigma}{}^{\Sigma}+
 \nonumber\\[.9ex]
&&{}+\frac{1}{8}\, \varepsilon^{\mu\nu\rho\sigma}\,
\theta^{\Lambda\alpha}\,B_{\mu\nu\,\alpha} \, \Big(
2\,\partial_{\rho} A_{\sigma\,\Lambda} + X_{MN\,\Lambda}
\,A_\rho{}^M A_\sigma{}^N
-\frac{1}{4}\,\theta_{\Lambda}{}^{\beta}B_{\rho\sigma\,\beta}
\Big)+
\nonumber\\[.9ex]
&&{} +\frac{1}{3}\,
\varepsilon^{\mu\nu\rho\sigma}X_{MN\,\Lambda}\, A_{\mu}{}^{M}
A_{\nu}{}^{N} \Big(\partial_{\rho} A_{\sigma}{}^{\Lambda}
+\frac{1}{4}  X_{PQ}{}^{\Lambda}
A_{\rho}{}^{P}A_{\sigma}{}^{Q}\Big)+
\nonumber\\[.9ex]
&&{} +\frac{1}{6}\,
\varepsilon^{\mu\nu\rho\sigma}X_{MN}{}^{\Lambda}\, A_{\mu}{}^{M}
A_{\nu}{}^{N} \Big(\partial_{\rho} A_{\sigma}{}_{\Lambda}
+\frac{1}{4}\, X_{PQ\Lambda}
A_{\rho}{}^{P}A_{\sigma}{}^{Q}\Big)	\,.	\label{boslag2}
\end{eqnarray}
The Chern-Simons terms in the last two lines generalize those in eq.\ (\ref{top}). On top of them, gauge invariance of the action requires the introduction of new topological terms, depending on the $B$-fields, which appear in the third line of (\ref{boslag2}). Notice that if the magnetic charges $\Theta^{\Lambda\,\alpha}$ vanish (i.e.\ we are in the electric frame), $B_\alpha$ disappear from the action, since the second line of (\ref{boslag2}) vanish as well as the $B$-dependent Stueckelberg term in $\mathcal{H}^\Lambda$. \par
The constraints (\ref{linear2}), (\ref{quadratic1}) and (\ref{quadratic2}) are needed for the consistent construction of the gauged bosonic action, which is uniquely determined. Just as discussed in Sect.\ (\ref{gaugingsteps}), they are also enough to guarantee its consistent supersymmetric completion through Step 4, which equally applies to this more general construction.\par\bigskip
Some comments are in order.
\begin{enumerate}[itemsep=1.5ex]
\item[i)]{The construction we are discussing in this Section requires the introduction of additional fields: $n_v$ magnetic potentials $A_{\Lambda\mu}$ and a set of antisymmetric tensors $B_{\alpha\,\mu\nu}$. These new fields come together with extra gauge-invariances (\ref{deltaA}), (\ref{deltaxi1}), (\ref{deltaxi2}), which guarantee the correct counting of physical degrees of freedom. As we shall discuss below these fields can be disposed of using their equations of motion.}
\item[ii)]{It is known that in $D$-dimensions there is a duality that relates $p$-forms to $(D-p-2)$-forms, the corresponding field strengths having complementary order and being related by a Hodge-like duality. In four dimensions vectors are dual to vectors, while scalars are dual to antisymmetric tensor fields. From this point of view, we can understand the 2-forms $B_\alpha$ as ``dual'' to the scalars in the same way as $A_{\Lambda}$ are ``dual'' to $A^\Lambda$. This relation can be schematically illustrated as follows:
    \begin{equation}
    \partial_{[\mu} B_{\nu\rho]}~\propto~ e\,\epsilon_{\mu\nu\rho\sigma}\partial^\sigma \phi~+~\dots\,.
    \end{equation}
   More precisely, we can write the non-local relation between $B_\alpha$ and $\phi^s$ in a $G$-covariant fashion as a Hodge-like duality between $\mathcal{H}^{(3)}_\alpha$ and the Noether current ${\bf j}_\alpha$ of the sigma model describing the scalar fields, associated with the generator $t_\alpha$:
  \begin{equation}
  \mathcal{H}_{\alpha\,\mu\nu\rho}\propto e\,\epsilon_{\mu\nu\rho\sigma}\,{\bf j}_\alpha^\sigma\;;\quad\;
  {\bf j}_\alpha^\mu\equiv \frac{\delta \Lagr_{\text{bos}}}{\delta \partial_\mu\phi^s}\,k_\alpha^s\,,\label{duaBphi}
  \end{equation}
  $k_\alpha^s$ being the Killing vector corresponding to $t_\alpha$. This motivated the choice of the 2-forms in the adjoint representation of $G$. In the gauged theory we will find a $G_g$-invariant version of (\ref{duaBphi}), see discussion below.}
\item[iii)]{It can be shown that the presence of the extra fields $B_\alpha$ and $A_{\Lambda}$ in the action is related to non-vanishing magnetic components $\Theta^{\Lambda\,\alpha}$ of the embedding tensor. In the electric frame in which $\Theta^{\Lambda\,\alpha}=0$, these fields disappear altogether from the Lagrangian and we are back to the gauged action described in Sect.\ \ref{gaugingsteps}.}
\item[iv)]{The kinetic terms in the Lagrangian only describe fields in the ungauged theory, while the extra fields enter topological terms or Stueckelberg-like couplings and satisfy first order equations, see discussion below. This feature is common to the $G$-covariant construction of gauged supergravities in any dimensions \cite{deWit:2004nw,deWit:2005hv,Samtleben:2005bp,deWit:2008ta}.}
\item[v)]{The dyonic embedding tensor $\Theta_M{}^\alpha$ determines a splitting of the $2n_v$ vector fields $A^M_\mu$ into the truly electric ones $A^{\hat{\Lambda}}_\mu$, which are singled out by the combination $A^M_\mu \Theta_M{}^\alpha$ and thus define the gauge connection. The remaining ones $\tilde{A}^M_\mu$ correspond to non-vanishing components of $Z^{M\,\alpha}$, that is to the components along which the Jacobi identity is not satisfied, see (\ref{nojacobi}). These latter vectors, of which there are at most $n_v$ independent, can be then written as $\tilde{A}^M_\mu=Z^{M\,\alpha} A_{\alpha\,\mu}$ and are ill-defined, since the corresponding field strengths do not satisfy the Bianchi identity. An other problem with the vectors $\tilde{A}^M_\mu$ is that they are not part of the gauge connection, but in general are charged under the gauge group, that is are minimally coupled to $A^{\hat{\Lambda}}_\mu$. These fields cannot therefore be consistently described as vector fields. However, this poses no consistency problem for the theory, since $\tilde{A}^M_\mu$ can be gauged away by a transformation (\ref{deltaxi1}), (\ref{deltaxi2}) proportional to $\Xi_\alpha$. In a vacuum, they provide the two degrees of freedom needed by some of the tensor fields $B_\alpha$ to become massive, according to the \emph{anti-Higgs} mechanism \cite{antiH1,Cecotti:1987qr}. In the electric frame, these vectors become magnetic ($A_{\hat{\Lambda}\,\mu}$) and disappear from the action. This phenomenon also occurs in higher dimensions: the vectors $\tilde{A}^M_\mu$ which do not participate in the gauge connection but are charged with respect to the gauge group, are gauged away by a transformation associated with some of the antisymmetric tensor fields which, in a vacuum, become massive.}
\item[vi)]{An important role in this construction was played by the linear constraint (\ref{linear2}), in particular by the property (\ref{linear22}) implied by it, which allowed to cancel the non-covariant terms in the gauge variation of $F^\Lambda$ by a corresponding variation of the antisymmetric tensor fields. It turns out that a condition analogous to (\ref{linear22}) represents the relevant linear constraint on the embedding tensor needed for the construction of gauged theories in higher dimensions \cite{deWit:2004nw,deWit:2005hv,Samtleben:2005bp,deWit:2008ta}.}
\end{enumerate}
\bigskip
Let us now briefly discuss the bosonic field equations for the antisymmetric tensor fields and the vectors. The variation of the action with respect to $B_{\alpha\,\mu\nu}$ yields equations (\ref{HG0}). By fixing the $\Xi_\alpha$-gauge freedom, we can gauge away the ill-defined vectors $\tilde{A}^M_\mu=Z^{M\,\alpha} A_{\alpha\,\mu}$ and then solve eqs.\ (\ref{HG0}) in $B_\alpha$ as a function of the remaining field strengths, which are a combination of the $F^{\hat{\Lambda}}$ only. Substituting this solution in the action, the latter will only describe the $A^{\hat{\Lambda}}_{\mu}$ vector fields and no longer contain magnetic ones or antisymmetric tensors. In other words by eliminating $B_\alpha$ through equations (\ref{HG0}) we effectively perform the rotation to the electric frame and find the action discussed in Sect.\ \ref{gaugingsteps}.\par
By varying the action with respect to $A^M_\mu$ we find the following equations:
\begin{equation}
\mathcal{D}_{[\mu}\Gd^M_{\rho\sigma]}=2\,e\,\mathbb{C}^{MN}\,\epsilon_{\mu\nu\rho\sigma}\,\mathcal{D}^\sigma \phi^s\Gm_{sr}\,k_N^r=2\,e\,\mathbb{C}^{MN}\,\epsilon_{\mu\nu\rho\sigma}\,{\bf j}_N^\sigma\,,\label{Max2}
\end{equation}
which are the manifestly $G$-covariant form of the Maxwell equations. The right-hand-side is proportional to the electric current
\begin{equation}
{\bf j}_N^\sigma ~\equiv~ \mathcal{D}^\sigma \phi^s\Gm_{sr}\,k_N^r
\=\Theta_N{}^\alpha\, \mathcal{D}^\sigma \phi^s\Gm_{sr}\,k_\alpha^r
\=\Theta_N{}^\alpha\,{\bf j}_\alpha^\sigma\,.
\end{equation}
If we contract both sides of (\ref{Max2}) with $\Theta_M{}^\alpha$, we are singling out the Bianchi identity for the fields strengths $F^{\hat{\Lambda}}$ of the vectors which actually participate in the minimal couplings. By using the locality condition on $\Theta$, we find:
\begin{equation}
\mathcal{D}_{[\mu}\Gd^M_{\rho\sigma]}\,\Theta_M{}^\alpha=2\,e\,\mathbb{C}^{MN}\,\Theta_M{}^\alpha\,\Theta_N{}^\beta\epsilon_{\mu\nu\rho\sigma}\,\mathcal{D}^\sigma \phi^s\Gm_{sr}\,k_\beta^r=0\,,\label{Bianchigauge}
\end{equation}
which are nothing but the Bianchi identities for $F^{\hat{\Lambda}}$. This is consistent with our earlier discussion, see eq.\ (\ref{BianchiFgauge}), in which we showed that the locality condition implies that the Bianchi identity for the gauge curvature have no magnetic source term, so that the gauge connection is well defined%
\footnote{
in our earlier discussion we showed that $\mathcal{D}\mathcal{H}^M\,\Theta_M{}^\alpha=\mathcal{D}F^M\,\Theta_M{}^\alpha=0$. This is consistent with eq.\ (\ref{Bianchigauge}) since on-shell $\mathcal{H}^M\Theta_M{}^\alpha=\Gd^M\Theta_M{}^\alpha$
}.\par
Now we can use the Bianchi identity (\ref{Bid1n}) to rewrite eq.\ (\ref{Bianchigauge}) as a dualization equation generalizing (\ref{duaBphi}). To this end, we consider only the upper components of (\ref{Bianchigauge}), corresponding to the field equations for $A_{\Lambda\,\mu}$:
\begin{equation}
Z^{\Lambda\alpha}\,\mathcal{H}_{\alpha\,\mu\nu\rho}=12\,e\,Z^{\Lambda\alpha}\,\epsilon_{\mu\nu\rho\sigma}\,
\mathcal{D}^\sigma \phi^s\Gm_{sr}\,k_\alpha^r\,.\label{Hduaga}
\end{equation}
When the gauging involves translational isometries \cite{deWit:2005ub}, $\phi^I\rightarrow \phi^I+c^I$, the above equations can be solved in the fields $A_\Lambda$ contained in the covariant derivative. This is done by first using the $\zeta$-gauge freedom associated with $A_\Lambda$ to gauge away the scalar fields $\phi^I$ acted on by the translational isometries. Eqs.\ (\ref{Hduaga}) are then solved in the fields $A_\Lambda$, which are expressed in terms of the remaining scalars, the vectors $A^\Lambda$ and the field strengths of the antisymmetric tensors. Substituting this solution in the action, we obtain a theory in which no vectors $A_\Lambda$ appear and the scalar fields $\phi^I$ have been effectively dualized to corresponding tensor fields $B_{I\,\mu\nu}$. The latter become dynamical and are described by kinetic terms.
These theories were first constructed in the framework of $\N=2$ supergravity in \cite{Dall'Agata:2003yr,D'Auria:2004yi}, generalizing previous results \cite{Louis:2002ny}.\par
The gauged theory we have discussed in this section features a number of non-dynamical extra fields. This is the price we have to pay for a manifest $G$-covariance of the field equations and Bianchi identities.
The embedding tensor then defines how the physical degrees of freedom are distributed within this larger set of fields, by fixing the gauge symmetry associated with the extra fields and solving the corresponding non-dynamical field equations (\ref{HG0}), (\ref{Hduaga}).

\paragraph{A view on higher dimensions.} As mentioned in point ii) above, there are equivalent formulations of ungauged supergravities in $D$-dimensions obtained from one another by dualizing certain $p$-forms $C_{(p)}$ (i.e.\ rank-$p$ antisymmetric tensor fields) into $(D-p-2)$-forms $C_{(D-p-2)}$ through an equation of the type:
\begin{equation}
dC_{(p)}={}^*dC_{(D-p-2)}~+~\dots\,.
\end{equation}
Such formulations feature in general different global symmetry groups. This phenomenon is called \emph{Dualization of Dualities} and was studied in \cite{Cremmer:1997ct}. The scalar fields in these theories are still described by a non-linear sigma model and in $D\ge 6$ the scalar manifold is homogeneous symmetric. Just as in four dimensions, the scalars are non-minimally coupled to the $p$-form fields (see below) and the global symmetry group $G$ is related to the isometry group of the scalar manifold and thus is maximal in the formulation of the theory in which the scalar sector is maximal, that is in which all forms are dualized to lower order ones. This prescription, however, does not completely fix the ambiguity related to duality in even dimensions $D=2k$, when order-$k$ field strengths, corresponding to rank-$(k-1)$ antisymmetric tensor fields $C_{(k-1)}$, are present. In fact, after having dualized all forms to lower-order ones, we can still dualize $(k-1)$-forms $C_{(k-1)}$ into $(k-1)$-forms $\tilde{C}_{(k-1)}$. This is the electric-magnetic duality of the four-dimensional theory, related to the vector fields, and also occurs for instance in six dimensions with the 2-forms and in eight dimensions with the 3-forms.\par
Duality transformations interchanging  $C_{(k-1)}$ with $\tilde{C}_{(k-1)}$, and thus the corresponding field equations with  Bianchi identities, are encoded in the group $G$, whose action on the scalar fields, just as in four dimensions, is combined with a linear action on the $k$-form field strengths ${F}_{(k)}$ and their duals $\tilde{F}_{(k)}$:
\begin{align}
g\in G \;:\quad
\begin{cases}
{F}_{(k)}\rightarrow {F}_{(k)}'=A[g]\,{F}_{(k)}+B[g]\,\tilde{F}_{(k)}\,,\cr \tilde{F}_{(k)}\rightarrow \tilde{F}_{(k)}'=C[g]\,{F}_{(k)}+D[g]\,\tilde{F}_{(k)}\,.
\end{cases}
\end{align}
As long as the block $B[g]$ is non-vanishing, this symmetry can only be on-shell since the Bianchi identity for the transformed ${F}_{(k)}$, which guarantees that the transformed elementary field $C_{(k-1)}'$ be well defined, only holds if the field equations $d\tilde{F}_{(k)}=0$ for $C_{(k-1)}$ are satisfied \cite{Tanii:1998px}:
\begin{equation}
d{F}_{(k)}'=A[g]\,d{F}_{(k)}+B[g]\,d\tilde{F}_{(k)}=B[g]\,d\tilde{F}_{(k)}=0\,.
\end{equation}
The field strengths ${F}_{(k)}$ and $\tilde{F}_{(k)}$ transform in a linear representation $\mathscr{R}$ of $G$ defined by the matrix:
\begin{equation}
g\in G\;\;\stackrel{\mathscr{R}}{\longrightarrow }\;\;\mathscr{R}[g]=\left(\begin{matrix} A[g] & B[g]\cr C[g] & D[g] \end{matrix}\right)\,.\label{RgD}
\end{equation}
Just as in four dimensions, depending on which of the $C_{(k-1)}$ and $\tilde{C}_{(k-1)}$ are chosen to be described as elementary fields in the Lagrangian, the action will feature a different global symmetry $G_{el}$, though the global symmetry group $G$ of the field equations and Bianchi identities remains the same.
The constraints on $\mathscr{R}$ derive from the non-minimal couplings of the scalar fields to the $(k-1)$-forms which are a direct generalization of those in four dimensions between the scalars and the vector fields%
\footnote{
the Hodge dual ${}^*\omega$ of a generic $q$-form $\omega$ is defined as:
\begin{equation}
{}*\omega_{\mu_1\dots \mu_{D-q}}=\frac{e}{q!}\,\epsilon_{\mu_1\dots\mu_{D-q}\nu_1\dots \nu_q}\,\omega^{\nu_1\dots \nu_q}\,,
\end{equation}
where $\epsilon_{01\dots D-1}=1$.\; One can easily verify that ${}^{**}\omega=(-)^{q(D-q)}\,(-)^{D-1}\,\omega$
},
see (\ref{bosoniclagr})
\begin{equation}
\Lagr_{{\rm kin},\,C}=
-\frac{e\varepsilon}{2k!}\left(\I_{\Lambda\Sigma}(\phi)\,F^\Lambda_{\mu_1\dots \mu_k}\,F^{\Lambda\,\mu_1\dots \mu_k}+\R_{\Lambda\Sigma}(\phi)\,F^\Lambda_{\mu_1\dots \mu_k}\,{}^*F^{\Lambda\,\mu_1\dots \mu_k}\right)\,,\label{kinC}
\end{equation}
where $\mu=0,\dots, D-1$\, and  \,$\Lambda,\Sigma=1,\dots, n_k$, being $n_k$ the number of $(k-1)$-forms $C_{(k-1)}$\, and \,$\varepsilon\equiv (-)^{k-1}$.\par\smallskip
The matrices $\I_{\Lambda\Sigma}(\phi),\,\R_{\Lambda\Sigma}(\phi)$ satisfy the following properties:
 \begin{equation}
 \I_{\Lambda\Sigma}=\I_{\Sigma\Lambda}<0\,,\quad\; \R_{\Lambda\Sigma}=-\varepsilon\,\R_{\Sigma\Lambda}\,.
 \end{equation}
Just as we did in four dimensions, see eq.\ (\ref{GF}), we define dual field strengths (omitting the fermion terms):
\begin{equation}
 \Gd_{\Lambda\,\mu_1\dots\,\mu_k}\equiv\varepsilon\,\epsilon_{\mu_1\dots\,\mu_k\nu_1\dots\nu_k}\,
 \frac{\delta \Lagr}{\delta F^\Lambda_{\nu_1\dots \nu_k}}
 \quad\Rightarrow\quad
 \Gd_\Lambda=-\I_{\Lambda\Sigma}\,{}^*F^\Sigma-\varepsilon\, \R_{\Lambda\Sigma}\,F^\Sigma\,,\label{defGk}
\end{equation}
and define the vector of field strengths:
\begin{align}
\mathbb{F}= (\mathbb{F}^M)\equiv\left(\begin{matrix} F^\Lambda \cr \Gd_\Lambda\end{matrix}\right)\,.
\end{align}
The definition (\ref{defGk}) can be equivalently written in terms of the \emph{twisted self-duality condition} \cite{Cremmer:1997ct}:
\begin{equation}
 {}^*\mathbb{F}=-\mathbb{C}_\varepsilon\,\mathcal{M}(\phi)\,\mathbb{F}\,,\label{TSDCD}
\end{equation}
which generalizes (\ref{FCMF}), where
\begin{equation}
\mathbb{C}_\varepsilon\equiv (\mathbb{C}^{MN})\equiv
\left(\begin{matrix}
\Zero & \Id   \cr
\varepsilon\,\Id & \Zero
\end{matrix}\right)
\,,\label{Ce}
\end{equation}
$\Id$, $\Zero$ being the $n_k\times n_k$ identity and zero-matrices, respectively, and
\begin{equation}
\mathcal{M}(\phi)= (\mathcal{M}(\phi)_{MN})\equiv
\left(\begin{matrix}(\I-\varepsilon\,\R\I^{-1}\R)_{\Lambda\Sigma} &
-(\R\I^{-1})_\Lambda{}^\Gamma\cr \varepsilon (\I^{-1}\R)^\Delta{}_\Sigma & \I^{-1\,
\Delta \Gamma}\end{matrix}\right)\,.\label{Me}
\end{equation}
The reader can easily verify that:
\begin{equation}
\mathcal{M}^T\,\mathbb{C}_\varepsilon\mathcal{M}=\mathbb{C}_\varepsilon\,.
\end{equation}
For $\varepsilon=-1$, which is the case of the vector fields in four dimensions, $\mathbb{C}_\varepsilon$ is the symplectic invariant matrix
and $\mathcal{M}$ is a symmetric, symplectic matrix. For $\varepsilon=+1$, which is the case of 2-forms in six dimensions,
$\mathbb{C}_\varepsilon$ is the ${\rm O}(n_k,n_k)$-invariant matrix and $\mathcal{M}$ a symmetric element of ${\rm O}(n_k,n_k)$.\par
The Maxwell equations read:
\begin{equation}
d\mathbb{F}=0\,.\label{MDd}
\end{equation}
In order for (\ref{RgD}) to be a symmetry of eqs.\ (\ref{TSDCD}) and (\ref{MDd}) we must have:
\begin{equation}
\mathcal{M}(g\star \phi)=\mathscr{R}[g]^{-T}\mathcal{M}( \phi)\mathscr{R}[g]^{-1}\,,
\end{equation}
and
\begin{equation}
\mathscr{R}[g]^{T}\mathbb{C}_\varepsilon \mathscr{R}[g]=\mathbb{C}_\varepsilon\,.
\end{equation}
This means that in $D=2k$ dimensions:
\begin{align}
\text{$k$ even}\,&:\quad\; \mathscr{R}[G]\subset {\rm Sp}(2n_k,\mathbb{R})\,,\nonumber\\
\text{$k$ odd}\,&:\quad\; \mathscr{R}[G]\subset {\rm O}(n_k,n_k)\,.
\end{align}
All other forms of rank $p\neq k-1$, which include the vector fields in $D>4$, will transform in linear representations of $G$. The corresponding kinetic Lagrangian only feature the first term of (\ref{kinC}), with no generalized theta-term ($\R=0$).\par
If we compactify Type IIA/IIB or eleven-dimensional supergravity on a torus down to $D$-dimensions, we end up with an effective ungauged, maximal theory in $D$ dimensions, featuring form-fields of various order. Upon dualizing all form-fields to lower order ones, we end up with a formulation of the theory in which $G$ is maximal, and is described by the non-compact real form ${\rm E}_{11-D(11-D)}$ of the group ${\rm E}_{11-D}$. Here we use the symbol ${\rm E}_{11-D(11-D)}$ as a short-hand notation for the following groups:
\begin{align}
D&=9 \;:\quad\; G={\rm E}_{2(2)}\equiv {\rm GL}(2,\mathbb{R})\,,\nonumber\\
D&=8 \;:\quad\; G={\rm E}_{3(3)}\equiv {\rm SL}(2,\mathbb{R})\times {\rm SL}(3,\mathbb{R})\,,\nonumber\\
D&=7 \;:\quad\; G={\rm E}_{4(4)}\equiv {\rm SL}(5,\mathbb{R})\,,\nonumber\\
D&=6 \;:\quad\; G={\rm E}_{5(5)}\equiv {\rm SO}(5,5)\,,\nonumber\\
D&=5 \;:\quad\; G={\rm E}_{6(6)}\,,\nonumber\\
D&=4 \;:\quad\; G={\rm E}_{7(7)}\,,\nonumber\\
D&=3 \;:\quad\; G={\rm E}_{8(8)}\,.
\end{align}
Only for $D\le 5$, ${\rm E}_{11-D(11-D)}$ is a proper exceptional group. The ungauged four-dimensional maximal supergravity was originally obtained from compactification of the eleven-dimensional one and dualization of all form-fields to lower order ones in \cite{Cremmer:1978ds}, where the ${\rm E}_{7(7)}$ on-shell symmetry was found.\par
In $D=10$ Type IIA and IIB theories feature different global symmetry groups: $G_{IIA}={\rm SO}(1,1)$ and $G_{IIB}= {\rm SL}(2,\mathbb{R})$, respectively.
The latter encodes the conjectured S-duality symmetry of Type IIB string theory. In this theory $G_{IIB}$ does not act as a duality group since the 5-form field strength is self-dual and is a $G_{IIB}$-singlet.\par
A $G$-covariant gauging \cite{deWit:2004nw,deWit:2005hv,Samtleben:2005bp,deWit:2008ta} is effected starting from the formulation of the ungauged theory in which $G$ is maximal and promoting a suitable global symmetry group of the Lagrangian $G_g\subset G$ to local symmetry. The choice of the gauge group is still completely encoded in a $G$-covariant embedding tensor $\Theta$:
\begin{equation}
\Theta\in \mathscr{R}_{v*}\times {\rm adj}(G)\,,\label{rtr}
\end{equation}
subject to a linear constraint, generalizing (\ref{linear22}), which singles out in the above product a certain representation $\mathscr{R}_\Theta$ for the embedding tensor, and a quadratic one expressing the $G_g$-invariance of $\Theta$.
In Table \ref{tab:T-tensor-repr} we give, in the various $D$-dimensional maximal supergravities, the representations $\mathscr{R}_\Theta$ of $\Theta$.\par
Just as in the duality covariant construction of the four-dimensional gaugings discussed above, one introduces all form-fields which are dual to the fields of the ungauged theory. All the form-fields will transform in representations of $G$ and dual forms of different order will belong to conjugate representations. In $D=2k$, in the presence of rank-$(k-1)$ antisymmetric tensors, this amounts to introducing the fields $\tilde{C}_{(k-1)\,\Lambda}$ dual to the elementary ones ${C}^\Lambda_{(k-1)}$, just as we did for the vector fields in four dimensions. Together they transform in the representation $\mathscr{R}$ discussed above. By consistency, each form-field is associated with its own gauge invariance. Only the fields of the original ungauged theory are described by kinetic terms, the extra fields enter in topological terms and in Stueckelberg-like combinations within the covariant field strengths. The latter, for a generic $p$-form field, can be schematically represented in the form (we suppress all indices)
\begin{equation}
F_{(p+1)}=\mathcal{D}C_{(p)}+Y_p[\Theta]\cdot C_{(p+1)}+\dots\,.
\end{equation}
where $Y_p[\Theta]$ is a constant \emph{intertwiner} tensor constructed out of $\Theta$ and of $G$-invariant tensors.
The gauge variation of the $p$-form has the following schematic expression:
\begin{equation}
\delta C_{(p)}=Y_p[\Theta]\cdot\Xi_{(p)}+\mathcal{D}\Xi_{(p-1)}+\dots\label{gfree}
\end{equation}
The embedding tensor defines, through the tensors $Y_p[\Theta]$, a splitting of the $p$-forms into physical fields and unphysical ones. The former will in general become massive by ``eating'' corresponding unphysical $(p-1)$-forms, while the latter, whose field strengths fail to satisfy the Bianchi identity, are in turn gauged away and become degrees of freedom of massive $(p+1)$-forms. The constraints on the embedding tensor and group theory guarantee the consistency of the whole construction.\par
Just as in the four-dimensional model discussed above, the embedding tensor defines the distribution of the physical degrees of freedom among the various fields by fixing the gauge freedom (\ref{gfree}) and solving the non-dynamical field equations. These $G$-covariant selective couplings between forms of different order, determined by a single object $\Theta$, define the so-called \emph{tensor hierarchy} and was developed in the maximal theories, in \cite{deWit:2005hv,Samtleben:2005bp,deWit:2008ta} as a general $G$-covariant formulation of the gauging procedure in any dimensions. In this formalism the maximal gauged supergravity in $D=5$ was constructed in \cite{deWit:2004nw}, generalizing previous works \cite{Gunaydin:1985cu,Andrianopoli:2000fi}. The general gauging of the six and seven -dimensional maximal theories were constructed in \cite{Bergshoeff:2007ef} and \cite{Samtleben:2005bp} respectively, extending previous works \cite{Pernici:1984fe}. In $D=8$ the most general gaugings were constructed in \cite{Bergshoeff:2003ri}. Finally, the $D=9$ gauged theory was studied in \cite{Cowdall:2000sq,Bergshoeff:2002mb,Hull:2002wg}. We refer to these works for the details of the construction in the different cases.

\begin{table}[t]
\centering
\renewcommand{\arraystretch}{1.7}
\begin{tabular}{ | M{0.5cm} | M{2cm} | M{3cm} | M{6cm} |}
\hline
$D$ &$G$& $H$ & $\Theta$  \\   \hline
7   & ${\rm SL}(5)$ & ${\rm USp}(4)$  & ${\bf 10}\times {\bf
  24}= {\bf 10}+\underline{\bf 15}+  \underline{\overline{\bf 40}}+
{\bf 175}$
\\
6  & ${\rm SO}(5,5)$ & ${\rm USp}(4) \times {\rm USp}(4)$ &
  ${\bf 16}\times{\bf 45} =
  {\bf 16}+ \underline{\bf 144} + {\bf 560}$ \\
5   & ${\rm E}_{6(6)}$ & ${\rm USp}(8)$ & ${\bf 27}\times{\bf
  78} =
  {\bf 27} + \underline{\bf 351} + \overline{\bf 1728}$  \\
4   & ${\rm E}_{7(7)}$ & ${\rm SU}(8)$  & ${\bf 56}\times{\bf 133} =
  {\bf 56} + \underline{\bf 912} + {\bf 6480}$   \\
3   & ${\rm E}_{8(8)}$ & ${\rm SO}(16)$ & ${\bf 248}\times{\bf 248} =
  \underline{\bf 1} + {\bf 248} + \underline{\bf 3875} +{\bf 27000}
  +  {\bf 30380}$
\\ \hline
\end{tabular}
\caption{\small
Decomposition of the embedding tensor $\Theta$ for maximal
supergravities in various space-time dimensions in terms of irreducible
${\rm G}$ representations
\cite{deWit:2002vt,deWit:2005hv}. Only the underlined representations
are allowed by supersymmetry. The R-symmetry group ${\rm H}$
is the maximal compact subgroup of ${\rm G}$.
}\label{tab:T-tensor-repr}
\end{table}


\newpage

\hypersetup{linkcolor=blue}
\phantomsection 
\addtocontents{toc}{\protect\addvspace{4.5pt}}
\addcontentsline{toc}{section}{References} 
\bibliographystyle{mybibstyle}
\bibliography{bibliografia} 

\end{document}

%% file: Structure.tex
\usepackage[utf8]{inputenc} 
\usepackage{lipsum} 

\usepackage[top=3cm, bottom=3.5cm, left=3cm, right=3cm,
           heightrounded,marginparwidth=1.5cm, marginparsep=1cm]{geometry} 

\usepackage[shortlabels]{enumitem} 

\usepackage[square,numbers,merge,comma,sort&compress]{natbib} 
\makeatletter
\def\NAT@spacechar{\,}  
\makeatother

\usepackage{amsmath,amssymb,amsfonts,amsthm}
\usepackage{mathtools} 
\usepackage{old-arrows}
\usepackage[scale=0.7]{miama} 
\usepackage{slashed,cancel}
\usepackage{comment}   
\usepackage{relsize}   
\usepackage{setspace}  
\usepackage{moresize}  
\usepackage{epsfig}
\usepackage{latexsym}
\usepackage{mathrsfs,calligra,aurical} 
\usepackage{calc}
\usepackage{float}
\usepackage{appendix}
\usepackage{xargs}
\usepackage{extarrows}
\usepackage{empheq}

\usepackage[blocks]{authblk}  

\usepackage{tikz}
\usetikzlibrary{shapes,arrows,scopes,decorations.pathmorphing,patterns,calc}

\usepackage[fulladjust]{marginnote}%

\usepackage{graphicx} 
\graphicspath{{Figures/}} 

\usepackage[labelsep=colon]{caption}  
\usepackage[labelsep=colon,aboveskip=10pt,belowskip=10pt]{subcaption}
\usepackage{array,multirow,makecell,booktabs}  
\newcolumntype{X}[2]{>{\centering\arraybackslash$}#1{#2\linewidth}<{$}}
\newcolumntype{R}[1]{>{\raggedleft\arraybackslash$}m{#1\linewidth}<{$}}
\newcolumntype{L}[1]{>{\raggedright\arraybackslash}m{#1\linewidth}}

\makeatletter
\renewcommand\mcell@classz{\@classx
   \@tempcnta \count@
   \prepnext@tok
   \@addtopreamble{
      \ifcase\@chnum
         \hfil
         \mcell@agape{\d@llarbegin\insert@column\d@llarend}\hfil \or
         \hskip1sp
         \mcell@agape{\d@llarbegin\insert@column\d@llarend}\hfil \or
         \hfil\hskip1sp
         \mcell@agape{\d@llarbegin \insert@column\d@llarend}\or
         \mcell@agape{$\vcenter
         \@startpbox{\@nextchar}\insert@column\@endpbox$}\or
         \mcell@agape{\vtop
         \@startpbox{\@nextchar}\insert@column\@endpbox}\or
         \mcell@agape{\vbox
         \@startpbox{\@nextchar}\insert@column\@endpbox}%
      \fi
      \global\let\mcell@left\relax\global\let\mcell@right\relax
    }\prepnext@tok}
\makeatother

\makeatletter
\renewcommand\section{\@startsection{section}{1}{\z@}%
                                   {-3.5ex \@plus -1.3ex \@minus -.7ex}%
                                   {2.3ex \@plus.4ex \@minus .4ex}%
                                   {\normalfont\Large\bfseries}}
\renewcommand\subsection{\@startsection{subsection}{2}{\z@}%
                                   {-2.3ex\@plus -1ex \@minus -.5ex}%
                                   {1.2ex \@plus .3ex \@minus .3ex}%
                                   {\normalfont\large\bfseries}}
\renewcommand\subsubsection{\@startsection{subsubsection}{3}{\z@}%
                                   {-2.3ex\@plus -1ex \@minus -.5ex}%
                                   {1ex \@plus .2ex \@minus .2ex}%
                                   {\normalfont\normalsize\bfseries}}
\renewcommand\paragraph{\@startsection{paragraph}{4}{\z@}%
                                   {1.75ex \@plus1ex \@minus.2ex}%
                                   {-1em}%
                                   {\normalfont\normalsize\bfseries}}
\renewcommand\subparagraph{\@startsection{subparagraph}{5}{\z@}
                                   {1.75ex \@plus1ex \@minus .2ex}%
                                   {-1em}%
                                   {\normalfont\normalsize\itshape}}
\makeatother

\usepackage{titletoc}

\addtocontents{toc}{\addvspace{-0.75em}}  

\titlecontents{section}
  [1.25em] {\addvspace{0.7em plus 0pt}\small}
  {\thecontentslabel\hspace{0.75em}}{}
  {\hspace{0.5em}\titlerule*[0.5em]{.}\contentspage}
  [\addvspace{0.0em plus 0pt}]

\titlecontents{subsection}
  [2.75em] {\addvspace{0.075em plus0pt}\fns}
  {\thecontentslabel\hspace{0.75em}}{\thecontentslabel\hspace{0.75em}}
  {\hspace{0.5em}\titlerule*[0.5em]{.}\small\contentspage}
  [\addvspace{0.075em plus 0pt}]

\usepackage{environ}
\makeatletter
\NewEnviron{subalign}[1]{
\begin{subequations}\label{#1}
\begin{align} \BODY \end{align}
\end{subequations}      }
\makeatother
\makeatletter
{\begingroup%
\setlength{\abovedisplayskip}{10pt plus 4pt minus 9pt}%
\setlength{\abovedisplayshortskip}{0pt plus 2pt minus 2pt}%
\setlength{\belowdisplayskip}{12pt plus 3pt minus 9pt}%
\setlength{\belowdisplayshortskip}{7pt plus 3pt minus 4pt}%
\begin{subequations}%
}%
{\end{subequations}\ignorespacesafterend%
\endgroup}%
\makeatother

\makeatletter
{\begingroup%
\setlength{\abovedisplayskip}{{#1}pt plus 2pt minus 9pt}%
\setlength{\abovedisplayshortskip}{0pt plus 0pt minus 2pt}%
\setlength{\belowdisplayskip}{{#2}pt plus 3pt minus 9pt}%
\setlength{\belowdisplayshortskip}{7pt plus 3pt minus 4pt}%
\begin{subequations}%
}%
{\end{subequations}\ignorespacesafterend%
\endgroup}%
\makeatother

\makeatletter
{\begingroup%
\setlength{\abovedisplayskip}{{#1}pt plus 3pt minus 9pt}%
\setlength{\abovedisplayshortskip}{0pt plus 3pt}%
\setlength{\belowdisplayskip}{{#2}pt plus 3pt minus 9pt}%
\setlength{\belowdisplayshortskip}{7pt plus 3pt minus 4pt}%
\begin{equation}%
}%
{\end{equation}\ignorespacesafterend%
\endgroup}%
\makeatother

\makeatletter
{%
\begin{equation}%
\begin{split}{#1}%
}%
{\end{split}%
\end{equation}\ignorespacesafterend%
}%
\makeatother

\usepackage{xcolor}
\definecolor{Green}{rgb}{0.05, 0.45, 0.25}
\definecolor{dogwoodrose}{rgb}{0.8, 0.1, 0.55}
\definecolor{RRed}{rgb}{0.7, 0.1, 0.525}

\usepackage{bm}  
\usepackage{dsfont}  

\DeclareMathAlphabet{\mathpzc}{OT1}{pzc}{m}{it}
\DeclareMathAlphabet{\mathcal}{OMS}{cmsy}{m}{n}
\DeclareSymbolFontAlphabet{\Scr}{rsfs}
\DeclareMathAlphabet{\mathbold}{U}{BOONDOX-ds}{m}{n}
\SetMathAlphabet{\mathbold}{bold}{U}{BOONDOX-ds}{b}{n}
\DeclareMathAlphabet{\mathcalboondox}{U}{BOONDOX-calo}{m}{n}
\SetMathAlphabet{\mathcalboondox}{bold}{U}{BOONDOX-calo}{b}{n}
\DeclareMathAlphabet{\mathbcalboondox}{U}{BOONDOX-calo}{b}{n}

\newcommand\eqlinkcol{RRed}



\usepackage[breaklinks=true,backref=page]{hyperref}
\hypersetup{
    bookmarks=true,         
    pdfmenubar=true,        
    pdffitwindow=false,     
    pdfpagemode={UseNone},
    pdfstartview={FitH},
    pdfauthor={},     
    pdfsubject={},   
    pdfcreator={},   
    pdfproducer={},  
    colorlinks=true,
    bookmarks=true,
    bookmarksnumbered=true,
    plainpages,
    a4paper,
    linktoc=page,
    citecolor=blue,
    filecolor=black,
    linkcolor=\eqlinkcol,
    urlcolor=Green,
}
\renewcommand*{\backref}[1]{}
\renewcommand*{\backrefalt}[4]{%
\ifcase #1 %
\relax
\or
~{\small [\textsc{p.~\fns{\!#2}}]}
\else
~{\small [\textsc{p.~\fns{\!#2}}]}%
\fi}

\usepackage{footnotebackref}
\usepackage{hypernat} 

\makeatletter
\newenvironment{sqcases}{%
  \matrix@check\sqcases\env@sqcases
}{%
  \endarray\right.%
}
\def\env@sqcases{%
  \let\@ifnextchar\new@ifnextchar
  \left\lbrack
  \def\arraystretch{1.2}%
  \array{@{}l@{\quad}l@{}}%
}
\makeatother

\newcolumntype{M}[1]{>{\centering\arraybackslash}m{#1}}

\def\N{\mathcal{N}}
\def\Ms{\mathscr{M}}
\def\Msi{\mathscr{M}_{\text{int}}}
\def\I{\mathcal{I}}
\def\R{\mathcal{R}}

\def\+{~+~}
\def\-{~-~}
\def\={~=~}
\def\*{{}^*}
\def\nn{\nonumber}
\def\nne{\nonumber\\}
\def\eps{\epsilon}

\def\GN{\mathrm{G}_{\textsc{n}}}
\def\Gm{\mathcalboondox{G}}
\def\LB{\mathscr{L}_{\text{bos}}}
\def\Lscal{\mathscr{L}_{\text{scal}}}
\def\Lgaug{\mathscr{L}_{\text{gauged}}}
\def\Lagr{\mathscr{L}}
\def\CC{\mathbb{C}}
\def\Lc{\textsl{L}}
\def\LL{\mathbb{L}}
\def\PP{\mathcal{P}}
\def\Ps{\mathscr{P}}
\def\Es{\mathscr{E}}
\def\W{\mathpzc{w}}
\def\Solv{\mathscr{S}}
\def\galg{\mathfrak{g}}
\def\halg{\mathfrak{H}}
\def\kalg{\mathfrak{K}}
\def\Tr{\mathrm{Tr}}
\def\M{\mathcal{M}}

\def\cD{\mathcal{D}}
\def\Gd{\mathpzc{G}}
\def\Gg{G_g}
\newcommand\Id{\mathds{1}}
\def\Zero{\mathbf{0}}
\def\Or{\mathcal{O}}
\def\Mscal{\mathscr{M}_{\rm scal}}
\def\Mink{{\mathbf{M_4}}^{\scalebox{0.6}{(1,3)}}}
\def\MsSK{\Ms_{\textsc{sk}}}
\def\MsQK{\Ms_{\textsc{qk}}}
\def\Rs{\mathscr{R}}
\def\Tb{\mathbb{T}}
\def\S{{\bf S}}
\def\Lambdae{{\Lambda_e}}

\def\dim{\mathrm{dim}}
\def\rank{\mathrm{rank}}
\def\Adj{\mathrm{adj}}
\def\AdS{\text{AdS}}
\def\Sp{\mathrm{Sp}}

\def\SU{\mathrm{SU}}
\def\SO{\mathrm{SO}}

\newcommand\fns{\footnotesize}

\newcommandx{\dmu}[1][1=\mu,usedefault]{\partial_{#1}}
\newcommand{\ov}[1]{{\overline{#1}}}
\newcommand{\und}[1]{{\underline{#1}}}
\newcommandx\AL[2][1=\Lambda,2=\mu,usedefault]{A^{#1}_{#2}}
\newcommandx\ALd[2][1=\Lambda,2=\mu,usedefault]{A_{{#1}\,{#2}}}
\newcommandx\ALe[2][1=\Lambdae,2=\mu,usedefault]{A^{#1}_{#2}}
\newcommandx\ALed[2][1=\Lambdae,2=\mu,usedefault]{A_{{#1}\,{#2}}}
\newcommandx\AM[2][1=M,2=\mu,usedefault]{A^{#1}_{#2}}
\newcommandx\XLe[1][1=\Lambdae,usedefault]{X_{#1}}
\newcommandx{\Th}[2][1=M,2=\alpha,usedefault]{\Theta_{#1}{}^{#2}}
\newcommandx{\T}[2][1=M,2=\alpha,usedefault]{\varmathbb{T}_{\und{#1}}{}^{\und{#2}}}
\newcommandx{\TT}[3][1=M,2=N,3=P,usedefault]{\varmathbb{T}_{\und{#1#2}}{}^{\und{#3}}}
\newcommandx{\LCu}[4][1=\mu,2=\nu,3=\rho,4=\sigma,usedefault]{\varepsilon^{{#1}{#2}{#3}{#4}}}
\newcommandx{\LCd}[4][1=\mu,2=\nu,3=\rho,4=\sigma,usedefault]{\varepsilon_{{#1}{#2}{#3}{#4}}}
\newcommandx{\Exc}[1][1=7,usedefault]{{\rm E}_{{#1}({#1})}}